\documentclass[structabstract]{aa}  

\usepackage{amssymb,amsmath}
\usepackage{graphicx}
\usepackage{xcolor}
\usepackage{url}
\usepackage{longtable}

\usepackage{txfonts}
\usepackage{natbib}
\bibpunct{(}{)}{;}{a}{}{,} % to follow the A&A style
\usepackage{ulem}
\usepackage{url}
\begin{document}
\title{Asteroseismology for ``\`{a} la carte'' stellar age-dating and weighing}
\subtitle{Age and mass of the CoRoT exoplanet host HD~52265}

   \author{Y. Lebreton
          \inst{1, 2}
          \and
          M.J. Goupil\inst{3}
          }
   \institute{Observatoire de Paris, GEPI, CNRS UMR 8111, F-92195 Meudon, France
         \and
             Institut de Physique de Rennes, Universit\'e de Rennes 1, CNRS UMR 6251, F-35042 Rennes, France \\
              \email{yveline.lebreton@obspm.fr}
         \and
             Observatoire de Paris, LESIA, CNRS UMR 8109, F-92195 Meudon, France
%             \email{yveline.lebreton@obspm.fr ; mariejo.goupil@obspm.fr}             
             }
   \date{Received , 2012; accepted , 2012}

% \abstract{}{}{}{}{} 
% 5 {} token are mandatory
 
  \abstract
  % context heading (optional)
{In the context of the space missions CoRoT, {\it Kepler}, Gaia, TESS, and PLATO, precise and accurate 
stellar ages, masses, and radii are of paramount importance. 
For instance, they are crucial for constraining scenarii of planetary formation and evolution.}
  %  aims heading (mandatory)
{We aim at quantifying how detailed stellar modelling 
can improve the accuracy and precision on age and mass of
individual stars. To that end, we adopt a multifaceted approach 
where we carefully examine how the number of observational 
constraints as well as the uncertainties on observations and on model input physics affect
the results of age-dating and weighing. }
   % methods heading (mandatory)
 {We modelled\thanks{Tables 4, 5, A.2, A.3, A.4, A.5 are also available in electronic form 
at the CDS via anonymous ftp to cdsarc.u-strasbg.fr (130.79.128.5) or via 
http://cdsweb.u-strasbg.fr/cgi-bin/qcat?J/A+A/}
in detail the exoplanet host-star HD~52265, 
a main-sequence, solar-like oscillator that CoRoT observed for four months. 
We considered different sets of observational constraints 
(Hertzsprung-Russell data, metallicity, various sets of seismic constraints). 
For each case, we determined the age, mass, and properties of HD~52265 inferred from stellar models, 
and we quantified the impact of the model input physics and free parameters. 
We also compared model ages with ages derived by empirical methods or 
Hertzsprung-Russell diagram inversion.}
  % results heading (mandatory)
{For our case study HD~52265, our seismic analysis provides an age $A=2.10-2.54$ Gyr, 
a mass $M=1.14-1.32\ M_\odot$, and a radius $R=1.30-1.34\ R_\odot$, which corresponds to 
age, mass, and radius uncertainties of $\sim{10}$, $\sim{7}$, and $\sim{1.5}$ per cent, respectively.
These uncertainties account for observational errors and current state-of-the-art stellar model uncertainties. 
Our seismic study also provides constraints on surface convection properties 
through the mixing-length, which we find to be $12-15$ per cent lower than the solar value. 
On the other hand, because of helium-mass degeneracy, the initial 
helium abundance is determined modulo the mass value. 
Finally, we evaluate the seismic mass of the exoplanet to be $M_\mathrm{p}\sin i=1.17-1.26 \ M_\mathrm{Jupiter}$,
much more precise than what can be derived by Hertzsprung-Russell diagram inversion.
}
  % conclusions heading (optional), leave it empty if necessary 
{ 
We demonstrate that asteroseismology allows us to substantially improve the age accuracy 
that can be achieved with other methods. 
We emphasize that the knowledge of the mean properties of stellar oscillations - such
as the large frequency separation- is not enough to derive accurate ages. 
We need precise individual frequencies to narrow the age scatter that is a result
of the model input physics uncertainties. 
      Further progress is required to better constrain the physics 
      at work in stars and the stars helium content. 
     Our results  emphasize the importance of precise  classical stellar parameters  and 
       oscillation frequencies such as will be obtained by the Gaia  and PLATO missions.
       }

%\footnote{Tables n and n' are also available in electronic form
%at the CDS via anonymous ftp to cdsarc.u-strasbg.fr (130.79.128.5)
%or via http://cdsweb.u-strasbg.fr/cgi-bin/qcat?J/A+A/}
 

\titlerunning{``\`{A} la carte'' stellar age-dating and weighing with asteroseismology}
\maketitle

%________________________________________________________________

\section{Introduction}
\label{Introduction}

Stellar ages are crucial input parameters in many astrophysical studies. 
For instance, the knowledge of the stellar formation rate and age-metallicity relation 
\citep{1999BaltA...8..203G,2000MNRAS.317..831H} is essential for understanding the formation and 
evolution of our Galaxy \citep{1993ASPC...49..125F}. In addition, precise ages of the oldest 
Galactic halo stars are essential to set a lower limit to the age of the Universe \citep{1998Sci...279..981W}. 
Moreover, the huge harvest of newly discovered exoplanets calls for accurate and precise  ages of their host stars, 
a crucial parameter for understanding  planet formation and evolution 
\citep{2011A&A...531A...3H}.

While stellar masses and/or radii can be measured directly for some particular stars -- masses and radii for members of binary systems,
radii for giant stars or bright dwarfs observable by interferometry -- stellar ages cannot be determined by direct measurements, 
 but can only be estimated or inferred. As reviewed by \citet{2010ARA&A..48..581S}, there are many methods to estimate
the age of a star according to its mass, evolutionary state, and configuration --  single star or star in a group. 
The ages of single main-sequence (MS) stars are often inferred from empirical indicators (activity or rotation)  
and/or from stellar model isochrones that are compared with observed classical parameters such as 
effective temperature $T_\mathrm{eff}$,  luminosity $L$ or 
surface gravity $\log g$, and metallicity $\mathrm{[Fe/H]}$. However, the precision and accuracy that can currently be  reached are
not good enough to precisely characterize exoplanets 
\citep{2011A&A...531A...3H}. Several error sources hamper single-star age-dating: errors on the observational data, internal error related to the 
age-dating method, and, for stellar-model-dependent methods, degree of realism of the models. 
Depending on the mass and evolutionary stage and on the method, the error on age may be in the range of 50 to more than 150 per cent 
\citep[see for instance][and references therein]{2009IAUS..258..419L}.

Indeed, stellar model outputs, as the age attributed to a given star, are quite 
sensitive to the physical inputs of the model calculation. For instance, 
the processes of transporting or mixing chemical elements, such as
 convection and overshooting, microscopic diffusion, and turbulent diffusion 
 induced by hydrodynamical instabilities have been found to have a major impact. Unfortunately, these
  processes are still only poorly described and often have to be parametrized.

Progress is made with the availability of  asteroseismic data provided by the high-precision photometric 
missions CoRoT \citep{2002esasp.485...17b} and {\it Kepler} \citep{2010ApJ...713L..79K}. 
Low-amplitude solar-like oscillations have been detected in many stars, and their frequencies have been 
measured with a precision typically of a few tenths of micro Hertz \citep[e.g.][]{2008Sci...322..558M,2013ARA&A..51..353C}. 
Seismic data have recently and very frequently been used to age-date and weigh stars.
Currently, two approaches are taken. The first one, ensemble asteroseismology,
 attempts to determine the mass and age of large sets of stars based on their
  mean seismic properties \citep{2014ApJS..210....1C}. In this approach, 
  interpolation in large grids of stellar models by different techniques
   provides the mass and age of the model that best matches the observations.
  The alternative approach, which is more precise, is the hereafter named 
 ``\`{a} la carte''
 modelling\footnote{We use ``\`{a} la carte'' in opposition to ``set meal'' to stress the point that models are specifically fashioned to study a case-study star.} \citep{2013eas63123}, that is, the detailed study of  specific stars, 
    one by one, also referred to as ``boutique''  modelling, (see e.g. D. Soderblom, 2013, invited review at the International Francqui  Symposium\footnote{\url{http://fys.kuleuven.be/ster/meetings/francqui}}). 
     This approach has been used to model CoRoT and {\it Kepler} stars
      \citep[see the reviews by ][]{2013ASPC..479..461B,2013ARA&A..51..353C}.
       Stars hosting exoplanets have been modelled for example by 
        \citet{2011ApJ...726....2G}, \citet{2012ASPC..462..469L}, \citet{2012A&A...543A..96E},  
	 \citet{2013eas63123}, and \citet{2013ApJ...766...40G}.

In the present study, we address the specific problem 
of quantifying the sources of inaccuracy that affect the estimates 
of age, mass, and radius of stars.
 In the past, \citet{1994ApJ...427.1013B} 
 addressed this problem theoretically, 
 while \citet{1995IAUS..166..135L} discussed it in 
 an early prospective study related to the preparation  
 of the Gaia-ESA mission.  
The need for this quantification has become even more crucial
  because  age, mass and radius  of exoplanet host-stars are key 
 to characterizing the planets and then to understanding their
  formation and evolution. This is therefore a prerequisite
   in the context of the space missions CoRoT, {\it Kepler}, and forthcoming Gaia, 
   TESS, and PLATO. Recently, \citet{2014A&A...561A.125V} used a 
  grid approach and a synthetic sample of solar-type MS stars 
  to carry out a theoretical investigation to identify and quantify the
 sources of biases on the mass and radius determinations. 
      Here we instead consider the \`{a} la carte approach
  to characterize a main {\small CoRoT}  target, \object{HD~52265}, as an illustrative 
  case-study. The G-type metal-rich star HD~52265 is a MS star that 
 hosts an exoplanet whose transit was not observable, 
    but {\small CoRoT} provided a rich solar-like oscillation spectrum 
    that was analysed by \citet{2011A&A...530A..97B} and \citet{2013PNAS..11013267G}.
HD~52265 has been modelled by \citet{2007A&A...471..885S} prior to its observation 
by {\small CoRoT}, and then by \citet{2012A&A...543A..96E} and
 \citet{2012A&A...544L..13L} on the basis of {\small CoRoT} data. 
The asteroseismic modelling by \citet{2012A&A...543A..96E} 
 was based on the large and small mean frequency separations
  (see  Sect.~\ref{obs}). It provided a seismic mass of 
  $1.24\pm 0.02\ M_\odot$, a seismic radius of $1.33\pm 0.02\ R_\odot$, 
  and a seismic age of $2.6\pm 0.2$ Gyr. Note that the error bars on 
  these values do not include the impact of the uncertainties on stellar 
  model inputs. The mass of the exoplanet was not evaluated either.
 
In this study, we characterize the star in terms of age, mass, radius, 
initial helium content, etc. To that end, we performed  
\`{a} la carte
modelling based on several classes of dedicated stellar models corresponding
 to different assumptions on the input physics and chemical composition, 
 and we used different sets of observational parameters to constrain 
 the models. We examined how the uncertainties on the 
 observational constraints and on the model input physics and free 
 parameters affect the results of stellar modelling.

In Sect. \ref{obs}, we review the observational data available for HD~52265. In Sect. \ref{LM}, 
we describe our methodology and choices for cases of interest. The results are detailed in Sect.\ref{results}, namely the  
range of age, mass, radius, initial helium content, and mixing-length parameter of the models. 
They show that using seismic constraints severely restricts these ranges. For comparison, 
Sect.  \ref{emp} discusses the empirical ages of the star, while Sect. \ref{iso} 
 presents ages obtained through isochrone placement in the Hertzsprung-Russell (H-R)
  diagram. We discuss the impact of the uncertainty on the mass of the host-star  on the mass of the exoplanet in Sect.~\ref{exoplanet}, 
and we draw some conclusions in Sect.  \ref{conclusion}.

%%%%%%%%%%%%%%%%%%%%%%%%%%%%%%%%%%%%%%%%%%%%%%%%%%%%%%%%%%%%%%%
\begin{table*}  %[t]
\caption{Main observational constraints for the modelling of HD~52265.}
\label{param}
\begin{tabular}{ccccccccc}
\hline\hline
  $T_\mathrm{{eff}}$ & $\log g$ & [Fe/H] & $L$ &  $\langle\Delta\nu\rangle$ & $\Delta\nu_\mathrm{asym}$ &$\langle d_{02}\rangle$ &$\langle r_{02}\rangle$ &$\langle rr_{01/10}\rangle$ \\
  {[K]} & [dex] & [dex] & [$L_\odot$] &   [$\mu$Hz] &  [$\mu$Hz] &   [$\mu$Hz] &--&-- \\
\hline
$6116\pm 110$ &  $4.32\pm 0.20$ & $0.22\pm 0.05$ &$2.053\pm 0.053$ & $98.13\pm 0.14$& $98.19 \pm 0.05$ & $8.20\pm 0.31$& $ 0.084\pm 0.003$&$ 0.033\pm 0.002$\\
\hline
\end{tabular}
\end{table*}
%%%%%%%%%%%%%%%%%%%%%%%%%%%%%%%%%%%%%%%%%%%%%%%%%%%%%%%%%%%%%%%

\section{Observational constraints for HD~52265}
\label{obs}

This section reviews the observational data that we used as constraints for the modelling of HD~52265.

\subsection{Astrometry, photometry, and spectroscopy}
\label{classic}

HD~52265 (HIP~33719) is a nearby single G0V star. According to its Hipparcos parallax $\pi{=}34.53{\pm}0.40$ mas \citep[][]{2007ASSL..350.....V}, 
it is located at ${{\approx}29}$ pc.
To model the star, we considered the observational data listed in Table\ \ref{param}.
In the literature, we gathered twenty spectroscopic determinations of the effective 
temperature $T_\mathrm{eff}$, surface gravity $\log g$, 
and metallicity $\mathrm{[Fe/H]}$ of HD~52265 reported since 2001.
 We adopted here the average of these quantities and derived the error bars 
 from the extreme values reported.
To derive the luminosity $L$, we used the parallax and the Tycho $V_\mathrm{T}$ magnitude, $V_\mathrm{T}{=}6.358{\pm}0.004$ mag, which we translated into the Johnson value, $V_\mathrm{J}=6.292\pm 0.005$ mag,  following \citet[][]{2002AJ....124.1670M}. The bolometric correction $\mathrm{BC}{=}{-}0.014\pm 0.012$ was derived from $T_\mathrm{eff}$,  $\log{g}$, and $\mathrm{[Fe/H]}$ using the tools developed by \citet[][]{2003AJ....126..778V}. 
The Stefan-Boltzmann radius corresponding to the  adopted values of $L$ and 
$T_\mathrm{eff}$ is $R_\mathrm{SB}=1.28 \pm 0.06\ R_\odot$.

\citet{2000ApJ...545..504B} detected an exoplanet through observed radial velocity (RV) variations of HD~52265. 
From the RV curve they derived the semi-amplitude $K$, orbital period $P$, eccentricity $e$, 
and semi-major axis $a$ of the orbit. From RV data and the Kepler third law,
 a lower limit on the mass of the planet can be inferred, via
  \begin{eqnarray}
  M_\mathrm{p} \sin i=M_\mathrm{star}^{2/3} K (1-e^2)^{1/2} (P/2\pi G)^{1/3},
 \label{mpsini}
  \end{eqnarray}
  where $i$ is the angle of inclination of the orbital plane on the sky 
  \citep[see e.g. ][]{2011exha.book.....p}. 
  This is discussed in more detail in Sect.~\ref{exoplanet}.

\subsection{CoRoT light-curve inferences and seismic constraints}

Ultra high-precision photometry of HD~52265, performed on-board {\small CoRoT} for four months,  
provided a light-curve carrying the solar-like oscillation signature \citep{2011A&A...530A..97B}. 

% for the bibliography, at the end
\subsubsection{Individual oscillation frequencies}
\label{absfreq}

HD~52265 shows a pressure-mode (p-mode) solar-like oscillation spectrum, in which \citet{2011A&A...530A..97B} 
identified 28 reliable low-degree p-modes of angular degrees $\ell=0, 1, 2$ and radial orders $n$ in 
the range $14$-$24$ (see their Table\ 4). The frequencies $\nu_{n, \ell}$ are in the range $1500$-$2550\ \mu$Hz. 
Because the data are of  high quality,  the precision on each frequency is of a few tenths of  $\mu$Hz.

In the present study, individual frequencies were used to constrain
 stellar models. Before turning to the problems related to the use of individual 
 frequencies, we briefly recall some basic properties of stellar 
 oscillations.

A formulation adapted from the asymptotic expansion by \citet[][]{1980apjs...43..469t} is commonly used to  interpret the 
observed low-degree oscillation spectra \citep[see e.g.][and references therein]{2013A&A...550A.126M}. It approximates 
the frequency of a p-mode of high radial order $n$ and angular degree $\ell{\ll}n$ as
\begin{eqnarray}
\label{asym}
%\nu_{n,\ell}{\simeq}\left( n+\frac{1}{2}\ell +\epsilon\right)\Delta \nu_\mathrm{asym}-\left(A\ell(\ell+1)-B\right) %\frac{\Delta \nu_\mathrm{asym}^2}{\nu_{n,\ell}}+...
\nu_{n,\ell}{\simeq}\left( n+\frac{1}{2}\ell
 +\epsilon\right)\Delta \nu_\mathrm{asym}-\ell(\ell+1)D_0,
\end{eqnarray}
where the coefficients $\epsilon$, $\Delta \nu_\mathrm{asym}$, and $D_0$ depend on the considered equilibrium state of the star.
In particular
\begin{eqnarray}
\label{asymstuff}
\Delta \nu_\mathrm{asym}{=} \left(2 \int\limits_{0}^{R} \frac{\mathrm{d}r}{c} \right)^{-1} \ \mathrm{and}\ \ D_0\approx -\frac{\Delta \nu_\mathrm{asym}}{4\pi^2 \nu_{n,\ell}} \int\limits_{0}^{R} \frac{\mathrm{d}c}{r},
\end{eqnarray}
where $c$ is the adiabatic sound speed 
${c}=(\Gamma_1 P/\rho)^{1/2}$ ($\Gamma_1$ is the first adiabatic
 index, $P$ the pressure, and $\rho$ the density). For a perfect gas, ${c}\propto(T/\mu)^\frac{1}{2}$ where $T$ is the temperature and $\mu$ the mean molecular weight.
 
 The quantity $\Delta \nu_\mathrm{asym}$ measures the inverse of the sound 
 travel time across a stellar diameter and is proportional to the square root of the mean density, while $D_0$
 probes the evolution status (and then age) through the sound speed gradient built by the
 chemical composition changes in the stellar core.
 
 The $\epsilon$ term weakly depends on $n$ and $\ell$ but  is highly sensitive 
to the physics of surface layers. The problem is that outer layers in solar-type oscillators 
are the seat of inefficient convection, a 3-D turbulent process, which is poorly understood. 
The modelling of near-surface stellar layers is uncertain and so are the related computed frequencies. 
These so-called near-surface effects are a main concern when using individual frequencies to constrain stellar models because they are at the origin of an offset between observed and computed oscillation frequencies. Some empirical recipes can be used to correct for this offset. 
This  is discussed in more detail in Sect.~\ref{oscillationcomputing}.
 
\subsubsection{$\nu_\mathrm{max}$, $\langle \Delta \nu \rangle$, and scaling relations}
\label{scaling}

From the oscillation power spectrum of HD~52265, \citet{2011A&A...530A..97B}  extracted the frequency at maximum amplitude $\nu_\mathrm{max}=2090\pm\  20\ \mu$Hz. This quantity is proportional to the acoustic cut-off frequency, itself related to effective temperature and surface gravity \citep[see e.g.][]{1994ApJ...427.1013B, 1995A&A...293...87K,2011A&A...530A.142B}. This yields a scaling relation that can be used to constrain the mass and radius of a star of known $T_\mathrm{eff}$,
\begin{eqnarray}
\label{scalingnumax}
\nu_\mathrm{max, sc}/\nu_\mathrm{max,\odot} {=}(M/M_\odot)(T_\mathrm{eff}/5777)^{-1/2} (R/R_\odot)^{-2},
\end{eqnarray}
 where  $\nu_\mathrm{max,\odot}=3050\ \mu$Hz is the solar value and the index $\mathrm{sc}$ stands for scaling.

The difference in frequency of two modes of same degree and orders that differ by one unit reads
\begin{eqnarray}
\label{indivsep}
\Delta \nu_{\ell}(n){=}\nu_{n,\ell}{-}\nu_{n-1,\ell},
\end{eqnarray}
and is named the large frequency separation.
We used the mean large frequency separation to constrain stellar models and calculated it in three different ways, 
 but each time the observational value and the model value were calculated accordingly.

First, the mean large frequency separation can be calculated as an average of the individual separations defined by Eq.~\ref{indivsep}.
We first calculated the mean separation $\langle\Delta\nu_\ell\rangle$ for each observed 
 angular degree ($\ell=0, 1, 2$) by averaging over the whole range of observed radial orders. We then obtained the overall mean  separation from $\langle\Delta\nu\rangle =\frac{1}{3}\sum\limits_{\ell=0}^2 \langle\Delta\nu_\ell\rangle$. Its value is reported in Table~\ref{param}.

Second, in the asymptotic approximation (Eq.~\ref{asym}), $\Delta \nu_{\ell}(n){\equiv}\Delta \nu_\mathrm{asym}$ 
is approximately constant regardless of the $\ell$ value. We carried out a weighted least-squares fit of the asymptotic relation, 
 Eq.~\ref{asym}, to the 28 identified frequencies and obtained the value of 
 $\Delta \nu_\mathrm{asym}$ given  in Table~\ref{obs}, as well as $D_0$ 
 ($1.43\pm 0.05\mu$Hz) and $\epsilon$ ($1.34 \pm 0.01$). The quoted error 
 bars were inferred from a Monte Carlo simulation.
  
Third,  as mentioned in Sect. \ref{absfreq},  $\langle \Delta \nu \rangle \propto \langle \rho\rangle^{1/2}$.
This yields a scaling relation that can be used to constrain stellar mass and radius \citep[see e.g.][]{1986apj...306l..37u, 1995A&A...293...87K},
\begin{eqnarray}
\label{scalingdeltnu}
\langle \Delta \nu \rangle_\mathrm{sc}/\langle \Delta \nu \rangle_\odot{=}(M/M_\odot)^{1/2} (R/R_\odot)^{-3/2},
\end{eqnarray}
where $\langle \Delta \nu \rangle_\odot=134.9\ \mu$Hz.

The inversion of Eqs.~\ref{scalingnumax} and 
\ref{scalingdeltnu}, with the observed values of  $\langle\Delta\nu\rangle$, 
$\nu_\mathrm{max}$, and $T_\mathrm{eff}$ of Table~\ref{param}, 
yields $R_\mathrm{sc}=1.33\pm 0.02\ R_\odot$,
  $M_\mathrm{sc}=1.25\pm 0.05\ M_\odot$, and 
  $\log\ \mathrm{g}_\mathrm{sc}=4.29\pm 0.01$ dex.

  \subsubsection{Small frequency separations and separation ratios}
\label{separ}  

The difference in frequency of two modes of degrees that differ by two units and orders that differ by one unit reads
\begin{eqnarray}
d_{\ell,\ \ell+2}(n){=}\nu_{n,\ell}{-}\nu_{n-1,\ \ell+2}
\end{eqnarray}
and is commonly referred to as the small frequency separation. 
 
Modes of $\ell=1$  are rather easy to detect, while  modes of $\ell=2$ are not always observed or are affected by large error bars.
This led \citet{2003A&A...411..215R} to propose to use the five points small separations $dd_{01}(n)$ and $dd_{10}(n)$ as diagnostics for stellar models.
They read
%\begin{eqnarray}
%d_{01}(n)= -\frac{1}{2}(\nu_{n-1, 1}-2\nu_{n, 0}+\nu_{n, 1})
%\end{eqnarray}
%\begin{eqnarray}
%d_{10}(n)= \frac{1}{2}(\nu_{n+1, 0}-2\nu_{n, 1}+\nu_{n, 0})
%\end{eqnarray}
\begin{eqnarray}
dd_{01}(n)= \frac{1}{8}(\nu_{n-1, 0}-4\nu_{n-1, 1}+6\nu_{n, 0}-4\nu_{n, 1}+\nu_{n+1, 0})\\
dd_{10}(n)= -\frac{1}{8}(\nu_{n-1, 1}-4\nu_{n, 0}+6\nu_{n, l}-4\nu_{n+1, 0}+\nu_{n+1, 1}).
\end{eqnarray}
According to the asymptotic relation (Eq.~\ref{asym}), $d_{02}(n)$ scales as $ {\approx}6D_0$,
  while  $dd_{01/10}(n)$ scales as $ {\approx} 2D_0$. Thus, they all probe the evolutionary status of stars.

Furthermore, \citet{2003A&A...411..215R} demonstrated that while the frequency separations are sensitive to near-surface effects, these effects nearly cancel in the frequency separation ratios defined as
\begin{eqnarray}
r_{02}(n)= d_{02}(n)/\Delta \nu_{1}(n) \\
rr_{01}(n)= dd_{01}(n)/\Delta \nu_{1}(n)\ ;\  rr_{10}(n)= dd_{10}(n)/\Delta \nu_{0}(n+1).
\end{eqnarray}

As detailed in the following, we used these frequency separations and ratios to constrain models of HD~52265.
We denote by $\langle d_{02}\rangle$, $\langle r_{02}\rangle$ and $\langle rr_{01/10}\rangle$,  the mean 
values of the small frequency separations and separation ratios. To calculate the mean values given in Table~\ref{obs}, 
we averaged over the whole range of observed radial orders. 

\subsubsection{Related seismic diagnostics}
\label{diagnostics}

\citet{1986apj...306l..37u} and \citet{1988IAUS..123..295C} proposed to use the pair ($\langle\Delta\nu\rangle$ and $\langle d_{02}\rangle$) as a diagnostic of age and mass of MS stars. To minimize near-surface physics the ($\langle\Delta\nu\rangle$ and $\langle r_{02}\rangle$) pair can be used instead \citep{2005MNRAS.356..671O}. 

\begin{figure}[!ht]
\begin{center}
\resizebox{\hsize}{!}{\includegraphics{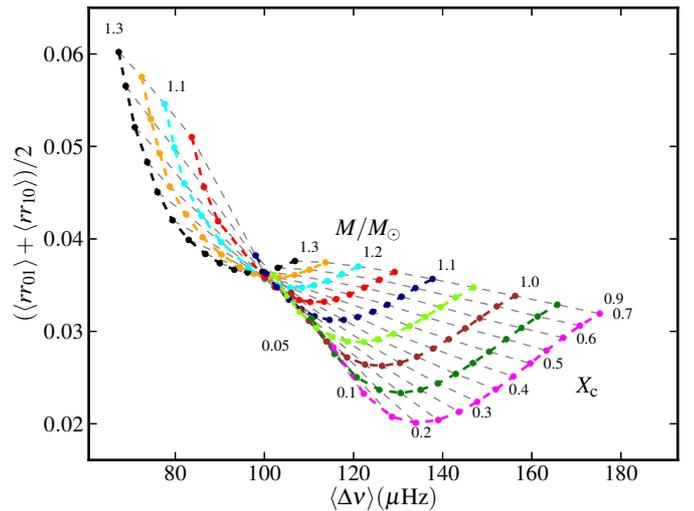}}
\caption{Asteroseismic diagram showing the run of $(\langle rr_\mathrm{{01}}\rangle+ \langle rr_\mathrm{{10}}\rangle)/2$ as a function of $\langle \Delta \nu\rangle$ for stars with masses in the range $0.9-1.3\ M_\odot$ during the MS. Models have been calculated with an initial helium abundance $Y=0.275$, solar metallicity $Z/X=0.0245$,  and mixing-length parameter $\alpha_\mathrm{conv}=0.60$. Evolutionary stages with decreasing central hydrogen abundance $X_\mathrm{c}$ are pinpointed. %\color{red}{zoom ? relation rr01, deltanu= f(M,Xc)}
}
\label{astdiag}
\end{center}
\end{figure}

The advantage of $\langle r_{02}\rangle$ is that it decreases regularly as evolution proceeds on the MS. But when only modes of $\ell=1$ are observed, it is interesting to consider the mean ratios $\langle rr_{01/10}\rangle$, which are also sensitive to age \citep[see e.g.][]{2005a&a...441..615m,2005a&a...441.1079m}.
 This is illustrated in Fig.~\ref{astdiag}, which shows the run of $(\langle rr_\mathrm{{01}}\rangle+ \langle rr_\mathrm{{10}}\rangle)/2$ as a function of $\langle \Delta \nu\rangle$ along the evolution on the MS of stars of different masses. For all masses, the $\langle rr_{01/10} \rangle$ ratio decreases at the beginning of the evolution on the MS to a minimum and then increases to the end of the MS. The lowest value is higher and occurs earlier as the mass of the star increases, 
 that is as a convective core appears and develops.

Deviations from the asymptotic theory are found in stars as soon as steep gradients of physical quantities build up. 
This occurs for instance at the boundaries of convective zones because of
 the abrupt change of energy transport regime. Such glitches affect the sound speed, and an oscillatory behaviour 
 is then visible in frequency differences \citep[see e.g. ][]{1990LNP...367..283G, 1994A&A...282...73A, 1994MNRAS.268..880R}. 
In \citet{2012A&A...544L..13L}, we used the oscillatory behaviour of the observed $rr_{01}(n)$ and $rr_{10}(n)$ ratios 
to estimate the amount of convective penetration below the convective envelope of HD~52265.

\subsubsection{Rotation period and inclination of the rotation axis}

\citet{2011A&A...530A..97B} derived a precise rotation period,
 $P_\mathrm{rot}=12.3{\pm}0.14$ days from the light curve.
  \citet{2013PNAS..11013267G} used seismic constraints to 
  estimate the inclination of the spin axis of HD~52265 
  and found $\sin i=0.59^{+0.18}_{-0.14}$. This allows us 
  to estimate the mass of the exoplanet detected by RV 
  variations \citep{2000ApJ...545..504B}, under the standard 
  assumption that the stellar rotation axis is the same as
   the axis of the exoplanet orbit (see Sect.~\ref{exoplanet}).

\section{Searching for an optimal model of HD~52265}
\label{LM}

We present the optimization method we used to obtain a model that best matches a selected set of 
observational constraints. 
We considered several choices of observational constraints, as explained in Sect.~\ref{modcase}. 
First, we chose a reference set of input physics for the modelling and, for this set,  
the optimization gave us a reference model for each selected case of observational constraints.
 In a second step, we carried out additional optimizations based on other possible choices for the input physics.  We then quantified the age and mass uncertainties, for instance by comparing the additional models to the reference one. Below we first present the inputs of the models (Sect.~\ref{physics}), then the calculation of the oscillation frequencies (Sect.~\ref{oscillationcomputing}), and finally the optimization method (Sect.~\ref{optimization}) and the choice of the sets of observational constraints (Sect.~\ref{modcase}) .

\subsection{Model input physics and chemical composition}
\label{physics}
We have modelled the star with the stellar evolution code
 Cesam2k \citep{2008Ap&SS.316...61M,2013A&A...549A..74M}. Following \citet{2008Ap&SS.316..187L}, 
 to ensure numerical accuracy, the models were calculated with
  $\approx 2000$ mesh points and $\approx 100$ time steps were
   taken to reach the optimized final model of the star.
Our aim has been to evaluate the effect of the choice of the model input physics on 
the inferred age, mass, and radius of HD~52265. We considered the input physics, and parameters 
described below and listed in Table~\ref{modelinputs}.

\begin{itemize}
\item \emph{Opacities, equation of state ({\small EoS}), nuclear reaction rates:}\ Our reference models  (set $A$, Table~\ref{modelinputs}) are based on the {\small OPAL05 EoS}  \citep[][]{2002ApJ...576.1064R} and on  {\small OPAL96} opacities \citep{1996ApJ...464..943I} complemented at low temperatures by {\small WICHITA} tables \citep{2005ApJ...623..585F}. We used the {\small NACRE} nuclear reaction rates \citep[][]{1999NuPhA.656....3A} except for the $^{14}N(p,\gamma)^{15}O$ reaction, 
 where we adopted the revised {\small LUNA} rate \citep{2004PhLB..591...61F}. 
To estimate uncertainties, we also calculated models based on the {\small OPAL01 EoS} (set 
 $H$) and the {\small NACRE} nuclear reaction rate for the $^{14}N(p,\gamma)^{15}O$ reaction (set $D$). 

\item \emph{Microscopic diffusion:}
Our reference models take into account the microscopic diffusion of helium and heavy elements including gravitational settling, thermal and concentration diffusion but no radiative levitation, following the formalism of \citet[][hereafter MP93]{MP93}. Note that \citet{2012A&A...543A..96E} showed that radiative accelerations do not affect the structure of the models of HD~52265. To estimate uncertainties, we calculated models without diffusion (set $E$), and in set $G$ models with diffusion,
 but treated with the \citeauthor{1969fecg.book.....B} formalism 
\citep[][hereafter B69]{1969fecg.book.....B}.
  In a test, we investigated the effect of mixing that results from the radiative 
diffusivity associated with the kinematic radiative viscosity $\nu_\mathrm{rad}$ 
(subset $A$ described in Appendix~\ref{appendice}). According to \citet{2002A&A...390..611M}, 
it can be modelled by adding an additional mixing with a diffusion coefficient $d_\mathrm{rad}=D_\mathrm{R}\times \nu_\mathrm{rad}$ with $D_\mathrm{R}\approx 1$. 
It limits gravitational settling in the outer stellar layers of stars with thin convective envelopes.

\item \emph{Convection:} Our reference is the {\small CGM} convection theory of \citet{1996ApJ...473..550C}, 
which includes a free mixing-length parameter {$\alpha_\mathrm{conv}=\ell/H_P$} ($\ell$ is the mixing length 
and $H_P$ the pressure scale height). To estimate uncertainties, we considered in set $B$ the {\small MLT} theory \citep{1958ZA.....46..108B}. 
When enough observational constraints were available, 
  the value of $\alpha_\mathrm{conv}$ was  derived from the optimization of the models. When too few
  observational constraints were available, the value of the mixing length had to be fixed. For the reference model, it  was fixed to the 
solar value $\alpha_\mathrm{conv, cgm}=0.688$ or $\alpha_\mathrm{conv, mlt}=1.762$, which results from the
 calibration of the radius and luminosity of a solar model with the same input physics 
 \citep[see e.g. ][]{2008Ap&SS.316...61M}. Other choices for the mixing length are considered in the study and presented 
 in Appendix~\ref{appendice}.

\item \emph{Core overshooting:} In reference models we did not consider overshooting. We explored its impact in alternate models where we assumed that the temperature gradient in the overshooting zone is adiabatic. A first option (set $I$) was to set the core overshooting distance to be $\ell_\mathrm{ov, c}{=}\alpha_\mathrm{ov}\times \min(R_\mathrm{cc}, H_{P})$, where $R_\mathrm{cc}$ is the radius of the convective core and $\alpha_\mathrm{ov}=0.15$. In a second option (set $J$), we adopted the \citet{1992A&A...266..291R} prescription, in which overshooting extends on a fraction of the mass of the convective core $M_\mathrm{cc}$, the mass of the mixed core being expressed as $M_\mathrm{ov, c}{=}\alpha_\mathrm{ov}\times M_\mathrm{cc}$ with $\alpha_\mathrm{ov}=1.8$. 

\item \emph{Convective penetration:} Convective penetration below the convective envelope 
is the penetration of fluid elements into the radiative zone due to their inertia. 
It leads to efficient convective heat transport with penetrative flows that establish a close to adiabatic temperature gradient, and to an efficient mixing of material in the extended region. A model for convective penetration has been designed by \citet{1991A&A...252..179Z}. In this model, the distance of fluid penetration into the radiative zone reads $L_p=\frac{\xi_\mathrm{PC}}{\chi_{P}} H_P$,
 where $\chi_P = (\partial \log \chi/\partial \log P)_\mathrm{ad}$ is the adiabatic 
derivative with respect to the pressure $P$ of the radiative conductivity 
$\chi =\frac{16 \sigma T^3}{3 \rho \kappa}$ ($\kappa$ and  $\sigma$ are the opacity and 
Stefan-Boltzmann constant, respectively). The free parameter $\xi_\mathrm{PC}$ is of 
the order of unity but has to be calibrated from the observational constraints. 
We adopted here either $\xi_\mathrm{PC}=0.0$ (reference set $A$, no penetration) or $\xi_\mathrm{PC}=1.3$ (set $K$), this latter best adjusts the oscillatory behaviour of the individual frequency separations $rr_{01/10}(n)$ of HD~52265 \citep{2012A&A...544L..13L}.

\item \emph{Rotation:} We did not include rotational mixing in our models, except in one test case (see subset $A$ in Appendix~\ref{appendice}), where we considered 
rotation and its effects on the transport of angular momentum and 
chemicals as described in \citet{2013A&A...549A..74M}. In that case, 
additional free coefficients enter the modelling: a 
coefficient $K_\mathrm{w}$  intervenes in the treatment of magnetic braking 
by stellar winds \citep[see Eq. 9 in ][]{2013A&A...549A..74M}, 
following the relation by \citet{1988ApJ...333..236K}. 
We adjusted $K_\mathrm{w}$ so that the final model has the observed 
rotation period. This is one option among many possible ones.
 A thorough study of the impact of rotation on the
modelling of HD~52265 will be presented in a forthcoming paper.

\item \emph{Atmospheric boundary condition:} 
The reference models are based on grey model atmospheres with the classical 
Eddington T-$\tau$ law. In set $F$, we investigated models based on the 
\citet{1993yCat.6039....0K} T-$\tau$ law. For consistency 
with these \citet{1993yCat.6039....0K} T-$\tau$ tables, in models $F$ convection is 
computed according to the {\small MLT} theory.

\item\emph{Solar mixture:}
We adopted the canonical {\small GN93} mixture  \citep{1993oee..conf...15G} as the reference, but considered the {\small AGSS09} 
solar mixture \citep{2009ARA&A..47..481A} in set $C$. The {\small GN93} mixture  $(Z/X)_\odot$ ratio is $0.0244$, while the  {\small AGSS09} mixture 
corresponds to $(Z/X)_\odot=0.0181$. 

\item \emph{Stellar chemical composition:}
The mass fractions of hydrogen, helium and heavy elements are denoted by $X$, $Y$, and $Z$ 
respectively. 
The present $(Z/X)$ ratio is related to the observed $\mathrm{[Fe/H]}$ value through 
$\mathrm{[Fe/H]}=\log(Z/X)-\log(Z/X)_\odot$. We took a relative error of $11$ per cent on $(Z/X)_\odot$ \citep{1989GeCoA..53..197A}. 
The initial $(Z/X)_0$ ratio is derived from model calibration, as explained below. 

For the initial helium abundance $Y_0$ we considered different possibilities. 
When enough observational constraints were available, 
  the value of $Y_0$ was  derived from the optimization of the models. When too few
  observational constraints were available, the value of $Y_0$ had to be fixed. For the reference
  model, we derived it from the helium-to-metal enrichment ratio 
$(Y_0 - Y_\mathrm{P})/{(Z- Z_\mathrm{P})}{=}{\Delta Y}/{\Delta Z}$, where $Y_\mathrm{P}$ 
and $Z_\mathrm{P}$ are the primordial abundances. 
We adopted $Y_\mathrm{P}{=}0.245$  \citep[e.g.][]{2007ApJ...666..636P}, $Z_\mathrm{P}{=}0.$ 
and, ${\Delta Y}/{\Delta Z}{\approx}2$, this latter from a solar model calibration 
in luminosity and radius. Other choices for $Y_0$ are considered in the study and presented 
 in Appendix~\ref{appendice}.

\item \emph{Miscellaneous:} The impact of  several alternate prescriptions  for the free parameters  
  is investigated and described in Appendix~\ref{appendice}.
\end{itemize}

%%%%%%%%%%%%%%%%%%%%%%%%%%%%%%%%%%%%%%%%%%%%%%%%%%%%%%%%%%%%%%%
\begin{table*}  %[t]
\caption{Summary of the different sets of input physics considered for the modelling of HD~52265. 
As detailed in the text, the reference set of inputs denoted by REF is based on OPAL05 EoS,  
OPAL96/WICHITA opacities, NACRE+LUNA  reaction rates (this latter only for $^{14}N(p,\gamma)^{15}O$), 
the CGM formalism for convection, the MP93 formalism for microscopic diffusion, 
the Eddington grey atmosphere, and the {\small GN93} solar mixture and includes neither overshooting
nor convective penetration or rotation. For the other cases we only indicated the input that is 
changed with respect to the reference. 
{The colours and symbols in column 3  are used in Figs. \ref{Allages} to \ref{rayon} and in Fig.~\ref{planet}, 
but note that the colour symbols used in Fig.~\ref{goodness} are unrelated}.}
\begin{tabular}{lll}
\hline\hline
Set & Input physics & Figure symbol/colour \\
\hline
$A$ & REF  &  circle, cyan \\
$B$ & convection MLT  & square,  orange\\
$C$ & AGSS09 mixture  &  diamond, blue\\
$D$ & NACRE for $^{14}N(p,\gamma)^{15}O$ & small diamond, magenta\\
$E$ & no microscopic diffusion  & pentagon, red\\
$F$ & Kurucz model atmosphere, MLT  & bowtie, brown\\%
$G$ & B69 for microscopic diffusion  & upwards triangle, chartreuse\\
$H$ & EoS OPAL01  & downwards triangle, purple\\
$I$ & overshooting $\alpha_\mathrm{ ov}{=}0.15 H_P$  & inferior, yellow\\%
$J$   & overshooting $M_\mathrm{ ov, c}{=}1.8\times{M_\mathrm{ cc}}$   & superior, gold  \\%
$K$  & convective penetration  $\xi_\mathrm{PC}{=}1.3 H_P$  & asterisk, pink\\
%L  & rotation  & \\%%
%M  & Renu  & \\%%
\hline\end{tabular}
\label{modelinputs}
\end{table*}
%%%%%%%%%%%%%%%%%%%%%%%%%%%%%%%%%%%%%%%%%%%%%%%%%%%%%%%%%%%%%%%

%%%%%%%%%%%%%%%%%%%%%%%%%%%%%%%%%%%%%%%%%%%%%%%%%%%%%%%%%%%%%%%
\begin{table*}  %[t]
\caption{Summary of the different cases considered for the modelling of HD~52265. A and M stand for age and mass.}
\label{cases}
\begin{tabular}{llll}
\hline\hline
Case & Observed  &Adjusted  & Fixed \\
\hline
$1$                     &  $T_\mathrm{{eff}}$, $L$, [Fe/H]                                                 & A, M, $(Z/X)_0$ & $\alpha_\mathrm{ conv}$, $Y_0$\\
$2$a, b, c &  $T_\mathrm{{eff}}$, $L$, [Fe/H], $\langle\Delta\nu\rangle$ & A, M, $(Z/X)_0$, $\alpha_\mathrm{ conv}$ & $Y_0$\\
$3$  &  $T_\mathrm{{eff}}$, $L$, [Fe/H], $\langle\Delta\nu\rangle$, $\nu_\mathrm{max}$ & A, M, $(Z/X)_0$, $\alpha_\mathrm{ conv}$, $Y_0$ & --\\
$4$      &  $T_\mathrm{{eff}}$, $L$, [Fe/H], $\langle\Delta\nu\rangle$, $\langle d_{02}\rangle$  & A, M, $(Z/X)_0$, $\alpha_\mathrm{ conv}$, $Y_0$ & --\\
$5$      &  $T_\mathrm{{eff}}$, $L$, [Fe/H], $\langle r_{02}\rangle$, $\langle rr_{01/10}\rangle$   & A, M, $(Z/X)_0$, $\alpha_\mathrm{ conv}$, $Y_0$ & --\\
$6$                    &  $T_\mathrm{{eff}}$, $L$, [Fe/H], $r_{02}(n)$, $rr_{01/10}(n)$ & A, M, $(Z/X)_0$, $\alpha_\mathrm{ conv}$, $Y_0$ & --\\
$7$                     &  $T_\mathrm{{eff}}$, $L$, [Fe/H], $\nu_{n, \ell}$                     & A, M, $(Z/X)_0$, $\alpha_\mathrm{ conv}$, $Y_0$ & --\\
\hline
\end{tabular}
\end{table*}
%%%%%%%%%%%%%%%%%%%%%%%%%%%%%%%%%%%%%%%%%%%%%%%%%%%%%%%%%%%%%%%

\subsection{Calculation of oscillation frequencies}
\label{oscillationcomputing}
We used the Belgium {\small LOSC} adiabatic oscillation code \citep{2008Ap&SS.316..149S} to calculate the frequencies. Frequencies and frequency differences were calculated for the whole range of observed orders and degrees. The observed and computed seismic indicators defined in Sect.~\ref{obs} were derived consistently.

As mentioned in Sect.~\ref{obs}, near-surface effects are at the origin of an offset between observed and computed oscillation frequencies. We investigated the impact of correcting the computed frequencies from these effects. For that purpose, in some models (presented in Section 3.4 below), we applied to the computed frequencies, the empirical corrections obtained  by \citet{2008ApJ...683L.175K} from the seismic solar model:
\begin{eqnarray}
\label{nearsurf}
\nu_{n, l}^\mathrm{mod, corr} = \nu_{n, l}^\mathrm{mod}+\frac{a_\mathrm{SE}}{r_\mathrm{SE}}\left(\frac{\nu_{n, l}^\mathrm{obs}}{\nu_\mathrm{max}}\right)^{b_\mathrm{SE}},
\end{eqnarray}
where $\nu_{n, l}^\mathrm{mod, corr}$ is the corrected frequency,  $\nu_{n, l}^\mathrm{mod}$ and $\nu_{n, l}^\mathrm{obs}$ are the computed and observed frequency, $b_\mathrm{SE}$ is an adjustable coefficient, $r_\mathrm{SE}$ is close to unity when the model approaches the best solution, and $a_\mathrm{SE}$ is deduced from the values of $b_\mathrm{SE}$ and $r_\mathrm{SE}$. We followed the procedure of \citet{2011A&A...527A..37B}. \citeauthor{2008ApJ...683L.175K} obtained a value of $b_\mathrm{SE ,\odot}=4.9$ when adjusting the relation on solar radial modes frequencies.
However, the value of $b_\mathrm{SE ,\odot}$ should depend on the input physics in the solar model
 considered. Indeed, \cite{2011A&A...535A..91D} obtained $b_\mathrm{SE, \odot}=4.25$ for
  a solar model computed with the Cesam2k code and adopting the {\small CGM} instead of the {\small MLT} 
  description of convection. Furthermore, $b_\mathrm{SE}$ can  differ from one star to another. 
  Under these considerations, we treated $b_\mathrm{SE}$ as a variable parameter of the modelling that
   we adjusted in the range $[3.5, 5.5]$ so as to minimize the differences between observed and computed individual frequencies \citep[see also][]{2012ApJ...749..109G}.

%-----------------------------------
\begin{figure}
      \resizebox{\hsize}{!}
	     {\includegraphics{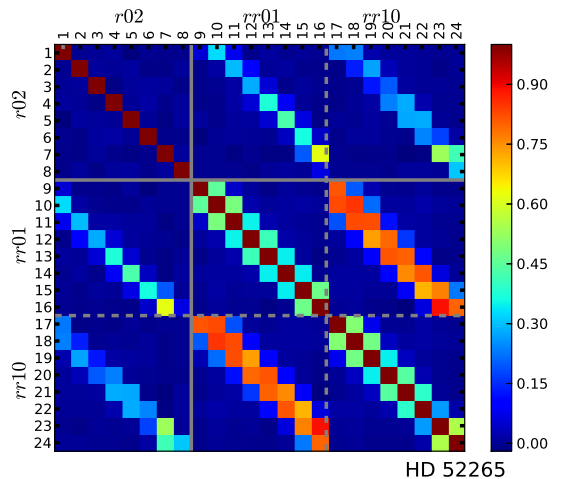}}
  \caption{
Elements of the correlation matrix of the observed ratios $r_{02}(n)$ and $rr_{01/10}(n)$. Lines (columns) 1 to 8 correspond to the $r_{02}(n)$ ratios ($n$ is in the range 17-24).  Lines (columns) 9 to 16 correspond to the $rr_{01}(n)$ ratios and lines (columns) 17 to 24 correspond to the $rr_{10}(n)$ ratios ($n$ is in the range 16-23). As expected, there are strong correlations between some of the data, in particular between the $rr_{01}(n)$ and $rr_{10}(n)$ ratios 
that have the same $n$ value or values of $n$ that differ by one unit.
    }
\label{correl}
\end{figure}
%-----------------------------------

\subsection{Model optimization}
\label{optimization}
We used the Levenberg-Marquardt minimization method in the way described by 
\citet[][]{2005a&a...441..615m} to adjust the free parameters of the modelling 
so that the models of HD~52265 match at best observations, within the error bars. 
 The goodness of the match is evaluated through a $\chi^2$-minimization. We calculated
\begin{eqnarray}
\label{chi2}
%\chi^2 = \frac{1}{N_\mathrm{obs}-1} \cdot\sum_{i=1}^{N_\mathrm{obs}} \frac{\left ( x_\mathrm{i, mod} - x_\mathrm{i, obs} \right ) ^2}{\sigma_\mathrm{i, obs}^2}
\chi^2 = \sum_{i=1}^{N_\mathrm{obs}} \frac{\left ( x_\mathrm{i, mod} - x_\mathrm{i, obs} \right ) ^2}{\sigma_\mathrm{i, obs}^2} \hspace{0.5cm} \mathrm{and}\hspace{0.5cm}\chi^2_\mathrm{R} = \frac{1}{N_\mathrm{obs}} \cdot\chi^2,
\end{eqnarray}
where $N_\mathrm{obs}$ is the total number of observational constraints considered, 
$x_\mathrm{i, mod}$ and  $x_\mathrm{i, obs}$ are the computed and observed values of 
the $i^\mathrm{th}$ constraint, respectively, and $\chi^2_\mathrm{R}$ is a reduced value.
 We distinguished  $\chi^2_\mathrm{classic}$ based on the classical parameters and 
 $\chi^2_\mathrm{seism}$ based on the seismic parameters.
The more observational constraints available, 
the more free parameters can be adjusted in the modelling process. 
If too few observational constraints are available, 
some free parameters have to be fixed (see below).
 Accordingly, seven optimization cases were considered,
  as listed in Table~\ref{cases} and described in Sect.~\ref{modcase} below. 

In the cases where we considered the constraints brought by individual separations ratios $r_{02}(n)$ and $rr_{01/10}(n)$, we had to take into account the strong correlations between the ratios. To evaluate the correlations, we drew random samples of 20,000 values of each individual frequency, assuming the errors on the frequencies are Gaussian, and then we calculated the corresponding ratios and the associated covariance matrix $C$, displayed in Fig.~\ref{correl}. In this case, the $\chi^2$ was calculated as
\begin{eqnarray}
\label{corrfreq}
\chi^2=  \sum_{i=1}^{N_\mathrm{obs}} \left ( x_\mathrm{i, mod} - x_\mathrm{i, obs} \right )^\mathrm{T}.C^{-1}. \left ( x_\mathrm{i, mod} - x_\mathrm{i, obs} \right ),
\end{eqnarray}
where $T$ denotes a transposed matrix \citep{2002nrc..book.....P}.

\subsection{Choice of the set of observational constraints}
\label{modcase}

We hereafter describe the various situations considered where $N_\mathrm{par}$ unknown parameters of a stellar model are adjusted to fit $N_\mathrm{obs}$ observational 
constraints. We considered the cases
 summarized in Table~\ref{cases}. For each case we made one model optimization with each set of input physics listed in Table~\ref{modelinputs}.
We recall that the mean values of the frequency separations --either 
observed or theoretical-- mentioned in the following were computed in the same way, as 
explained in Sect.~\ref{diagnostics}.

\begin{enumerate} 
\item \emph{Case $1$: Age and mass from classical parameters.}

In this case, we assumed that the classical parameters alone are constrained by observation 
($T_\mathrm{{eff}}$, present surface [Fe/H], $L$). We determine the mass $M$, age $A$, and initial
 metal-to-hydrogen ratio $(Z/X)_0$ required for the model to match these constraints.
 Since this gives three unknowns for three observed parameters, we made assumptions on 
 the other inputs of the models, mainly the initial helium abundance $Y_0$, 
 the mixing length $\alpha_\mathrm{ conv}$, and overshooting parameter $\alpha_\mathrm{ ov}$. 
 As a reference, we assumed that $Y_0$ can be derived from the helium-to-metal enrichment
  ratio ${\Delta Y}/{\Delta Z}=2$. We took $\alpha_\mathrm{ conv, \odot}{=}0.688$ from solar model 
  calibration and $\alpha_\mathrm{ ov}{=}0.0$. Other models, presented in Appendix~\ref{appendice},
  consider extreme values of $Y_0$ (the primordial, minimum allowed value, $0.245$), 
  of $\alpha_\mathrm{ conv}$ ($0.550, 0.826$, i.e. a change in $\alpha_\mathrm{conv, \odot}$ by 20 per cent), 
  and of $\alpha_\mathrm{ ov}$ ($0.30$).

\item \emph{Cases $2$a, b, c: Age and mass from large frequency separation $\langle \Delta \nu \rangle$ and classical parameters.}

In case $2$, we assumed that only $\langle \Delta \nu \rangle$  is known
together with the classical parameters. We then adjusted the mixing-length parameter together with the mass, 
age, and initial metal-to-hydrogen ratio (four unknowns, four constraints). We still had to fix the initial helium abundance from ${\Delta Y}/{\Delta Z}$. We considered  three sub-cases.

In sub-case $2$a, we did not  explicitly calculate the frequencies of the  model but derived 
 the model's large frequency separation from the scaling relation (Eq.~\ref{scalingdeltnu}) and compared this with the observed 
mean large frequency separation $\langle\Delta\nu\rangle$ of Table~\ref{param}. % computed as a mean as explained in Sect.~\ref{diagnostics}. 
In sub-case $2$b, we estimated the theoretical $\langle \Delta \nu \rangle$ from the computed set of individual frequencies and compared it with the observed 
$\langle\Delta\nu\rangle$.
In sub-case $2$c, we adjusted the computed $\Delta \nu_\mathrm{asym}$ (Eq.~\ref{asym}) 
to the observed value in Table~\ref{param}. In cases $2$b and $2$c, we corrected the model frequencies for near-surface effects according to Eq.~\ref{nearsurf}. %, with $b_\mathrm{SE, \odot}=4.9$. 

\item \emph{Case $3$: Age and mass from scaled values of $\langle \Delta \nu \rangle$ and $\nu_\mathrm{max}$, 
and classical parameters.} 

This case is similar to case $2$a with the additional constraint on $\nu_\mathrm{max}$ from the scaling relation (Eq.~\ref{scalingnumax}). Frequencies are not explicitly calculated in this case. 

\item \emph{Case $4$: Age and mass from large frequency separation $\langle \Delta \nu \rangle$, 
mean small frequency separation  $\langle d_{02} \rangle$,  and classical parameters.} 

In this case the frequencies are explicitly calculated and corrected for near-surface
 effects according to Eq.~\ref{nearsurf}. %, with $b_\mathrm{SE, \odot}=4.9$.
The model $\langle \Delta \nu \rangle$  and
 $\langle d_{02} \rangle$ are compared with the observed values of Table~\ref{param}.

\item \emph{Cases $5$ and $6$: Age and mass from frequency separations ratios --$r_{02}$, $rr_{01/10}$-- and classical parameters.}

In both cases the frequencies are explicitly calculated.
In case $5$, the mean values of $r_{02}$ and $rr_{01/10}$ were calculated and 
compared with the observed values of Table~\ref{param}.
In case $6$, we used the observed individual ratios $r_{02}(n)$, $rr_{01/10}(n)$ to constrain the models. 
Since the use of $r_{02}$, $rr_{01/10}$ allows us to minimize the impact of near-surface effects \citep{2003A&A...411..215R}, we used uncorrected ratios. On the other hand, we always accounted for observed data correlations. However, a model neglecting these correlations was calculated, 
 which shows that they do not affect the results very much (see Appendix~\ref{appendice}). In the appendix, we also discuss 
 the point made by \citet{2013A&A...560A...2R} on the correct way to extract from a model the frequency separation ratios that are to be compared with observed ones. 

\item \emph{Case $7$: Age and mass from individual frequencies $\nu_{n, \ell}$ and classical parameters.} 

In this case, we considered the full set of $28$ frequencies as constraints and corrected the model frequencies according to Eq.~\ref{nearsurf},   where we adjusted $b_\mathrm{SE}$ to minimize the $\chi^2$. In Appendix~\ref{appendice}, we evaluated the differences obtained when no correction for
 near-surface effects was applied or when the surface effects are corrected using the \citet{2008ApJ...683L.175K} solar value of the $b_\mathrm{SE}$ parameter ($b_\mathrm{SE, \odot}=4.90$).
\end{enumerate}

%-------------------------------------------------------------------------------
\begin{table*}  %[t]
\caption{Model results for the reference physics (set $A$, 
different cases, see Sect.~\ref{LM} and Tables~\ref{modelinputs} 
and \ref{cases}). The uncertainties result from the Levenberg-Marquardt
 minimization procedure (i.e. diagonal terms of the covariance matrix of 
 inferred parameters). No uncertainty is indicated when the parameter has 
 not been inferred but fixed (to few observational constraints as in
  cases $1$ and $2$a, b, c). Note that for the sake of homogeneous 
  tables here and in Appendix~\ref{appendice}, we give a column 
  listing the values of $\alpha_\mathrm{ov}$ and $\xi_\mathrm{PC}$ 
  although there are equal to $0$ in reference set $A$.}
\begin{tabular}{lccccccccl}
\hline\hline
Model &   Age (Gyr) & $M/M_\odot$ & $ (Z/X)_0$ &  $Y_0$&  $\alpha_\mathrm{conv}$ &  $\alpha_\mathrm{ov}$ / $\xi_\mathrm{PC}$ & $b_\mathrm{SE}$ & $a_\mathrm{SE}/r_\mathrm{SE}$& $\chi^2_\mathrm{R, classic}$ / $\chi^2_\mathrm{R, seism}$\\
\hline
$A1$& 2.90$\pm$ 1.09& 1.18$\pm$ 0.02&0.0483$\pm$0.0051&0.311&0.688&0.00/0.00& --&--&${6.4\ 10^{-7}}$/\ \ --\\
$A2$a& 2.38$\pm$ 0.88& 1.19$\pm$ 0.02&0.0493$\pm$0.0049&0.312&0.596$\pm$0.100&0.00/0.00& --&--&${1.3\ 10^{-1}}$/${7.9\ 10^{-5}}$\\
$A2$b& 1.81$\pm$ 0.80& 1.20$\pm$ 0.02&0.0491$\pm$0.0052&0.312&0.559$\pm$0.093&0.00/0.00&5.5&-4.5/1.00&${1.2\ 10^{-1}}$/${2.6\ 10^{-4}}$\\
$A2$c& 1.97$\pm$ 0.90& 1.20$\pm$ 0.02&0.0486$\pm$0.0053&0.311&0.575$\pm$0.086&0.00/0.00&5.5&-4.3/1.00&${2.3\ 10^{-1}}$/${1.6\ 10^{-3}}$\\
$A3$& 1.98$\pm$ 1.05& 1.23$\pm$ 0.01&0.0487$\pm$0.0055&0.298$\pm$0.020&0.565$\pm$0.102&0.00/0.00&--&--&${2.6\ 10^{-1}}$/${1.7\ 10^{-2}}$\\
$A4$& 2.08$\pm$ 0.27& 1.13$\pm$ 0.03&0.0509$\pm$0.0055&0.350$\pm$0.021&0.581$\pm$0.024&0.00/0.00&5.5&-4.7/1.00&${2.4\ 10^{-3}}$/${4.2\ 10^{-5}}$\\
$A5$& 2.17$\pm$ 0.32& 1.25$\pm$ 0.05&0.0467$\pm$0.0061&0.280$\pm$0.038&0.671$\pm$0.093&0.00/0.00&--&--&${3.3\ 10^{-7}}$/${4.0\ 10^{-3}}$\\
$A6$& 2.21$\pm$ 0.11& 1.22$\pm$ 0.02&0.0502$\pm$0.0024&0.299$\pm$0.011&0.599$\pm$0.031&0.00/0.00&--&--&${1.7\ 10^{-1}}$/${8.5\ 10^{-1}}$\\
$A7$& 2.17$\pm$ 0.02& 1.27$\pm$ 0.00&0.0486$\pm$0.0007&0.274$\pm$0.001&0.601$\pm$0.004&0.00/0.00&4.2&-5.1/1.00&${5.3\ 10^{-1}}$/${1.7\ 10^{0}}$\\
\hline
\end{tabular}
\label{resultsA1}
\end{table*}

%\addtocounter{table}{1}

%%%%%%%%%%%%%%%%%%%%%%%%%%%%%%%%%%%%%%%%%%%%%%%%%%%%%%%%%%%%%%%
\begin{table*}  %[t]
	\caption{Model restitution of the chosen observational constraints plus quantities of interest. Models listed here were optimized with the reference physics 
(set $A$, different cases, see Sect.~\ref{LM} and Tables~\ref{modelinputs} and \ref{cases}). As explained in the text, in cases $2$b, $2$c, $4$, and $7$  the optimization considers individual frequencies and separations that were corrected for near-surface effects while in cases $5$ and $6$ the optimization is based on un-corrected separation ratios 
	$r_{02}$ and $rr_{01/10}$. In case $1$ no seismic constraints are considered. In cases $1$, $5$, and $6$, we nevertheless chose to list below a corrected value of $\langle\Delta\nu\rangle$, i.e. a value corrected for surface effects a posteriori,  after the optimized model was obtained. In cases $2$a and $3$, the scaling values of $\langle\Delta\nu\rangle$ are given, while in case $2$c, we list $\Delta\nu_\mathrm{asym}$. Models without reference options are presented in the appendix.}
\begin{tabular}{lcccccccccccccl}
\hline\hline
Model &   $T_\mathrm{{eff}}$ &$L$ & [Fe/H] & $\log g$ &  $R$& 
 $\langle\Delta\nu\rangle$ & $\nu_\mathrm{max}$& $\langle r_{02}\rangle$ &$\langle rr_{01/10}\rangle$  &$X_C$ & ${\Delta Y}/{\Delta Z}$& $M_\mathrm{{cc}}$ & $R_\mathrm{{zc}}$ & $M_\mathrm{{p}}\sin i$  \\
 & {[K]} & [$L_\odot$] & [dex] & [dex] & [$R_\odot$]  &   [$\mu$Hz] &  [$\mu$Hz] & -- &   --&-- & --&  [$M_\star$] &  [$R_\star$] &[$M_\mathrm{{Jupiter}}$] \\
\hline
$A1$&6116.&2.053& 0.22&4.29&1.28&101.33&2127.& 0.074& 0.033&0.28& 2.0&0.032&0.767&1.17$\pm$0.03\\
$A2$a&6050.&2.063& 0.22&4.28&1.31& 98.13&2061.& 0.081& 0.034&0.33& 2.0&0.024&0.798&1.18$\pm$0.03\\
$A2$b&6053.&2.063& 0.22&4.28&1.31& 98.14&2089.& 0.088& 0.035&0.40& 2.0&0.019&0.819&1.19$\pm$0.03\\
$A2$c&6067.&2.082& 0.21&4.28&1.31& 98.26&2083.& 0.086& 0.034&0.38& 2.0&0.021&0.813&1.19$\pm$0.03\\
$A3$&6024.&2.066& 0.22&4.28&1.32& 98.15&2086.& 0.086& 0.034&0.39& 1.5&0.017&0.813&1.20$\pm$0.03\\
$A4$&6110.&2.055& 0.22&4.28&1.28& 98.13&2035.& 0.084& 0.035&0.33& 3.3&0.035&0.809&1.14$\pm$0.03\\
$A5$&6116.&2.053& 0.22&4.32&1.28&104.16&2253.& 0.083& 0.033&0.38& 1.0&0.011&0.785&1.22$\pm$0.05\\
$A6$&6043.&2.081& 0.23&4.29&1.32& 98.29&2095.& 0.083& 0.034&0.37& 1.5&0.021&0.801&1.20$\pm$0.03\\
$A7$&6020.&2.102& 0.23&4.29&1.34& 98.29&2124.& 0.084& 0.034&0.39& 0.8&0.014&0.800&1.23$\pm$0.03\\
\hline
\end{tabular}
\label{resultsA2}
\end{table*}
%\addtocounter{table}{1}

\section{Results of \`{a} la carte  stellar modelling}
\label{results}

The inputs and results of the different models presented in Sect.~\ref{LM} are listed in 
Tables~\ref{resultsA1} and \ref{resultsA2} for the reference models (set $A$, Table~\ref{modelinputs}).
  Table~\ref{resultsA1} lists the quantities that are common main inputs of stellar models and 
  can be determined by the optimization -or not- depending on the number of available 
  observational constraints (see Table~\ref{cases}). Table~\ref{resultsA2} lists quantities that are common
   outputs of a model calculation, some of them may also be observational constraints. 
For clarity, we chose to present more results in Appendix~\ref{appendice}. 
In Tables~\ref{modelinputsA}, \ref{resultsAA1}, and \ref{resultsAA2}, the results of the model
 optimization for different options in calculating reference models of set $A$ are given. 
 In Tables~\ref{resultsB1} and \ref{resultsB2}, the results of the model optimization for 
 the input physics listed in Table~\ref{modelinputs} are presented.
% {The data in Tables  ~\ref{resultsA1}, ~\ref{resultsA2}, ~\ref{resultsAA1}, ~\ref{resultsAA2},
% ~\ref{resultsB1}, and ~\ref{resultsB2} are available at 
% \url{http://mygepi.obspm.fr/~lebreton/Modeles/CESAM.html}, in ASCII format.}

The reduced values ($\chi^2_\mathrm{R}$) are also given to show the goodness of the match. Depending on the optimization case, there may be orders of magnitude
 differences in the $\chi^2_\mathrm{R}$-values. For instance, in case $1$, 
 the $\chi^2_\mathrm{R}$-values are very low because it is quite easy to find a 
 model that matches the classical parameters alone. Hence, {in case $1$, 
  for a given set of input physics and fixed free parameters, 
 the optimization provides a very precise solution but the stellar model may not be accurately determined.}
The adopted values of the fixed parameters such as the mixing length may not be appropriate for the studied star; that introduces biases that
can have strong impact on the results, as discussed below. {Conversely, in the other cases, seismology sets severe constraints on the models,
which results in higher $\chi_\mathrm{R}^2$-values. The result is then less precise but the stellar model 
is more accurate.}  Note that at this point we listed all models regardless of their $\chi_\mathrm{R}^2$-value.

\subsection{Observational constraints restitution}

\begin{figure*}
\resizebox{\hsize}{!}
{\includegraphics{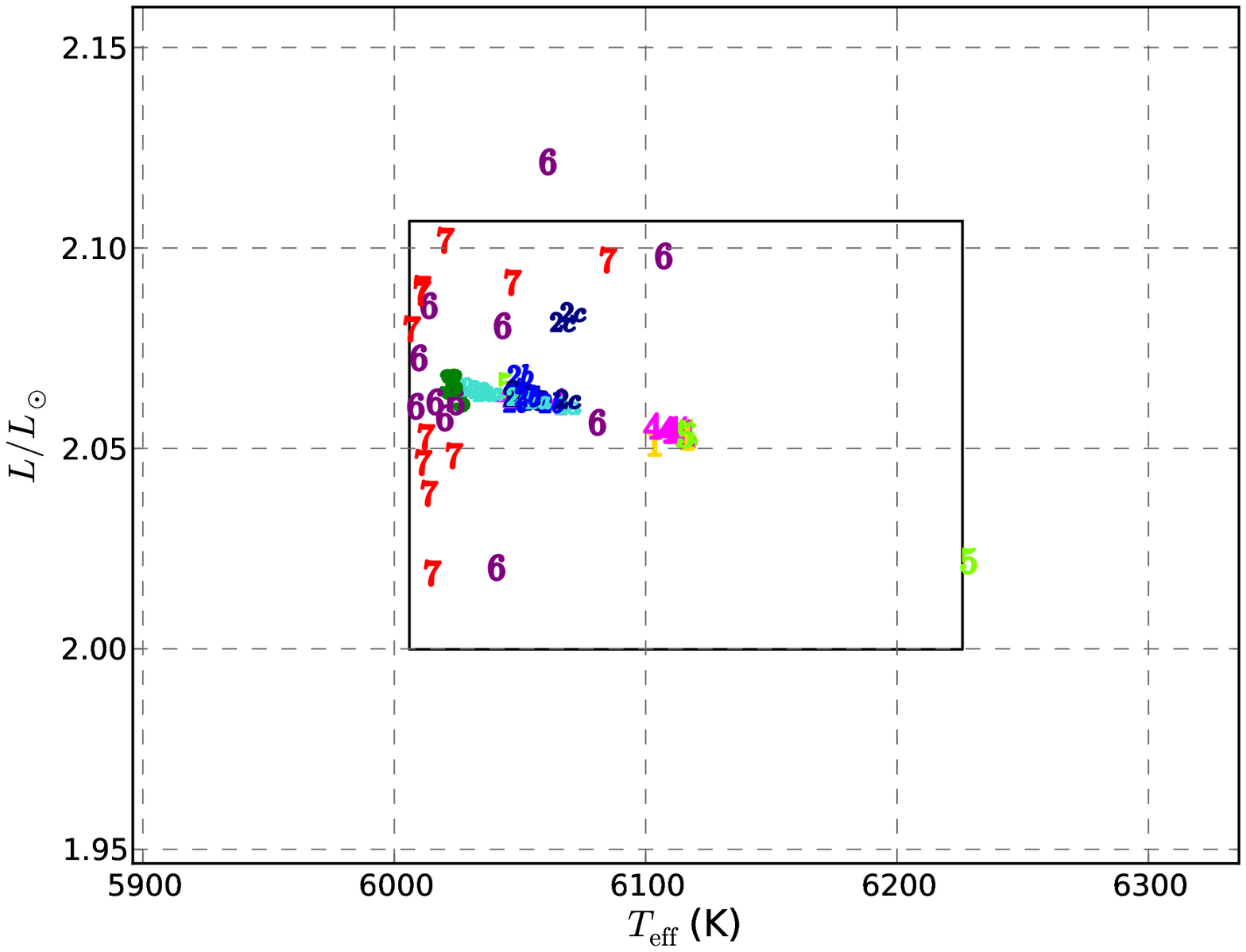}\includegraphics{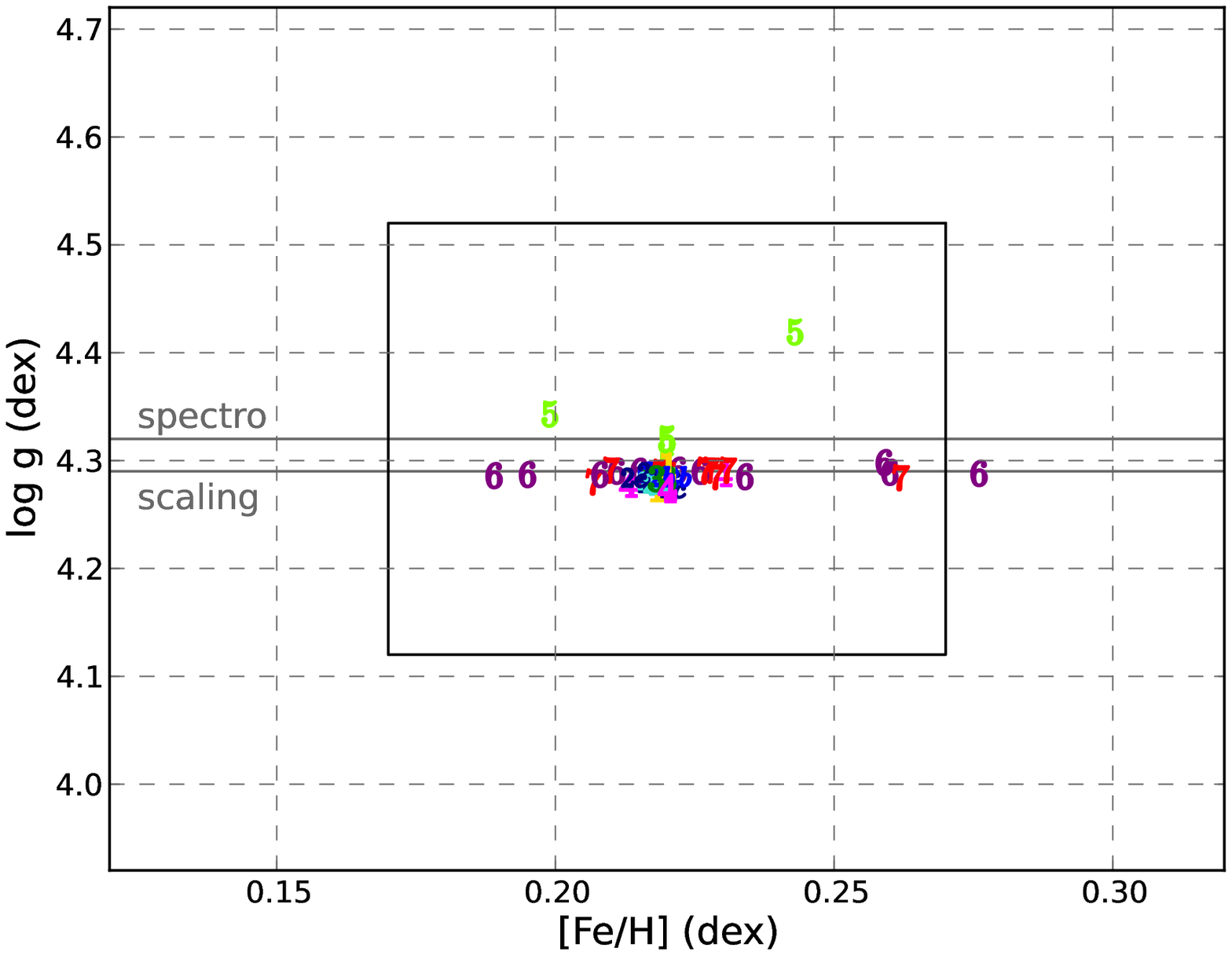}}
\resizebox{\hsize}{!}
{\includegraphics{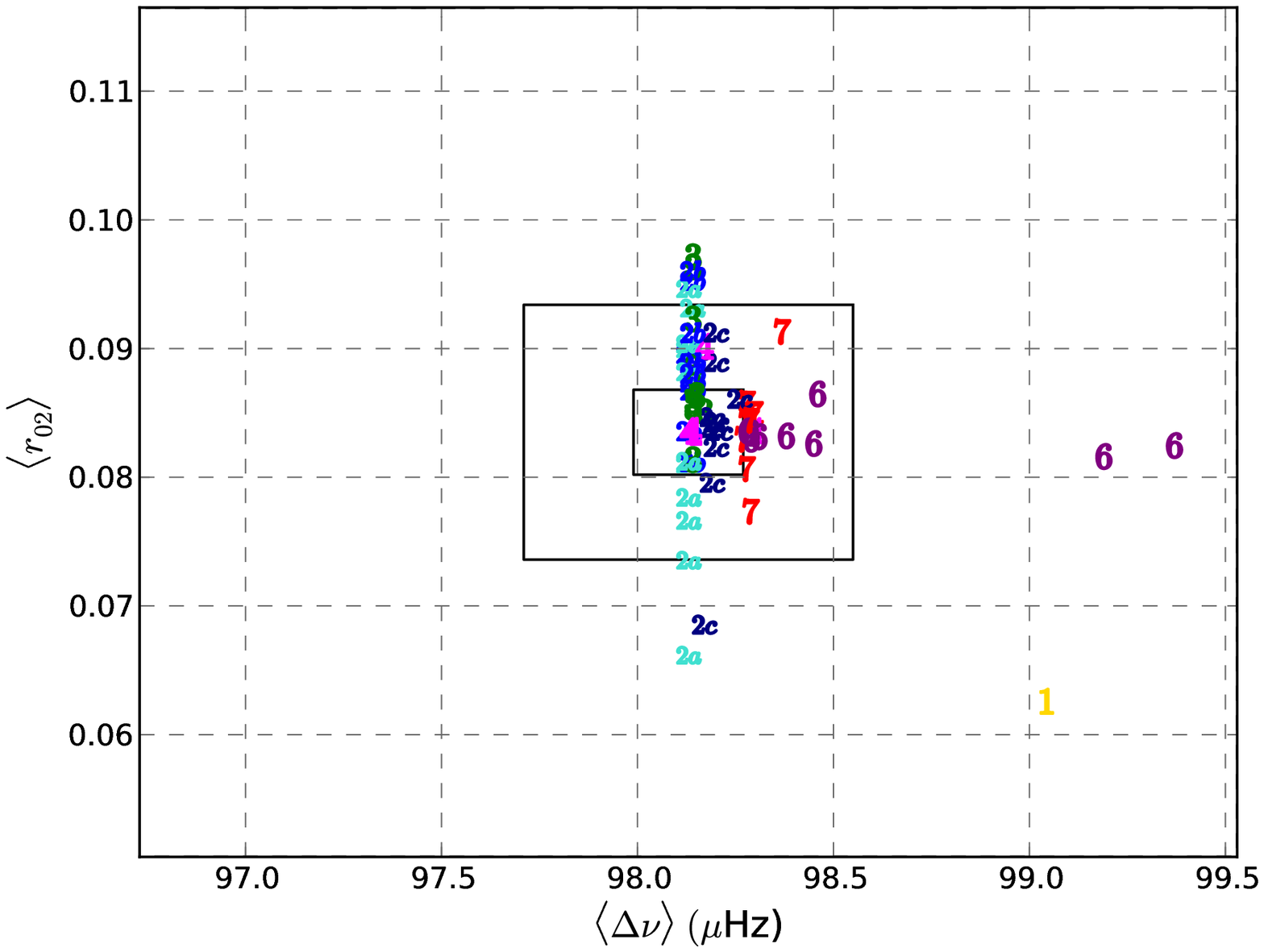}\includegraphics{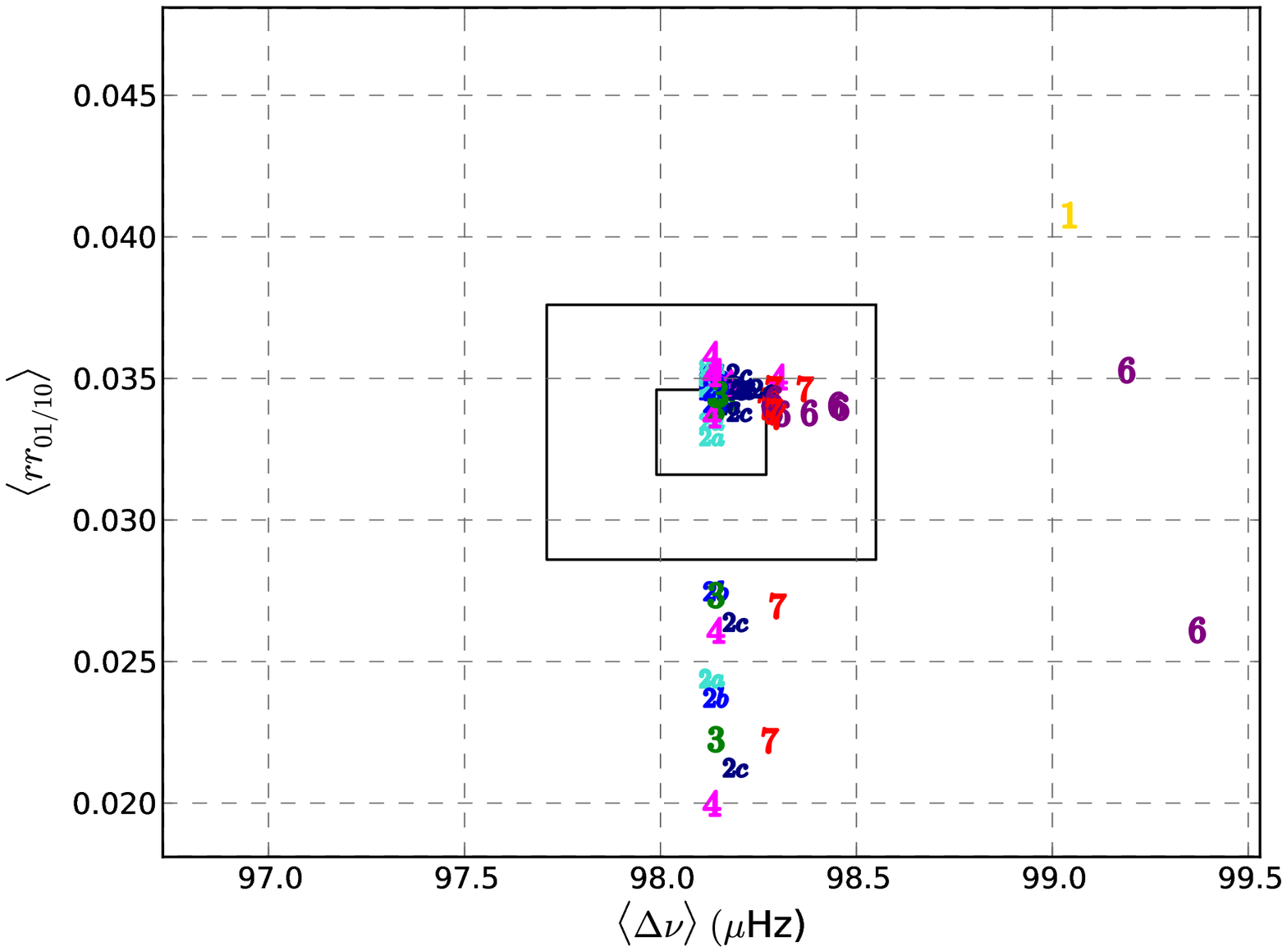}}
\caption{
Restoration of the observational constraints in the models. Upper figures are for classical parameters: H-R diagram (left), [Fe/H]-$\log$ g plane (right). The inner rectangle delimits the $1\sigma$ observational error bars, and the boundaries of the whole plotting area represent $2\sigma$ errors. Horizontal lines give the $\log$ g values of spectroscopy and seismic scaling. Lower figures are for seismic indicators: $\langle \Delta \nu \rangle$-$\langle r_{02}\rangle$ plane (left) and $\langle \Delta \nu \rangle$-$\langle rr_{01/10}\rangle$ plane (right). Here, the rectangles delimit  $1$ and $3 \sigma$ error bars, and the boundaries of the whole plotting area are  $10\sigma$ errors. For each point, the symbol corresponds to the model number, as explained in Table~\ref{cases}. {We used different colours to highlight the different cases (sets of observational constraints as defined in Table~\ref{cases}) at the basis of the modelling. Note that these
colours are unrelated to the colours defined in Table~\ref{modelinputs}  and used in Figs. \ref{Allages} to \ref{rayon}, and in Fig.~\ref{planet}.}
    }
\label{goodness}
\end{figure*}

{To each of the nine sets of observational constraints presented in Table~\ref{cases} correspond 11 different models 
optimized with the input physics listed in Table~\ref{modelinputs}. Figure~\ref{goodness} illustrates how the observational constraints of HD~52265 
listed in Table~\ref{param} are reproduced by these models.} 

{First, we examine the classical observables $T_\mathrm{{eff}}$, $L$, [Fe/H], and $\log$ g. We considered two diagrams, a $T_\mathrm{{eff}}$--$L$ and
a  [Fe/H]--$\log$ g diagram, as plotted in the upper left and right panels of  Fig. ~\ref{goodness}. For the purpose of readability
we distinguish each set in Table~\ref{cases} but not the different input physics in Table~\ref{modelinputs}. These latter are discussed in the next sections. 
The total size of these diagrams is the size of the 2$\sigma$ error box
on the classical parameters, while the inner box identifies the  1$\sigma$ error box. The figures show that all models satisfy 
the classical constraints at the 2$\sigma$ level 
with only two models lying outside the $1 \sigma$ error box on $T_\mathrm{{eff}}$, $L$, [Fe/H], and $\log$ g.} 
These outliers are model $E6$ ([Fe/H]=0.28, no microscopic diffusion) and model $K6$ ($L/L_\odot= 2.12$, convective penetration). 
We point out that models optimized with seismic data {(especially the case $6$ and $7$ models)} show a trend towards the lower range of the observed spectroscopic effective 
temperature. {This supports the claim for a redetermination of the observed $T_\mathrm{{eff}}$  in the light of the seismic surface gravity, which is robust, as discussed 
below.}

The seismic value for $\log$ g is found to be mostly independent of the model input physics (see
also Sect.~\ref{radius}) and differs from the spectroscopic one by $0.03$ dex. 
Note that two models have a $\log$ g in the upper range of the spectroscopic 
error bar, that is higher and more different than the seismic value. These are the
 two models of case $5$ with overshooting ($I5$ and $J5$). However, 
 as discussed below, these models are not considered in the following 
 (model $I5$ has a very low initial helium content, much lower than the
  primordial value, while model $J5$ has a high $\chi_\mathrm{R, seism}^2$). 
 
 {Second, we examine the seismic mean observables $\langle \Delta \nu \rangle$, $\langle r_{02}\rangle$, 
  and $\langle  rr_{01/10} \rangle$. We considered two diagrams, a $\langle \Delta \nu \rangle$--$\langle r_{02}\rangle$ and
 a  $\langle \Delta \nu \rangle$--$\langle rr_{01/10}\rangle$ diagram, as plotted in the lower left and right panels of  Fig. ~\ref{goodness}.
 Here, the total size of these diagrams is the size of the 10$\sigma$ error box
 on the seismic parameters, while two inner boxes delimit the $1\sigma$ and $3\sigma$  error boxes.} 
 We note that many models lie outside
 the $1 \sigma$ error box of  $\langle \Delta \nu \rangle$, $\langle r_{02}\rangle$, 
 and $\langle  rr_{01/10} \rangle$. {Some models are not even in the plotted areas.} In most of these cases, the seismic data have not been used as model constraints.
{ In particular, no model of case $1$ matches the seismic constraints.
 This shows that a model that only matches the classical parameters can indeed be very far from 
 a seismic model.}

We also note that several models in Tables~\ref{resultsA2},
 \ref{resultsAA2}, \ref{resultsB2} have $\nu_\mathrm{max}$-values 
well outside the $1\sigma$ range of the observational determination 
of \citet{2011A&A...530A..97B} (we recall that case $3$ models have been  optimized on $\nu_\mathrm{max}$).
This can be explained by the fact that the error bar on $\nu_\mathrm{max}$ given 
by \citeauthor{2011A&A...530A..97B} is an internal error that is quite small 
($1$ per cent). As shown by \citet{2013JPhCS.440a2031B}, the method adopted
 to infer $\nu_\mathrm{max}$ from the power spectra affects its value. 
 Furthermore, one can expect differences between the  observational value
  of $\nu_\mathrm{max}$ and the theoretical value, which obeys the scaling relation \citep[e.g.][]{2013ASPC..479...61B}. 

In the following, to derive the age and mass of HD~52265 and their uncertainties, 
we considered only the models that satisfy $\chi^2_\mathrm{R, classic}<1$. Because of the high precision of the frequencies, 
no model with $\chi^2_\mathrm{R, seism}<1$ was found. We therefore only kept models with
 $\chi^2_\mathrm{R, seism}<2$.

\subsection{Age of HD~52265}
\label{Age}

\begin{figure*}
      \resizebox{\hsize}{!}
	     {\includegraphics{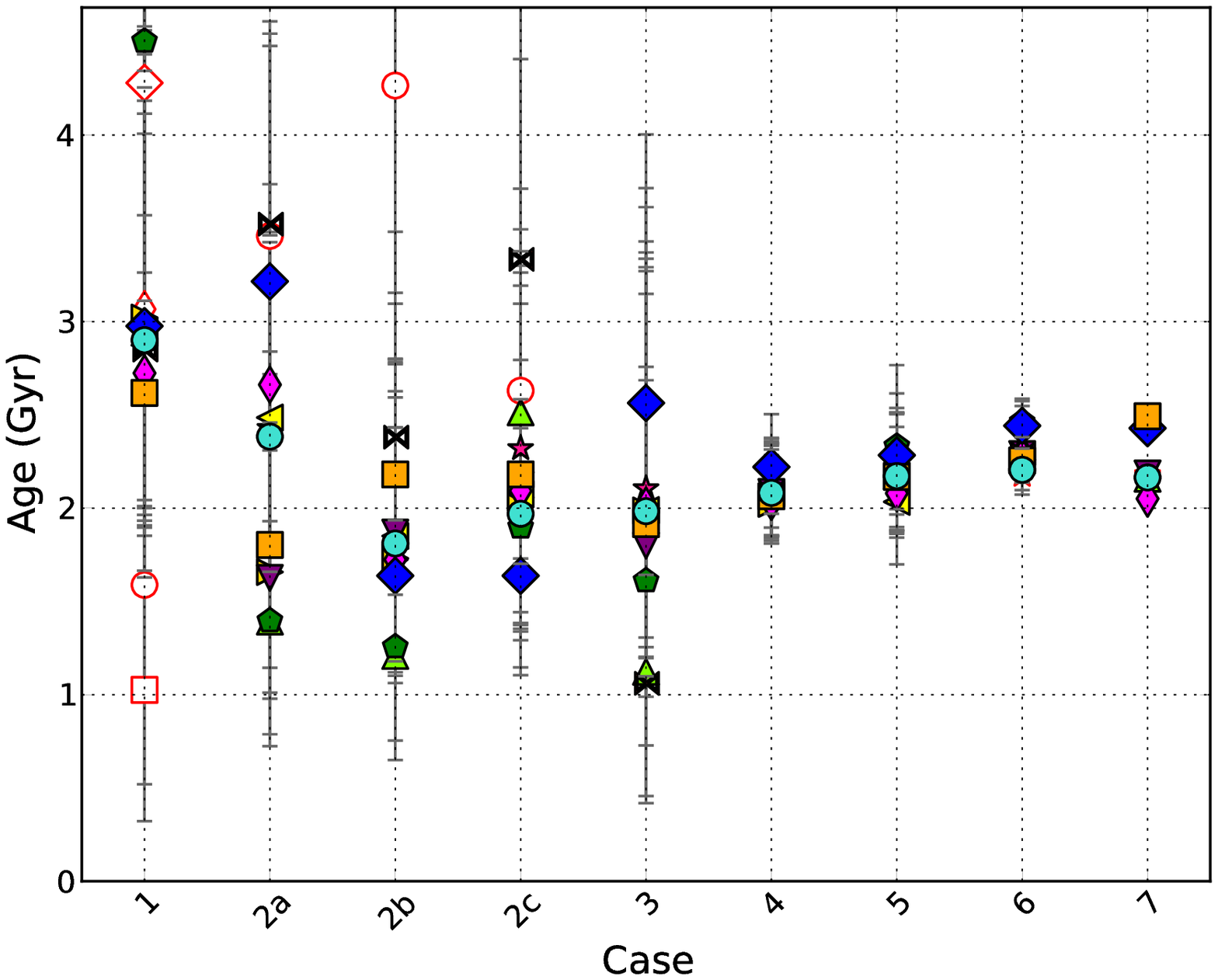}\includegraphics{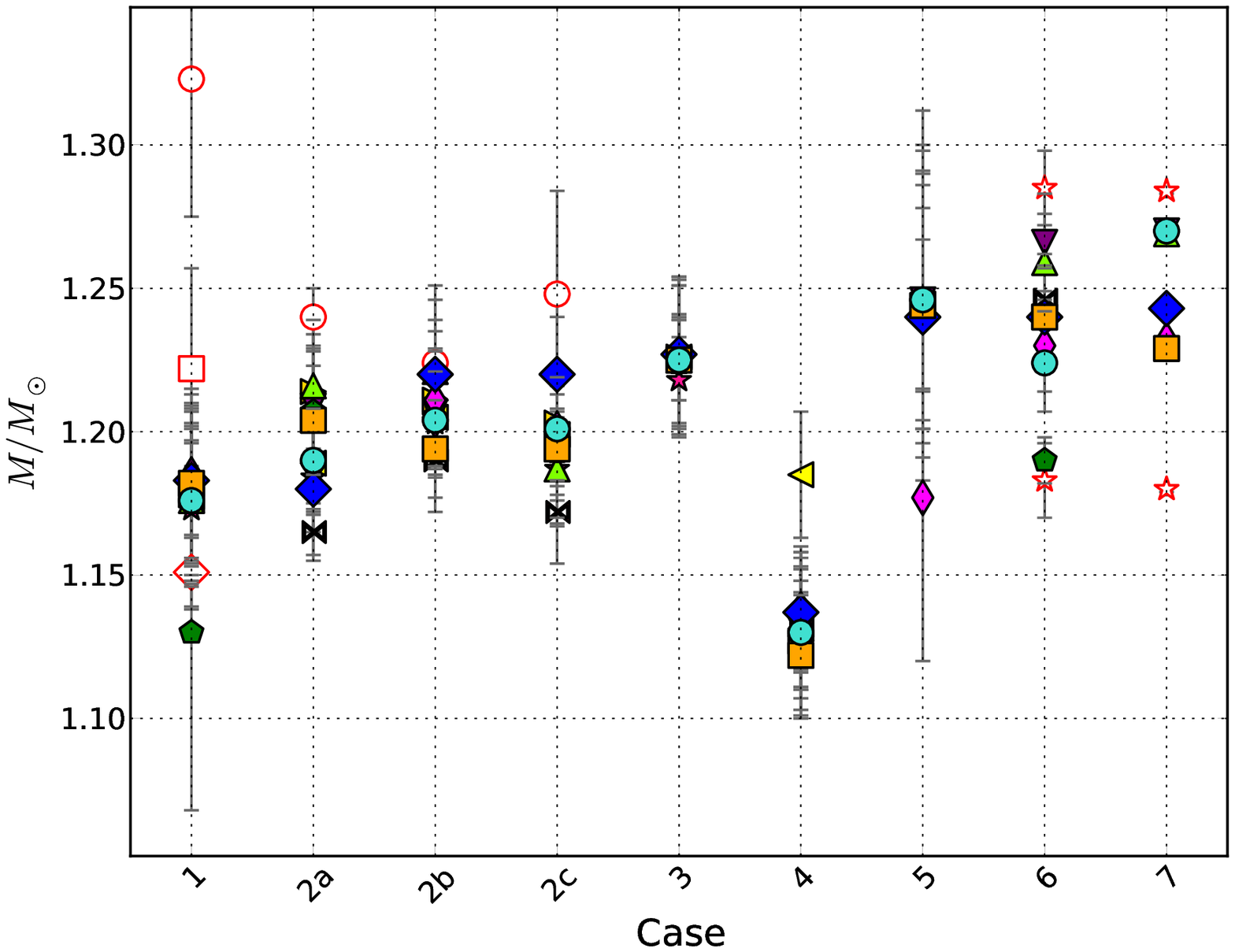}}
	       \caption{Ranges of ages (left) and masses (right) derived from stellar model optimization for HD~52265. In the abscissae, we list
	        the case numbers as defined in Table~\ref{cases}. For each case, several model optimizations can be identified according to the symbols and colours indicated in Table~\ref{modelinputs}. In addition, open red symbols are for additional models of set $A$ described in Table~\ref{modelinputsA} of Appendix~\ref{subsets}: circles are for different, low $Y_0$ values, square and diamond are for low and high $\alpha_\mathrm{conv}$ values, small diamonds for models 
	        with large core overshooting. Red stars illustrate the $Y_0$-$M$ degeneracy in cases $6$ and $7$, but the inferred range is the same for all cases.
    }
\label{Allages}
\end{figure*}

Figure~\ref{Allages} (left panel) shows for each case in Table~\ref{cases} 
the age of  the optimal model for a given set of physics in Table~\ref{modelinputs}.

 In case $1$ ($2$), $Y$ and  $\alpha_\mathrm{conv}$ ($Y$) 
 could not be inferred and had to be fixed by the modeller.  
 Therefore, for these two cases, we calculated additional models
  with somewhat extreme choices of $Y$ and  $\alpha_\mathrm{conv}$. In case $1$, we also included a model with much 
  overshooting ($\alpha_\mathrm{ov}=0.30$). 
  These additional models, included in Fig.~\ref{Allages},  are presented in Appendix~\ref{subsets}.

%\addtocounter{figure}{1}

In case $1$ (column 1 in Fig.~\ref{Allages}), 
there is a large scatter in the ages of HD~52265 obtained 
with different sets of input physics when no seismic observations are available,
that is ${\approx\pm}60$ per cent with respect to the reference age. 
 This is the usual situation of age-dating from classical parameters
  $L$, $T_\mathrm{eff}$ and surface [Fe/H]. The values of $Y$,
   $\alpha_\mathrm{conv}$, and $\alpha_\mathrm{ov}$ had to be fixed and 
   different ages result from different choices. 
   In particular, for a change of  $\alpha_\mathrm{conv}$ of $20$ per cent around 
   the solar calibrated value, the age changes by more than $50$ per cent.
At the highest boundary of the age interval, the oldest models are those without diffusion
and the model  with a long mixing length $\alpha_\mathrm{conv}=0.826$. The youngest  
models are models with the lowest --primordial-- initial helium abundance ($Y_\mathrm{P}=0.245$) 
 and the model with low $\alpha_\mathrm{conv} = 0.55$. The  ages of the other models are concentrated   
 in a narrow age interval, $2.6$-$3.0$ Gyr, about that of reference model $A$.
   
    We point out that if the error bars on the classical parameters 
    were to be reduced, as will be the case after the Gaia-ESA 
    mission \citep[see e.g.][ and references therein]{2012MNRAS.426.2463L}, 
    the error bar of an individual age determination with a given set of 
    input physics would be reduced, but 
{ the scatter associated with the use 
    of different input physics would remain the same
    unless significant  advance in stellar modelling is made.}

 In cases $2$a, b, and c, where the large frequency separation is included 
as a model constraint, the age scatter is smaller than in case $1$. 
Unlike case $1$, the ages of the 
optimal models computed with different input
physics and free parameters span the whole range of 
the scattered interval. Indeed, the values of the 
inferred mixing length differ from one case to another and still span a wide range 
[$0.466$-$0.656$] for $\alpha_\mathrm{conv, cgm}$ for case $2$a, for instance, 
as can be seen in Tables~\ref{resultsA1}, \ref{resultsAA1}, and \ref{resultsB1}. 
Moreover, the initial helium $Y_0$ slightly changes in the optimization because $\Delta Y/\Delta Z$ is fixed but $Z/X$ is adjusted in the optimization.
%For instance, for case $2$a, $Y_0$ is in the range $0.27 -0.31$. 
Note  that  for given input physics and free parameters 
(cases $2$a, $2$b, $2$c  for set $A$  in Tables~\ref{resultsA1}), the age is significantly 
modified depending on the way the mean large frequency separation is computed; the changes are correlated with changes in the inferred mixing-length values. 
   { The scatter in the inferred mixing-length values, hence on the age,
is smaller when $\langle\Delta\nu\rangle$ is calculated 
explicitly (i.e. not from the scaling relation) using the stellar models (cases $2$b and $2$c)}. 

In case $3$, the age scatter is slightly smaller than in case $2$a. 
As already pointed out in \citet[]{2013eas63123}, this is because the additional constraint on $\nu_\mathrm{max}$ 
does not add much more knowledge on the age of the star. 

Cases $4$, $5$, $6$, and $7$ all take into account seismic constraints directly sensitive 
to age, that is either the small frequency separation $d_{02}$, the frequency separation 
ratios $r_{02}$, $rr_{01/10}$, or the individual frequencies. The spectacular consequence 
is a reduction of the age scatter, as can be seen in Fig.\ref{Allages}.

For case $4$, considering the different possible options for the input physics of the stellar models,
  our criterion  $\chi^2_\mathrm{R, seism} \le 2$ excludes the 
model without microscopic diffusion (model $E4$).  Accordingly, the ages  
range between $2.02 \pm 0.22 $ Gyr (model $J4$) and $2.22 \pm 0.27$ Gyr (model $C4$). 
 This yields an age of $2.15\pm 0.35$ Gyr, that is an age uncertainty of {$\sim \pm 16$} per cent. 

For case $5$, considering the different possible options for the input physics of 
the stellar models, two models ($I5$, $J5$) were excluded because 
their initial helium abundance $Y_0$ was found to be much lower than 
the primordial value $Y_\mathrm{P}$. 
 Accordingly, the ages range between $2.08 \pm 0.25 $ Gyr (model $D5$) and
 $2.33 \pm 0.40$ Gyr (model $E5$). This yields an age of $2.28\pm 0.45$ Gyr, 
which means an age uncertainty of $\sim\pm 20$ per cent. It is possible to reduce this scatter even more. 
Helioseismology has shown that microscopic diffusion must be included 
if one considers the Sun. This must also be true for solar-like stars like HD~52265: 
its mass is only slightly larger than the solar one and it has an extended convective envelope.  
 Excluding the model without diffusion then 
yields  an age in the range $2.08 \pm 0.25 $ Gyr (model $D5$) - $2.28 \pm 0.31 $
  Gyr (model $C6$), that is, an age of $2.21\pm 0.38$ Gyr, i.e. an age uncertainty of $\pm 17$\%. 
 Hence the main cause of scatter on the high side of the age interval here is microscopic diffusion
  (model $E5$) followed by the solar mixture (model $C5$). The low side
  of the age interval comes from the change of nuclear reaction rates
   (model $D5$). {This shows that we start to reach the quality level of data that enables 
  testing the microscopic physics in stars other than the Sun.}

For case $6$, considering the different possible options for the input physics of the stellar models, 
our criteria $\chi^2_\mathrm{R, classic} \le 1$ and $\chi^2_\mathrm{R, seism} \le 2$ exclude models $I6$, $J6$ and $K6$. Accordingly, the ages range between $2.21 \pm 0.11$ Gyr (model $A6$) and $2.46 \pm 0.08$ Gyr  (model $E6$). 
 This yields an age of $2.32\pm 0.22$ Gyr, that is an age uncertainty $\sim\pm 9.5$\%.  
The main cause of scatter here is microscopic diffusion closely followed by the choice of  the solar  mixture: {\small GN93} (model $A6$) versus {\small AGSS09} (model $C6$).
The range for the mixing-length value is [0.588, 0.606].
This range is considerably narrower than the
 one usually taken a priori to compute stellar models.  
Compared with the values we obtained from a calibration of a solar model with the input physics of reference set 
$A$ ($\alpha_\mathrm{conv, cgm, \odot}=0.688\pm 0.014$),
the values obtained for HD~52265 are lower than solar by $12-15$ per cent.
The results for the initial helium abundance are discussed in Sect.~\ref{MYdeg} below.
{We point out that the range of ages obtained in case $6$ is quite close to that obtained in case $5$.
This can be understood by the fact that, as shown in Fig.~\ref{astdiag}, 
the mean value of $\langle rr_{01/10}\rangle$ is already a good indicator of the evolutionary state, 
at least if the classical parameters are also used to constrain the stellar mass and if the convective core is small. 
We therefore reach a similar accuracy in cases $5$ and $6$, but the precision on
the individual ages is better in case $6$, which is more constrained by the use of individual ratios.
Furthermore, as discussed in Sect.~\ref{seismicprop}, in case $6$, 
the individual values $rr_{01/10}$ provide additional information on the star, unrelated to age.
Their oscillatory behaviour, which is related to steep gradients of the sound speed, is
an invaluable asset to characterize the depth of the convective envelope and the 
thermal and chemical structure at its radiative-convective interface \citep{2012A&A...544L..13L}.  
However, HD 52265 is a case study, with a small
convective core, and we can expect that larger convective cores whose size is crucial for the age-dating are better
characterized by individual ratios than by their mean values.}

For case $7$, considering the different possible options for the input physics of the stellar models, our criterion  $\chi^2_\mathrm{R, seism} \le 2$ excludes several models ($E7$, $I7$, $J7$, $K7$).
%, $A$7-YMnoSE, $A$7-noSE, $A$7-ov). 
Accordingly, the ages range between $2.05 \pm 0.02 $ Gyr (model $D7$) and $2.49 \pm 0.02$ Gyr (model $B7$). This yields an age of $2.27\pm 0.24$ Gyr,
that is an age uncertainty $\sim \pm 9.5$\%. The upper boundary of the age interval is due to the choice of the {\small MLT} approach  for the convective transport 
(model $B7$) closely followed by the choice of the solar mixture (model $C7$). The lower boundary is due to changes in the nuclear rates (model $D7$).

To summarize, the best precision and accuracy, in the context of the present input physics, is 
therefore obtained in  case $4$ -constrained by the mean values of the large and small frequency separations, 
case $6$ -constrained by the individual values of the frequency separation ratios, and case $7$ -constrained by individual frequencies. 
However, as stressed above, 
cases $4$ and $7$ suffer from the caveat that the individual frequencies were corrected for surface effects. 
As can be seen from model $A7$-$\mathrm{noSE}$ in Appendix~\ref{appendice}, 
the age is increased by more than $40$ per cent in a model optimized without correcting for these effects. 
{ We therefore consider that case $6$ is the optimal choice of observational constraints 
regarding the age of the star \citep[see also the discussion by][]{2013ApJ...769..141S}}.

Additional models, based on different choices for the optimization (correction from surface 
effects, correlations between the seismic ratios, etc.), do not imply drastic changes in
 the optimized age (see Appendix~\ref{subsets}).

\emph{Degeneracy between age and mixing length:}
Figure~\ref{YM}, left panel, shows the relation between the
 convection parameter $\alpha_\mathrm{conv}$ and age of the optimized models.
  There is a possible trend for $\alpha_\mathrm{conv}$ to increase with age. 
From a linear regression we derived $\alpha_\mathrm{conv}/\alpha_\mathrm{conv, \odot}\simeq 0.13\times A +0.59$,
 where the age is in Gyr. This trend can be understood as follows: higher ages on the MS would imply
   lower $T_\mathrm{eff}$ and larger radii.  
The radius of a star is smaller for a more efficient convective energy transport. Smaller radii, hence
  higher $\alpha_\mathrm{conv}$ --related to more efficient convection-- 
  are therefore needed to bring $T_\mathrm{eff}$ back into the observational range. However, we checked that the ages determined by model optimization in cases $6$ and $7$ are not affected by this possible degeneracy because allowed variations of $\alpha_\mathrm{conv}$ along the regression line are limited by the constraints on $T_\mathrm{eff}$ and [Fe/H].

\subsection{Mass and initial helium abundance}
\label{MYdeg}
\begin{figure*}
      \resizebox{\hsize}{!}
	     {\includegraphics{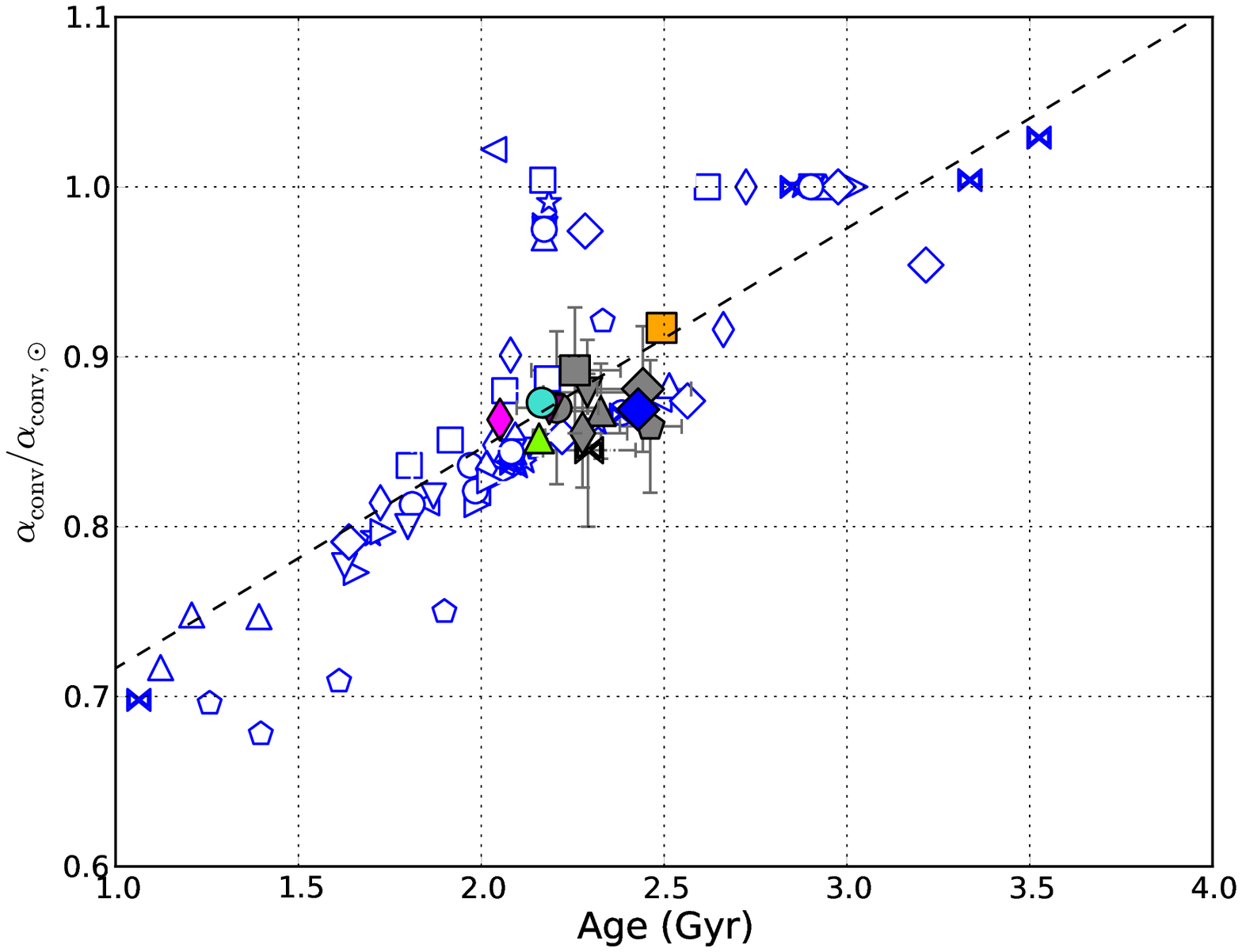}\includegraphics{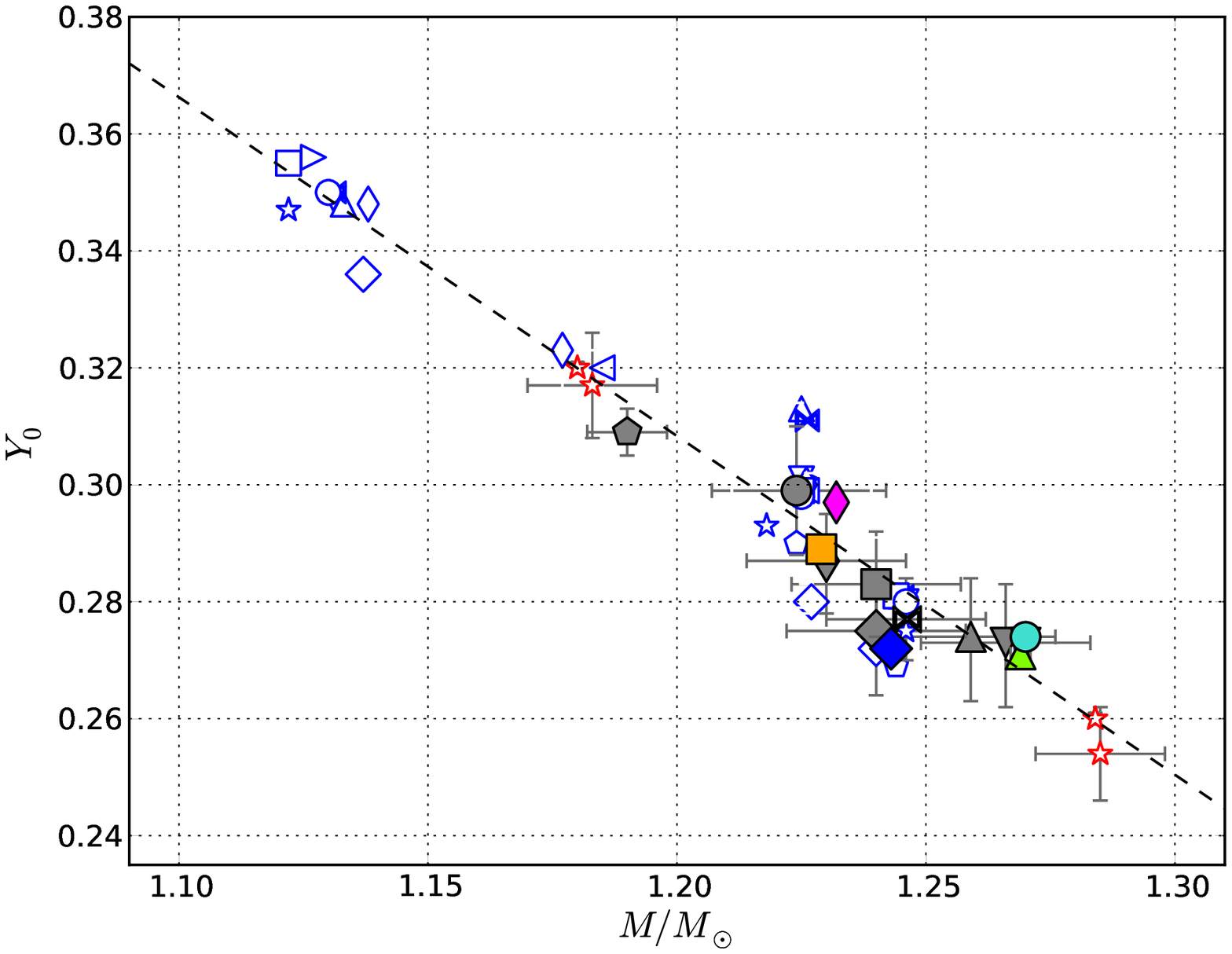}}
	       \caption{
	       		 \emph{Left}:  		     
	       relation 
	       		 between the convection parameter $\alpha_\mathrm{conv}$
	       		 and the age.
	       To compare models, we expressed $\alpha_\mathrm{conv}$ 
	        in units of the solar value, 
	       which, as mentioned in Sect.~\ref{physics}, mainly depends on the 
	       convection theory used in the model, i.e. 
	        $\alpha_\mathrm{conv, \odot, cgm}=0.688$ and $\alpha_\mathrm{conv, \odot, mlt}=1.762$.
	       The regression line is 
	       $\alpha_\mathrm{conv}/\alpha_\mathrm{conv, \odot}\simeq 0.13\times A+0.59$ ($A$ in Gyr.). 
	       		    Colour symbols are for case $7$ models 
	       		    (symbols and colours are listed in 
	       		    Table~\ref{modelinputs}), grey symbols
	       		     for case $6$, and open blue for other cases.
	       \emph{Right}: relation between the initial helium abundance and
	        the mass of HD~52265, as inferred from model optimization. 
		The regression line is $Y_0\simeq-0.58\times M/M_\odot + 1.00$.
		 Open red stars illustrate the impact of the $Y_0$-$M$ 
		 degeneracy  (Table~\ref{modelinputsA}). 
		     }
\label{YM}
\end{figure*}

Figure~\ref{Allages}, right panel shows for each case in Table~\ref{cases}
 the mass of  the optimal model for a given set of physics in Table~\ref{modelinputs}. 
 Like the age, the mass is better constrained when seismic data are taken into account, 
 in particular when seismic constraints explicitly sensitive to mass are used (large frequency separation, frequency at maximum power as in cases $2$ and $3$). 

Optimization of models of case $3$ relies on the scaling-relations 
(Eqs.~\ref{scalingnumax} and ~\ref{scalingdeltnu}). These models
 therefore have the peculiarity that their mass and radius are tightly fixed
 because scaling relations provide quite { precise} values of the mass and radius. 
 On the one hand, the accuracy of the derived values for the mass and radius depend on the { accuracy} of the scaling relations.
As proposed in Sect.~\ref{Age}, models of case $6$, optimized with the frequency separation 
   ratios, provide the best age but since they do not constrain surface layers, they do not explicitly fix the mass. 
 On the other hand, the ages of models of case $7$ are less secure because they are affected by the correction for surface effects. 
 Nevertheless, as can be seen from model $A7$-$\mathrm{noSE}$ in Appendix~\ref{subsets}, correction for surface effects only slightly modifies the optimized mass.
  { Thus, we suggest that models of  case $7$, optimized with the individual 
    frequencies, are probably more suitable for mass determination.}

\emph{Degeneracy between mass and initial helium abundance:}
Figure~\ref{YM}, right panel, shows the relation between the initial helium abundance 
$Y_0$ and mass $M$ of the optimized models. Like in \citet{2012A&A...538A..73B},
 a clear anti-correlation between $Y_0$ and $M$ is found. 
  The lower $Y_0$, the higher $M$.
 From a linear regression, we derived $Y_0\simeq-0.58\times M/M_\odot + 1.00 $ 
with a scatter about this mean value of less than $\pm 0.02$.

 This $Y_{0}$-$M$ degeneracy
 agrees with what is expected from
    homology relations \citep[see e.g.][]{cg68}. For a MS star in the 
 domain of mass of HD~52265, the luminosity varies as 
 $L{\propto}\mu^{7.5} M^{5.5}R^{-0.5}$ with
  $R{\propto}\mu^{0.55} M^{0.73}$ ({\small CNO} cycle) or 
  $R{\propto}\mu^{-0.43} M^{0.14}$ (pp chain). In addition, for a
   fully ionized gas, $\mu{\approx}4/(8-5Y-6Z)$, which increases
    with $Y$. Therefore, for a given observed luminosity -fixed 
    in the optimization process- there is a range of pairs 
    ($Y_0$, $M$) that lead to the same $L$ value.
The $Y_0$-$M$ degeneracy may severely hamper the determination of 
the mass of HD~52265, as shown by the results of cases $4$ and $5$. 
A large scatter in the mass value is obtained. However, low-mass models such as those found in case $4$ 
have high values of $Y_0$ (i.e. higher than $0.33$), which are hardly acceptable.
More reasonable $Y_0$ values (i.e. lower than $0.30$) would yield models with  masses larger than $1.20 M_\odot$. 

For set $A$, cases $6$ and $7$, we calculated additional optimized models with different ($Y_0$, $M$) pairs. Results are given in Appendix~\ref{subsets} and appear as open red stars in the different figures. 
For our preferred case $7$, taking into account the $Y_0$-$M$ degeneracy and keeping $Y_0$ in the range $0.26$-$0.32$, we found reference models of set $A$ with masses in the range $1.18-1.28\ M_\odot$ and ${\Delta Y}/{\Delta Z}$ in the range $0.4-2.3$. The scatter in mass around the central value is of $\sim 5$ per cent. 
%If we further assume that metal-rich stars probably have helium values higher than the solar one, the mass scatter is narrower ($1.18-1.28\ M_\odot$).
In addition, for a given model, changing the physics would induce a mass scatter of about $0.04\ M_\odot$ ($\sim 3$ per cent). We note that, as is found in solar modelling, models $C$ optimized with the {\small AGSS09}  solar mixture show a trend towards lower $Y_0$, related to their lower metallicity. Other impacts
 of the $Y_0$-$M$ degeneracy are discussed below.

\subsection{Radius and surface gravity}
\label{radius}

Figure~\ref{goodness} (top, left) shows that the range of surface gravities 
of the models is very narrow, which means that $\log$ g is very well 
determined by the modelling and is hardly sensitive to the input physics 
{\citep[see also the determinations of the seismic surface gravities of {\it Kepler} stars by][]{2012ApJ...749..152M,2012ApJ...748L..10M}}.
 For HD~52265, taking into account the $Y_0$-$M$ degeneracy, 
 seismic models of cases $6$ and $7$ have $\log$ g in the range 
  $4.28-4.32$, which improves the precision
   on $\log$ g by a factor of ten with respect to spectroscopy. We point out that
    for this star, the central value of the spectroscopic
     $\log$ g ($4.32\pm 0.20$ dex) agrees well with 
     both the value inferred from the scaling relation 
     ($4.29\pm 0.01$ dex) and our seismic optimized value
      ($4.30\pm 0.02$). This would not be the case for all the stars.
The potential of asteroseismology to improve 
       the determination of surface gravity and therefore of 
        spectroscopic parameters ($T_\mathrm{eff}$, [Fe/H] ) has already been demonstrated. 
	For instance, it has been applied to the spectroscopic analysis of two CoRoT targets by \citet{2013A&A...552A..42M},
	  and proposed for the calibration of $\log$ g of Gaia stars by \citet{2013MNRAS.431.2419C}.

\begin{figure}
      \resizebox{\hsize}{!}
	     {\includegraphics{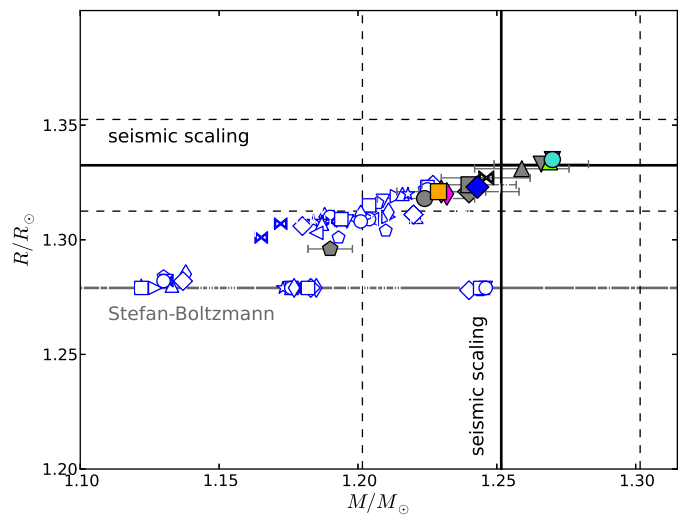}}
	       \caption{Relation between the radius and the mass of HD~52265, 
	       as inferred from model optimization. 
	       Colours are for optimized models of case 
	       $7$ of Table~\ref{cases} for different input physics, 
	       as listed in Table~\ref{modelinputs}. Models for case 
	       $6$ are plotted in grey, other cases are shown with open blue symbols. 
Bold solid horizontal and vertical black lines represent the values
of the mass and radius derived from the scaling relations (Sect.~\ref{scaling}). The
corresponding uncertainties are displayed with dashed lines. The radius
obtained using the Stefan-Boltzmann relation (Sect.~\ref{classic}) is 
shown as a triple-dot dashed line. 
Three groups of models lie on the Stefan-Boltzmann line: 
case $1$, $4$, and $5$ models, as discussed in the text.
    }
\label{rayon}
\end{figure}

The results for the star radius are shown in Fig.~\ref{rayon}. Interestingly, 
all models have a radius in-between the Stefan-Boltzmann radius $R_\mathrm{SB}=1.28 \pm 0.06\ R_\odot$ 
and the scaling radius  $R_\mathrm{sc}=1.33\pm 0.01\ R_\odot$. 
Three groups of models lie on 
the Stefan-Boltzmann line. There are case $1$ models --which is
 expected because they were optimized with only the classical parameters-- 
 and nearly all models of case $4$ and $5$. In the optimization process,
  these latter reproduce the observed effective 
  temperature and luminosity of HD~52265 best.
Case $6$ and case $7$ models with different input physics and accounting for the the $Y_0$-$M$ 
degeneracy, have seismic radii in the range $1.30-1.34\ R_\odot$. This 
represents a precision of $\approx \pm 1.5$ per cent on the radius, which is a good improvement on
 what can be obtained with the Stefan-Boltzmann law ($\approx \pm 5$ per cent) for HD~52265. 
%The $Y_0$-$M$ degeneracy slightly alters this result but the precision remains better than $\sim 2$ per cent.

\subsection{Internal structure}
\label{internal}
HD~52265 has a very small convective core and a convective envelope. 
For instance, in model $A7$, the convective core 
has a mass $M_\mathrm{cc}\sim 0.014\  M_\star$ and a 
radius $R_\mathrm{cc}\sim 0.045\  R_\star$, while the radius 
at the basis of  the convective envelope is 
$R_\mathrm{zc}\sim 0.80\ R_\star$. These quantities can 
be quite different in other models, which are nevertheless seismically equivalent. 
In particular, the $Y_0$-$M$ degeneracy has a major impact on the core mass. For instance,
 changing $Y_0$ from $0.26$ to $0.32$ changes $M_\star$ from $1.28$ to $1.18 M_\odot$, 
 $M_\mathrm{cc}$ from $0.006$ to $0.023\  M_\star$, and $R_\mathrm{cc}$ from $0.035$ to $\sim 0.053\  R_\star$. 
 The depth of the convective envelope is unaffected.

Models optimized with rather high, currently accepted or predicted values of core overshooting (sets $I$ and $J$)
 show quite high values of $\chi^2_\mathrm{R, seism}$, indicating that overshooting is probably ruled out for this star. 
 We investigated this point in depth by performing an optimization where we also adjusted the overshooting parameter 
 (models $A6$-ov and $A7$-ov, in the appendix). We found that quite low overshooting is indeed preferred, with $\alpha_\mathrm{ov}$ in the range $0.0-0.04$.
In principle, seismology has the potential to distinguish between different values of overshooting even for small cores
 through the signature left  in the oscillation spectrum by the convective core \citep[see the recent works by][and references therein]{2013ApJ...769..141S, 2013MNRAS.tmp.2970B}. Such diagnostics are beyond the scope of this paper. However, we made some additional tests that indicate that
  in the case of HD~52265, the seismic data are probably not precise enough to allow us to infer the size of the mixed core precisely or accurately.

\subsection{Seismic properties}
\label{seismicprop}
\begin{figure*}
      \resizebox{\hsize}{!}
	     {\includegraphics{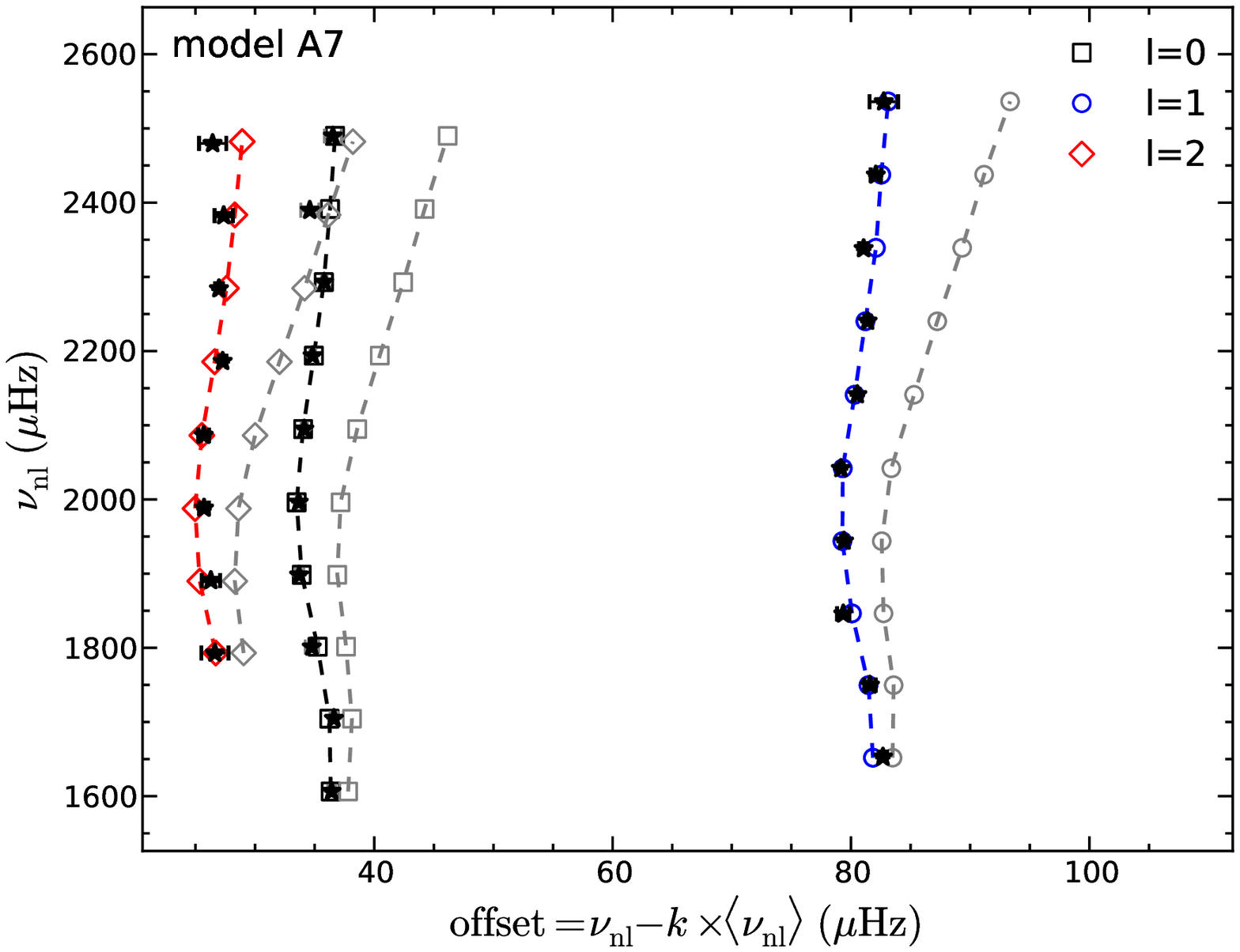}\includegraphics{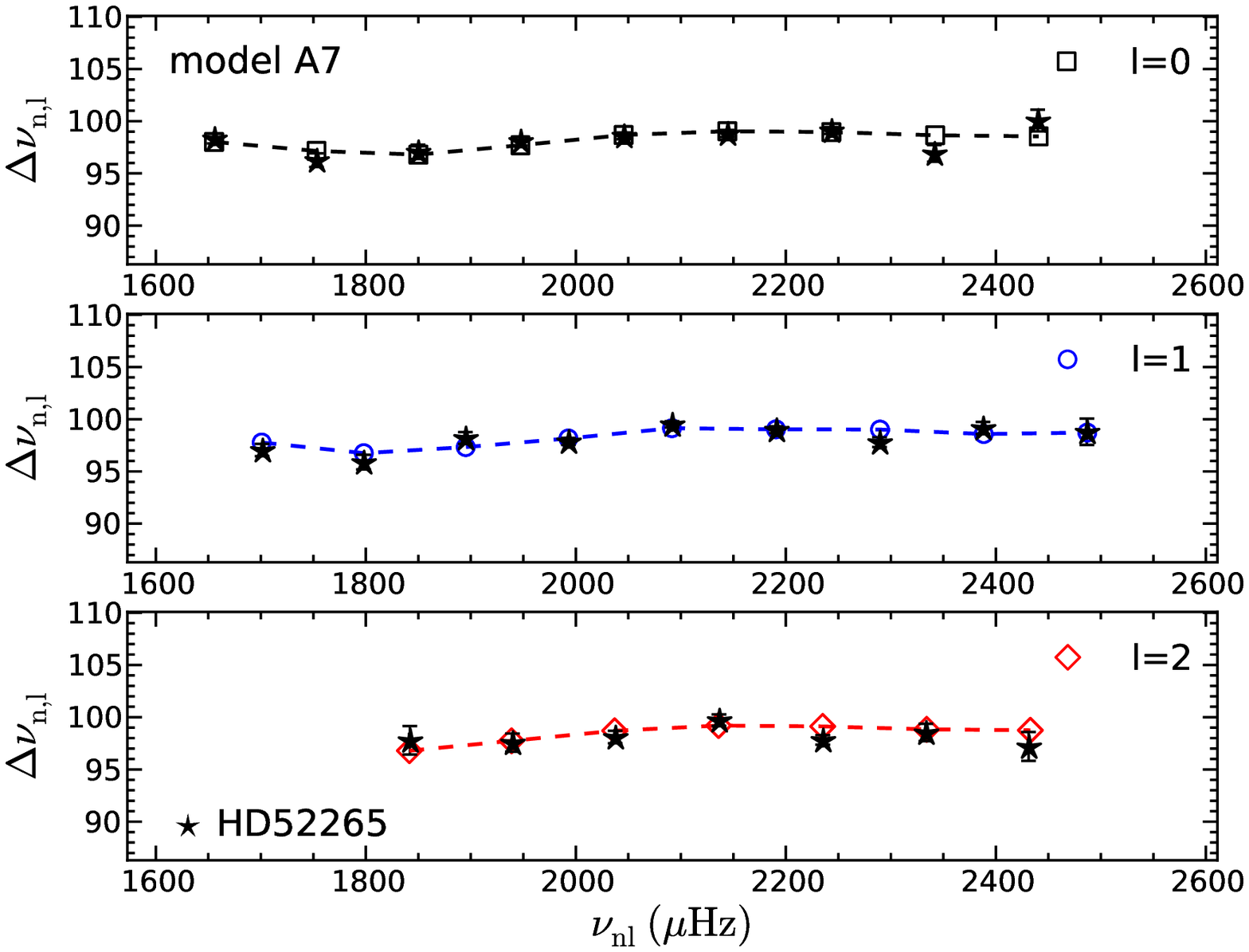}}
\resizebox{\hsize}{!}
	     {\includegraphics{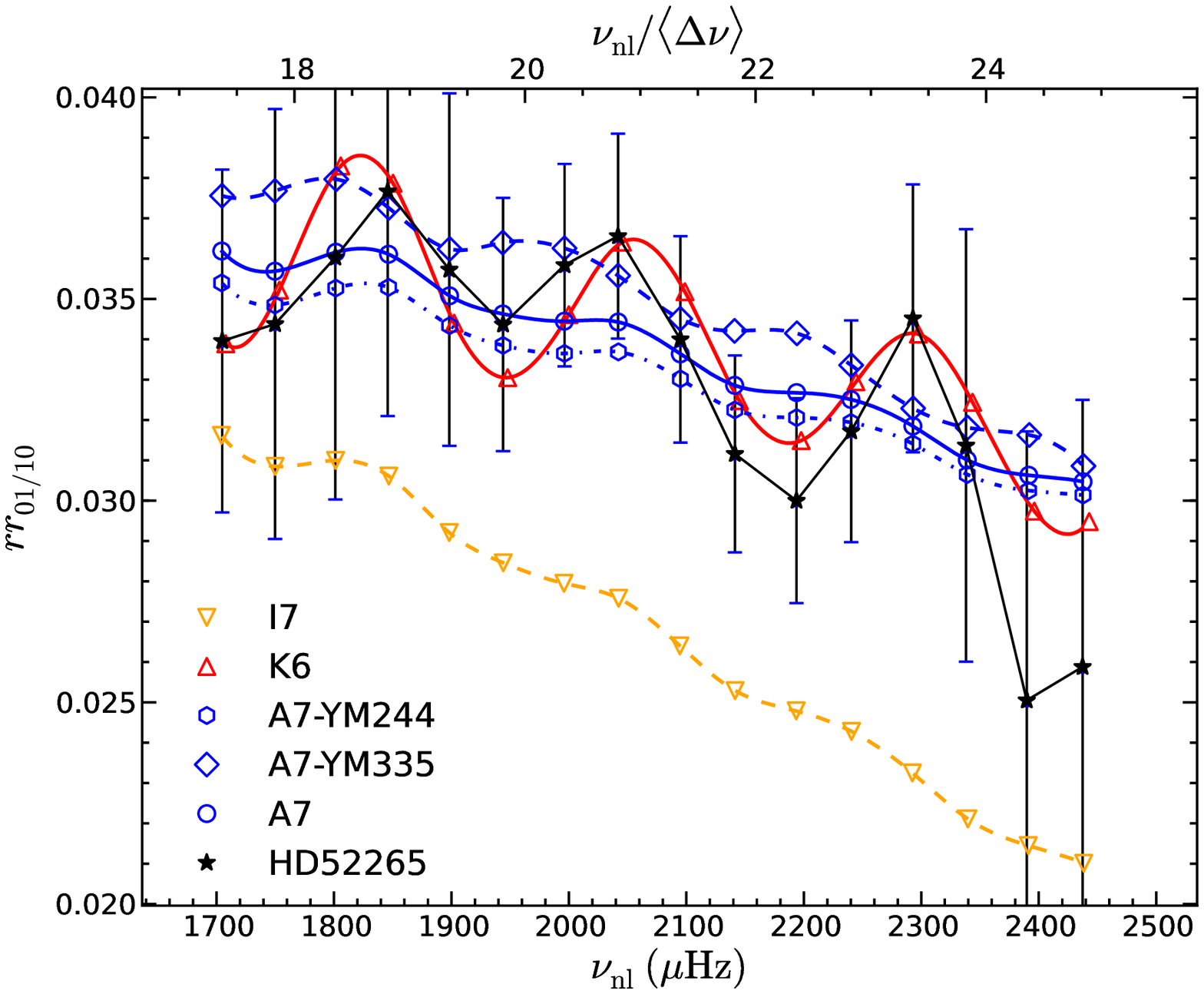}\includegraphics{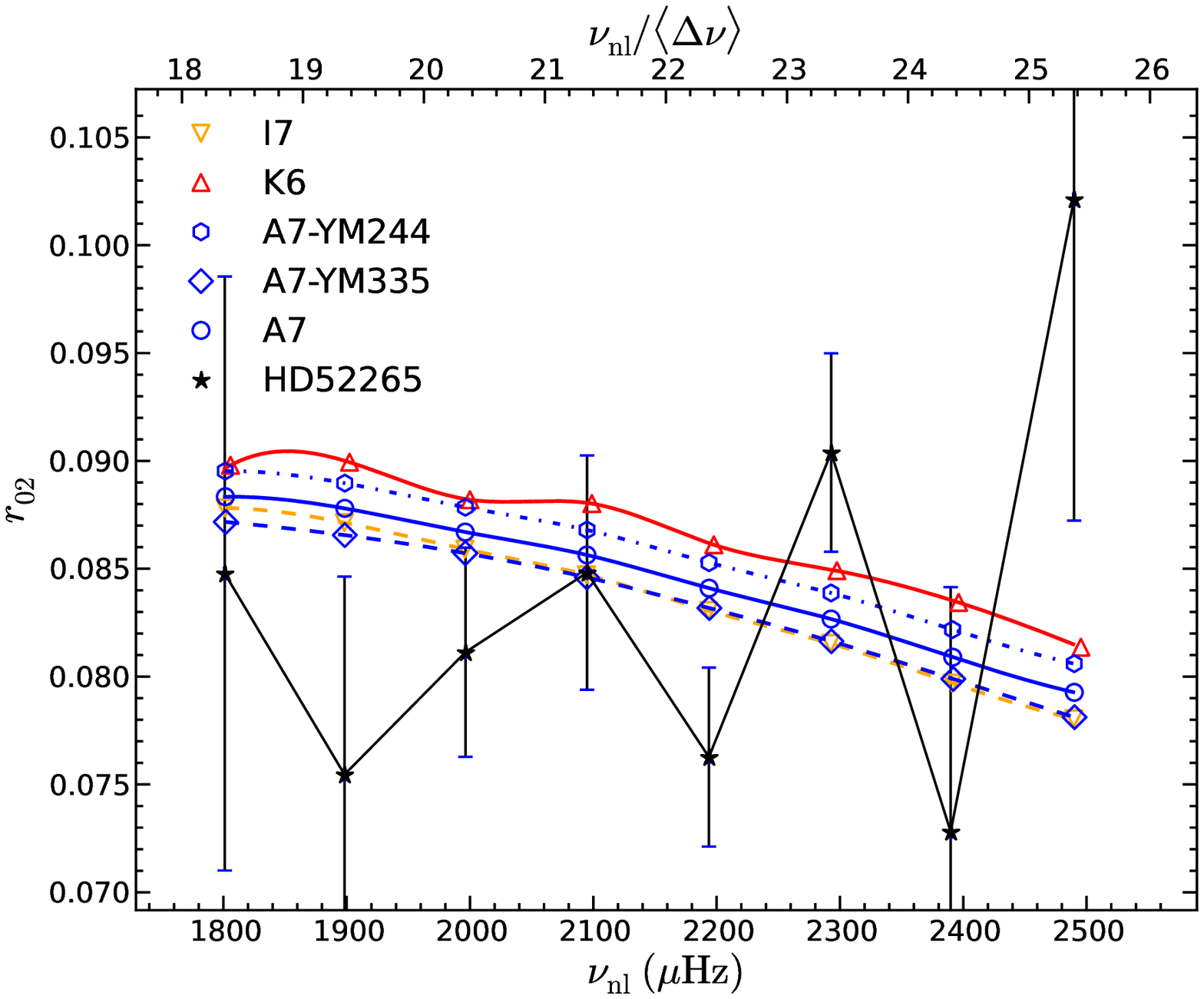}}
	     \caption{
	     {Seismic properties of a selected set of models. 
	     \textsl{Top left panel:} \'echelle diagram for the best model $A7$, optimized on the basis of the individual frequencies and including -or not- the correction for surface 
	     effects. 
	     Star symbols represent the observed frequencies, while black squares, blue circles, and red diamonds denote
	      the model frequencies for angular degrees $\ell= 0, 1$, and $2$.
	      In grey, we indicate the corresponding data before correction for surface effects. 
	      \textsl{Top right panel:}  comparison of observed large frequency separations for case $A7$ (corrected frequencies). The symbols are the same as in the top right figure.
	      \textsl{Bottom left panel:} comparison of the observed frequency separation ratios $rr_{01/10}(n)$ for a selected set of models, including the best model $A7$ 
	      (continuous blue line).
	      Results for models $A7$-$YM$-244 (pentagons) and $A7$-$YM$335 (diamonds) illustrate the effect of the $Y_0$-$M$ degeneracy. 
	      Results for model $K6$ (continuous red line) show the effect of including convective penetration below the convective envelope, 
          while results for model $I7$ (dashed 
	      orange line)
	      show the effect of a moderate overshooting of the convective core.
	      	      \textsl{Bottom right panel:} same as in the bottom left figure, but for the  $r_{02}(n)$ frequency separation ratios.}
    }
\label{sismicfig}
\end{figure*}

{In Fig.~\ref{sismicfig}, we show how the stellar models succeed -or not- in matching observed oscillation frequencies and frequency separations.
Since a thorough examination of seismic properties of all the models is beyond the scope of this paper, we selected some models.
The top left panel shows the \'echelle diagram corresponding to the model of case $A7$, 
optimized on the basis of the individual frequencies. When surface effects are corrected for, the model succeeds rather well in reproducing the \'echelle diagram
for a value of the adjustable parameter $b_\mathrm{SE}=4.2$ (Eq. \ref{nearsurf}) compatible with the solar value obtained by  \citet{2011A&A...535A..91D}  
with quasi-similar input physics  ($b_\mathrm{SE, \odot}=4.25$). 
On the other hand, in the high-frequency range, un-corrected models do not match observations.
In this respect, models $A6$ (frequencies not corrected, not plotted) and $A7$ (with uncorrected frequencies) give similar results. 
Furthermore, the top right panel shows that model $A7$ reproduces the observed individual large frequency separations
$\Delta \nu_{\ell}(n)$ quite well.
 The bottom left panel shows the comparison of the observed and model frequency separation ratios $rr_{01/10}(n)$. 
 Model $A7$ reproduces the mean slope of the variation of the ratios rather well, but not the oscillatory behaviour. 
 As shown by \citet{2012A&A...544L..13L}, this behaviour in HD 52265
 is reproduced in models
 that include convective penetration below the convective envelope, like the model of set $K$,
  which is similar to set $A$ but with $\xi_\mathrm{PC}=1.3$. 
 The figure also shows the effect of the $Y_0$-$M$ degeneracy on the diagram. 
 The larger the helium abundance, the higher the $rr_{01/10}(n)$ ratios. However, with the present accuracy on the data, 
 it is hard to distinguish the models with different ($Y_0$, $M$) values. 
 Finally, we plotted a model of set $I$ that takes into account a moderate amount of core 
 overshooting ($\alpha_\mathrm{ov}=0.15$). As already pointed out in Section \ref{internal}, the overshooting amount cannot be very large since even a 
 moderate amount of  overshooting is ruled out by the present data. The bottom right panel shows the fit of the $r_{02}(n)$ ratios. 
 In this case, regarding the precision on the data, it is difficult to distinguish the models.}

\section{Ages from other methods}

We estimate below the age of HD~52265, on the basis of  other age-dating methods (empirical or H-R diagram inversion).
We compare the resulting ages with the age inferred from 
\`{a} la carte  stellar modelling.
%``sur mesure'' stellar modelling.

\begin{figure}
      \resizebox{\hsize}{!}
	     {\includegraphics{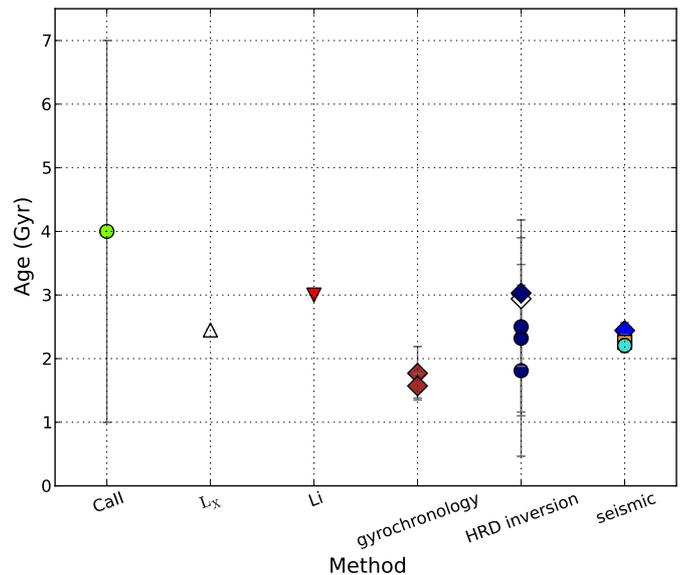}}
  \caption{
Age estimates for HD~52265. Columns CaII, $L_\mathrm{X}$, Li and gyrochronology  give empirical estimates based on the $R^\prime_\mathrm{ HK}$ index (circle), 
the lower limit from X-luminosity (upwards triangle), the upper limit from lithium surface abundance (downwards triangle) and, the gyrochronology (diamonds). 
The column HRD inversion shows estimates based on inversion of isochrones with circles for Padova isochrones and diamonds for BaSTI isochrones, full symbols for Bayesian methods, and empty symbol for $\chi^2$-minimisation, see text.
The column  seismic shows the seismic determination for 
\`{a} la carte models of case $6$ (Table~\ref{cases} and Fig.~\ref{Allages})   }
\label{otherages}
\end{figure}

\subsection{Empirical ages}
\label{emp}

\subsubsection{Activity}
The chromospheric activity and age of solar-type dwarfs appear to be anti-correlated. Empirical relations allow us to rely the Ca{\textsc{II}} H \& K emission index 
$R^\prime_\mathrm{ HK}=L_\mathrm{ HK}/L_\mathrm{ bol}$ to age \citep[see e.g.][for a recent calibration]{2008ApJ...687.1264M}. For HD~52265, values of $\log 
R^\prime_\mathrm{ HK}$ listed in the literature are in the range $[-5.02, -4.59]$. These low values indicate very low chromospheric activity. Using the 
\citeauthor{2008ApJ...687.1264M} $R^\prime_\mathrm{ HK}$-age relation, we derived an age of $4.0\pm 3.0$ Gyr. The ages can also be roughly estimated from the  
\citeauthor{2008ApJ...687.1264M} relation between the fractional X-ray emission $R^\prime_\mathrm{ X}=L_\mathrm{ X}/L_\mathrm{ bol}$ and age. 
For HD~52265, \citet{2008ApJ...687.1339K} derived an upper limit, $L_\mathrm{ X}<28.28$, which provides a lower age limit of $2.5$ Gyr.
Clearly, such empirical calibrations are too coarse to provide a reliable age of evolved stars with low chromospheric activity. Indeed, as recommended recently by 
\citet{2013A&A...551L...8P}, the use of chromospheric activity as a stellar clock should be limited to stars younger than about $1.5$ Gyr.

\subsubsection{Photospheric lithium abundance}

At the surface of low-mass stars, the lithium abundance can be depleted when the convective zone reaches the shallow regions where Li is destroyed by nuclear 
reactions at $T{\approx}2.5\times 10^6$ K or when mixing processes carry Li from the basis of the convective zone to the nuclear-burning region. A relation between the 
Li-abundance, effective temperature, and age is observed (but not fully understood).  We used the Li abundance curves as a function of $T_\mathrm{{eff}}$ published by 
\citet{2005A&A...442..615S}  for clusters of different ages and derived a lower limit of  $2.5$ Gyr on the age of  HD~52265 from several published values of its surface Li 
abundance ($\log \epsilon_\mathrm{Li} \in[1.67, 2.88]$). With $\log \epsilon_\mathrm{Li} =2.40\pm 0.06$ from \cite{ 2010MNRAS.403.1368G}, we found an age $<3$ 
Gyr. 

\subsubsection{Gyrochronology} 

 In the course of their evolution, solar-type stars lose angular momentum via magnetic braking due to their mass loss. It leads to a decrease of their rotation rate, first 
 quantified by \citet{1972ApJ...171..565S}. Gyrochronology, as proposed by \citet{2007ApJ...669.1167B}, is a new method  to derive the age of solar-type stars via an 
 empirical relation linking their rotation period, colour, and age, $t_\mathrm{Myr}^n = P_\mathrm{ days}\times a^{-1} \times ((B-V)-c)^{-b}$ where $a$, $b$, and $c$, are 
 constants. The constants were calibrated on the Sun, nearby field stars and clusters by \citet{2007ApJ...669.1167B} and then revised by \citet{2008ApJ...687.1264M}. We 
 applied these relations to HD~52265  ($P_\mathrm{rot}{=}12.3{\pm}0.14$ days, see Sect.~\ref{obs}) and found an age of $1.57\pm 0.19$ Gyr with 
 \citeauthor{2007ApJ...669.1167B}'s values of  $a$, $b$, and $c$ and $1.77\pm 0.42$ Gyr with the values of \citeauthor{2008ApJ...687.1264M}.

\subsection{Age from pre-calculated sets of model isochrones}
\label{iso}

This method consists of placing a star in an H--R diagram and of interpreting its position by means of a grid of pre-calculated theoretical isochrones or evolutionary tracks. It 
is widely used to age-date large samples of stars for Galactic evolution studies \citep[see e.g.][and references therein]{2011A&A...530A.138C}.  Different inversion techniques 
can be used to extract the stellar age (and mass) from theoretical isochrones. However, in several regions of the H--R diagram, for instance when the star is close to the 
zero-age main-sequence (ZAMS) or at turn-off, the morphology of the isochrones is complex and leads to severe age degeneracy. To cope with these problems, 
\citet{2004MNRAS.351..487P} {and \citet{2005a&a...436..127j}} proposed to take a Bayesian approach, with several priors, in particular, one on the initial mass function.

Bayesian inversion techniques using the Padova isochrones give ages of $2.5{\pm}1.4$ Gyr \citep{2009A&A...501..941H} and $2.32{\pm}1.16$ Gyr 
\citep{2011A&A...530A.138C}. We also used the \citet[][]{2002A&A...391..195G} and \citet{2006A&A...458..609D} {\small PARAM} web 
interface\footnote[1]{\url{http://stev.oapd.inaf.it/cgi-bin/param}} and found an age of $2.81{\pm}1.49$ Gyr.
On the other hand, from the use of BaSTI isochrones \citep{2004ApJ...612..168P}, 
\citeauthor{2011A&A...530A.138C} obtained $3.03{\pm}1.15$ Gyr. Still with 
BaSTI isochrones, using the tools developed by \citet[][]{Celine}, 
we inferred $2.94$ Gyr from a $\chi^2$-minimisation. As can be seen in Fig.~\ref{otherages}, 
the ages obtained cover a wide range, $0.5{-}4.2$ Gyr, because of the different isochrone 
grids and inversion methods used. {For the star HD 52265, which is approximately half-way on its MS, 
we expect  the isochrone inversion technique and the optimization performed in case $1$ of the present study to be equivalent in terms of
 precision -for a given set of input physics-  because they are both based on the observational constraints on the classical parameters. 
 However, the isochrone grid to be used for the inversion has to be dense enough both in mass and chemical composition. 
 On the other hand, for stars lying in regions of degeneracy in the H-R diagram, priors should be included in the optimization 
 process in case $1$ to deal with multiple possible solutions \citep[see e.g.][]{2005a&a...436..127j}.}

\subsection{Comparison of ages from different methods}

In Fig.\ref{otherages}, ages of HD~52265 obtained from different methods are compared. As discussed before, the ages from the $R^\prime_\mathrm{ HK}$ index,  the 
X-luminosity, and the lithium surface abundance are not reliable for this star. The ages from gyrochronology are very precise but not accurate because gyrochronology is an 
empirical method that relies on calibrations (on solar, nearby stars, and cluster ages). As pointed out by D. Soderblom (2013, invited review talk at the International Francqui 
Symposium), seismic ages combined with precise rotation periods as provided by the {\it Kepler} or CoRoT missions will give the potential to more fully test and calibrate 
gyrochronology. Finally, there is a large scatter in the ages derived from H-R diagram inversion. This scatter is similar to that obtained in case $1$ stellar modelling, 
when no seismic constraints are available. The \`{a} la carte
seismic age-dating that we obtained in the present study is by far the most precise.

\section{Age and mass of the exoplanet orbiting HD~52265}
\label{exoplanet}

A first estimate of the mass of the exoplanet orbiting 
HD~52265 was proposed by \citet{2000ApJ...545..504B} using their RV 
measurements and Eq.~\ref{mpsini}. From a grid of stellar models,
\citeauthor{2000ApJ...545..504B} inferred the mass 
   of the star, first by placing the star in a colour-magnitude diagram 
   at solar metallicity, and then by correcting for the metallicity of
    the star --which they estimated to be $\mathrm{[Fe/H]}=0.11$, using 
    stars of published mass and metallicity. They inferred a stellar 
    mass of  $1.13 \pm 0.03 M_\odot$ (note that the error bar here is
     an internal error bar and does not take into account the uncertainties 
     of the stellar models) and deduced that the exoplanet has a mass 
     $M_\mathrm{p} \sin i=1.13\ M_\mathrm{Jupiter}$. They did not give 
     uncertainties on this determination, nor the values for the age 
     of the host-star.
   
Using the RV data reported by \citet{2000ApJ...545..504B} and 
isochrone fits to derive the mass of the host-star, 
\citet{2013PNAS..11013267G} estimated the minimum mass 
of the exoplanet to be $M_\mathrm{p, min}=1.09 \pm 0.11\ M_\mathrm{Jupiter}$ 
(i.e. $\sin i=1$ in Eq.\ref{mpsini}). Here, again, the error bar does not account for the uncertainties 
     of stellar models. Furthermore, with their measure 
of the inclination of the spin axis of the star ($\sin i=0.59^{+0.18}_{-0.14}$),
 assumed to be the axis of the planetary orbit as well, they estimated
  the mass of the exoplanet to be
   $M_\mathrm{p}=1.85^{+0.52}_{-0.42}\ M_\mathrm{Jupiter}$.
    Note that the dominant source of error is the $\sin i$ error.

\begin{figure}
      \resizebox{\hsize}{!}
	     {\includegraphics{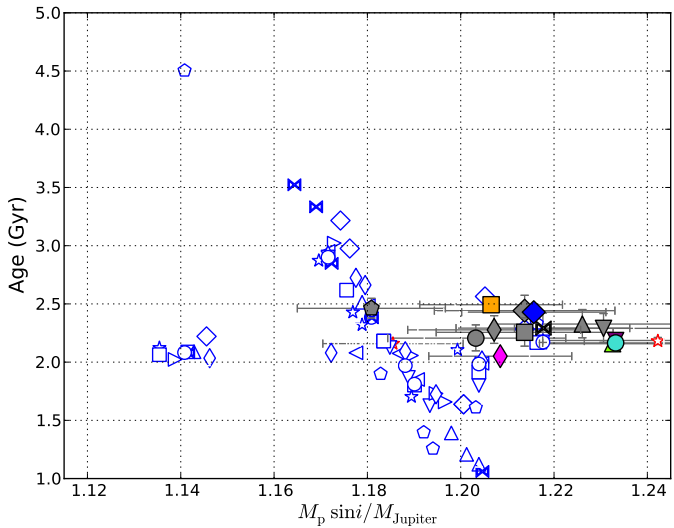}}
  \caption{
Age and mass of the exoplanet  $M_\mathrm{p} \sin i$ inferred 
from the mass of its host star. Colours are for optimized models
 of case $7$ of Table~\ref{cases} for different input physics, 
 as listed in Table~\ref{modelinputs}. Models for case $6$ are plotted
 in grey, other cases are shown with open blue symbols. The group of models with low mass and age of $\sim 2$ Gyr
 are case $4$ models, to be rejected because of their high initial helium abundance (see text).
 }
\label{planet}
\end{figure}

We have re-estimated the exoplanet mass on the basis of the range of mass of the host-star discussed above.
Figure~\ref{planet} shows the range of mass of the exoplanet as a function of its age, assumed to be the age of the host-star, 
for the different physical options and optimization sets considered.
 When considering all cases, the scatter is quite large. In particular, models of case $4$, which correspond to  low stellar mass,  
 form the bunch of points at low planet mass and at age around $2$ Gyr, but note that these models have to be excluded 
 because of their very high initial helium abundance. On the other hand, the use of seismic constraints in cases $6$  and $7$ 
 considerably narrows the range of ages of the star while they mostly favour higher masses with respect to other cases (see Figs.~\ref{Allages}). 
 Hence, these optimized models predict a higher mass for the exoplanet. Taking into account the $Y_0-M$ degeneracy 
 and excluding models $E$ without microscopic diffusion, the exoplanet mass $M_\mathrm{p} \sin i$ is found to be 
 in the range $1.16-1.26\ M_\mathrm{Jupiter}$, where we included the error budget due to the error on the host-star mass 
 optimization and on the RV data characterizing the exoplanet orbit. The scatter around the central value is therefore $\sim \pm 4$ per cent. 
 These values of the mass are higher ($\approx 7$ per cent) than the value of \citet{2000ApJ...545..504B}. 
Furthermore, with the extreme values of $\sin i$ given by \citet{2013PNAS..11013267G}, the mass of the exoplanet would be in 
the range $1.5-2.8\ M_\mathrm{Jupiter}$. The scatter is slightly larger than that in the result 
by \citeauthor{2013PNAS..11013267G} because we accounted for the uncertainties in the stellar model inputs. 
We therefore confirm that the companion of HD~52265 is a planet, not a brown dwarf.

%---------------------------------------------------

% % % % % % % % % % % % % % % % % % % % % % % % % % % % % % % % % % % % % % % % % % % % % % % % % % % %
\section{Conclusions}
\label{conclusion}

The optimization of models of the star HD~52265 using both classical and seismic
 observational data provides strong constraints on the age, mass, 
 radius, and surface gravity of the star, and on the mass of its exoplanet.
  Taking into account the full information provided by seismic observations 
  and considering the current uncertainties that affect the calculation of stellar models, 
  we found an age $A=2.10-2.54$ Gyr, a mass $M/M_\odot=1.14-1.32$, a 
  radius $R/R_\odot=1.30-1.34$, and a surface gravity $\log\ \mathrm{g}=4.28-4.32$. 

We stress that in the present case, 
  the mass and radius  given by the scaling relations agree  quite well with the result of the full
  modelling. This can be explained by the fact that the effective temperature used in the scaling
  relations  is accurately  determined by spectroscopy  and does correspond to the evolutionary
  stage of the star for its metallicity.
   
  The mass of the exoplanet is found to be in the range
   $M_\mathrm{p}\sin i=1.16$-$1.26\ M_\mathrm{Jupiter}$. This represents 
   considerable progress with respect to what can be achieved without seismic data
    or when using only mean values inferred from seismic observations, as the mean
     frequency separation and frequency at maximum power. An important point is that 
     while the age, mass, and radius of the star can be rather well estimated from classical 
     and mean seismic parameters after the input physics of the models are fixed, uncertainties on the input physics produce a large 
     scatter of the results. We demonstrated here that taking full profit of seismic data 
     -in particular considering the frequency separation ratios- considerably reduces this scatter by allowing one 
     to estimate free parameters of the models.

 The full optimization performed here allowed us 
 to better estimate the uncertainties and identify their 
 origin, in particular concerning the age. 

When no seismic constraints are available, the main cause for the age scatter 
and inaccuracy  are the values of the mixing length and initial helium abundance that belong 
to a wide interval of possible values. Seismic constraints allowed us
 to estimate these free parameters of stellar models, 
 as the mixing-length parameter for convection and the core
  overshoot and convective penetration parameters. 
 As a result,  the  interval of possible values for 
  $\alpha_\mathrm{conv}$,  [0.59, 0.61], is considerably narrowed. 
  These values are lower  by $12-15$ per cent than the solar values 
  obtained by a solar  calibration  using the same input physics.
 This provides an observational constraint on the convective transport 
 efficiency  that can be
compared with the results of $3$-D numerical simulations. The initial 
 helium abundance could also be obtained but there remains a 
  degeneracy between helium and mass. To remove this degeneracy, 
  more precise oscillation frequencies are required. As studied by \citet{2007MNRAS.375..861H}, 
  it would allow us to detect an oscillatory 
  behaviour of the frequencies that results from changes in the 
  adiabatic exponent in the second helium ionization zone, a seismic diagnostic of the helium abundance. 
  When the mixing length and initial helium abundance are unlocked, we find that the
  main cause of the age scatter is the choice of the solar mixture. 
  { When this problem is solved, we will then be at the level of testing nuclear reaction
     rates and internal transport  processes, as demonstrated here.}
    
Currently, the availability of precise observational frequencies for close well-known stars
 only concerns a small number of stars among which very few host an exoplanet.
These calibrators are key to better understanding the physics that govern stellar interiors.
A crucial need for the future is to increase the number of these calibrators.
In this context, the ESA-PLATO high-precision photometry mission is really needed \citep{2013arXiv1310.0696R}. 
It will provide essential information that complements the classical data that will be provided, 
with ultra high accuracy, by the Gaia-ESA mission, and interferometry and spectroscopic ground-based observations. 
Obtaining numerous and precise asteroseismic data is also a necessary step to make towards 
the full characterization of the age, mass, and radius of exoplanets.
 It is only {with this effort} that we will achieve a new understanding of the interiors, 
 habitability, and formation scenario of exoplanets and exoplanetary systems.

\begin{acknowledgements}
This research has made use of the SIMBAD database,
operated at CDS, Strasbourg, France and of the NASA's Astrophysics Data System. We warmly thank K\'{e}vin Belkacem for his comments on the manuscript.
\end{acknowledgements}

% for the bibliography, at the end
\bibliographystyle{aa} % style aa.bst
\bibliography{yl} % your references Yourfile.bib

\begin{appendix}
\section{Models with different input physics or optimization details}
\label{appendice}

%%%%%%%%%%%%%%%%%%%%%%%%%%%%%%%%%%%%%%%%%%%%%%%%%%%%%%%%%%%%%%%
\begin{table}  %[t]
\caption{Specifications of additional optimized models assuming different prescriptions for the modelling.}
\begin{tabular}{llll}
\hline\hline
Set & Case & Particularities & Figure symbol \\
\hline
$A$ & $1$-$Y$~/~$2$-$Y$               &   $Y=0.25~/~0.27$  &  open red circle \\
$A$ & $1$-$\alpha$0.550  &   $\alpha_\mathrm{conv}=0.550$  &  open red square \\
$A$ & $1$-$\alpha$0.826  &   $\alpha_\mathrm{conv}=0.826$  &  open red diamond \\
$A$ & $1$-ov0.30     &   $\alpha_\mathrm{ov}=0.30$  &  open red pentagon \\
$A$ & $1$, $2$a, $7$-$\nu$rad        &     &  not plotted \\
$A$ & $5$-allfreq             &   -  &  not plotted \\
$A$ & $6$-nocorrel            &   -  &  not plotted \\
$A$ & $6$-interR             &   -  &  not plotted \\
$A$ & $6$, $7$-ov             &   optimized overshooting  &  not plotted \\
$A$ & $6$, $7$-$YM$             &   $Y_0$-$M$ degeneracy  &  open red star \\
$A$ & $7$-noSE             &  no SE corrections  &  not plotted \\
$A$ & $7$-bSE4.9             &   solar SE corrections  &  not plotted \\%$A$ & 7-YM-noSE             &   -  &  not plotted \\
$A$ & $7$-pms             &   -  &  not plotted \\
$A$ & $7$-rot             &  rotation  &  not plotted \\%$A$ & 7-YM-ov             &   -  &  not plotted \\
\hline\end{tabular}
\label{modelinputsA}
\end{table}
%%%%%%%%%%%%%%%%%%%%%%%%%%%%%%%%%%%%%%%%%%%%%%%%%%%%%%%%%%%%%%%

\subsection{Optimization of set $A$ models with alternate prescriptions}

In the following, we present other optimization models, all based on the reference physics of set 
$A$ (Table~\ref{modelinputs}). These models were optimized following Table~\ref{cases}, 
but with different approaches or choices of free parameters, as described below 
and listed in Table~\ref{modelinputsA}. The results of the models are listed in Tables~\ref{resultsAA1} and \ref{resultsAA2}.

\begin{enumerate}
\item {\em Cases $1-Y$~/~$2-Y$}

As explained in the main text, in cases $1$ and $2$a, b, and c, the initial helium content $Y$ could 
not be adjusted because of the lack of observational constraints. Here we investigate 
the consequence of not deriving $Y$ from the $\Delta Y/\Delta Z=2$ enrichment law 
(as was done in cases $1$ and $2$ of Table~\ref{cases}). We sought a solution with the lowest possible 
initial helium content, never lower than the primordial abundance. 
These choices have an important impact on the results and are therefore discussed in the main text. 
We point out that in case $A1$-$Y$ it was possible to find a solution with a primordial helium abundance. 
This is because no seismic constraints were used in this case. On the other hand, 
in cases $A2$a$-Y$, b$-Y$ , and c$-Y$, 
we had to increase the helium abundance above the primordial one 
because of the seismic constraints introduced,
to find a solution that agreed 
with both the seismic and the classical observational constraints 
(we found that models with low $Y$ either have a too low temperature or a too high luminosity).

\item {\em Cases $1$-$\alpha$0.550, $1$-$\alpha$0.826}

In case $1$, $\alpha_\mathrm{conv}$ could not be adjusted because of the lack of observational constraints 
and, as often seen in papers, the solar value $\alpha_\mathrm{conv, \odot}$ was used. 
Here, we investigated the impact of other choices and we sought a solution for acceptable, 
extreme values of $\alpha_\mathrm{conv}$. Numerical 2-D simulations of convection suggest that 
$\alpha_\mathrm{conv}$ might differ by  a few tenths of dex from the solar value 
\citep[see e.g.][]{1999A&A...346..111L}. We investigated changes of $\alpha_\mathrm{conv}$ of 
20 per cent around the solar value that correspond to ${\alpha_\mathrm{conv, min}}=0.550$ 
and $\alpha_\mathrm{conv, max}=0.826$. These choices have an important impact on the results
 and are therefore discussed in the main text.

\item {\em Cases $1$-ov30}

No overshooting was assumed in the reference models of case $1$. 
Here, to estimate its impact, we chose a rather high value, i.e. $\alpha_\mathrm{ov}=0.30$. 
The impact on case $1$ results is discussed in the main text.

\item {\em Cases $1$, $2$, $7$-$\nu$rad}

In these models, the impact of mixing that results from the radiative diffusivity associated with the 
kinematic radiative viscosity is investigated. Following \citet{2002A&A...390..611M}, we added 
an extra mixing diffusion coefficient $d_\mathrm{rad}=D_\mathrm{R}\times \nu_\mathrm{rad}$ 
with $D_\mathrm{R}=1$ that limits gravitational settling in the outer stellar layers of stars with 
thin convective envelopes. As can be seen in Tables~\ref{resultsAA1} and \ref{resultsAA2}, 
the impact is weak and is not discussed further.

\item {\em  Case $5$-allfreq} 

In the list of frequencies extracted by \citet{2011A&A...530A..97B}, $31$ frequencies were given, of 
which $28$ were flagged as secure. In this model, we considered the $31$ values
 and found that the impact is weak and that the results remain inside the uncertainty range we gave in the main text.

\item {\em Case $6$-$YM$, $7$-$YM$}

For each optimization case, there are a 
range of initial helium-mass doublets ($Y_0$, $M$) that provide seismically
equivalent optimized models. We investigated the $Y_0-M$ degeneracy by searching 
for optimized models with different values of $Y_0$ and $M$. 
As discussed in the main text, the impact on age is weak but 
a range of possible masses of HD~52265 is found.

\item {\em  Case $6$-nocorrel}

As explained in the main text, in case $6$, we took into account the correlations between 
the frequency separation ratios and calculated the $\chi^2$ from Eq.~\ref{chi2}. 
In this model the correlations are not considered and the $\chi^2$ is evaluated from Eq.~\ref{corrfreq}. 
The impact is weak and is not  discussed further.

\item {\em Case  $6$-interr} 

\citet{2013A&A...560A...2R} recently claimed that model-fitting by searching for a best fit of 
observed and model separation ratios at the same radial orders $n$
is incorrect, and that the correct procedure is to compare the model ratios interpolated to the observed 
frequencies. We followed this recommendation here. The impact is weak and the results remain in the 
uncertainty range we gave in the main text.

\item {\em  Case  $7$-rot}  

As explained in the main text, rotation and its effects on the transport of angular momentum and 
chemicals was treated as in \citet{2013A&A...549A..74M}, and we tuned the $K_\mathrm{w}$ coefficient that enters  
the treatment of magnetic braking by winds to match the observed rotation period. 
The impact on age and mass is weak for this rather evolved star. 
A thorough study of the effects of rotation on HD 52265 will be presented in a forthcoming paper.

\item {\em Case  $7$-noSE and case  $7$-bSE4.9}.

As explained in the main text, in case $7$, we corrected the individual frequencies for the surface 
effects with the \citet{2008ApJ...683L.175K} empirical prescription and calibrated the 
$b_\mathrm{SE}$ parameter in Eq.~\ref{nearsurf} to achieve the best match between observed and modelled frequencies.
In model 7-bSE4.9, we used the solar value of $b_\mathrm{SE}$ calibrated by \citet{2008ApJ...683L.175K} 
and found that the impact on the results is weak. 
In model 7-noSE, we did not correct frequencies for the surface effects. 
The impact is strong, in particular on age. This is discussed in the main text.

\item {\em  Case  $7$-pms}. 

The main models were calculated by starting the computation at the zero-age main sequence. 
This model was evolved from the pre-main sequence. The impact is weak.

\item {\em  Cases  $6$ and $7$-ov}. 

In the models presented in the main text, overshooting was either neglected (sets $A-H$, $K$) or fixed (sets $I$, $J$). 
Since there are enough observational constraints to add overshooting and convective penetration 
as additional free parameters in cases $6$ and $7$, we considered this possibility here. 
We therefore also optimized the values of $\alpha_\mathrm{ov}$ and $\xi_\mathrm{PC}$. 
As also discussed in the main text, we found that low values of $\alpha_\mathrm{ov}$ 
are favoured (range $0.00-0.04$) and rather high values of $\xi_\mathrm{PC}$ (range $0.90-1.25$). 
This latter result, related to the oscillatory behaviour of the frequencies, 
close to the convective envelope agrees with the conclusions of \citet{2012A&A...544L..13L}.

\end{enumerate}

\begin{table*}  %[t]
\caption{Same as Table~\ref{resultsA1}, but for different optimization options
 (see Sect.~\ref{LM} and Tables~\ref{modelinputsA} and \ref{cases}).}
\begin{tabular}{lccccccccl}
\hline\hline
Model &   Age (Gyr) & $M/M_\odot$ & $ (Z/X)_0$ &  $Y_0$&  $\alpha_\mathrm{mlt}$ &  $\alpha_\mathrm{ov}$ / $\xi_\mathrm{PC}$ & $b_\mathrm{SE}$ & $a_\mathrm{SE}$/$r_\mathrm{SE}$& $\chi^2_\mathrm{R, classic}$ / $\chi^2_\mathrm{R, seism}$\\
\hline
$A1$-$Y$    & 1.59$\pm$ 2.17& 1.32$\pm$ 0.05&0.0447$\pm$0.0051&0.245&0.688&0.00/0.00& --&--&${1.0\ 10^{-19}}$/\ \ --\\
$A1$-al0.550& 1.03$\pm$ 1.47& 1.22$\pm$ 0.03&0.0467$\pm$0.0066&0.309&0.550&0.00/0.00& --&--&${1.4\ 10^{-13}}$/\ \ --\\
$A1$-al0.826& 4.28$\pm$ 0.78& 1.15$\pm$ 0.01&0.0471$\pm$0.0049&0.310&0.826&0.00/0.00& --&--&${4.1\ 10^{-5}}$/\ \ --\\
$A1$-ov0.3  & 3.07$\pm$ 1.35& 1.18$\pm$ 0.03&0.0487$\pm$0.0054&0.312&0.688&0.30/0.00& --&--&${1.8\ 10^{-17}}$/\ \ --\\
$A1$-Renu   & 2.91$\pm$ 1.30& 1.18$\pm$ 0.04&0.0481$\pm$0.0052&0.311&0.688&0.00/0.00& --&--&${9.7\ 10^{-6}}$/\ \ --\\
$A2$a-Renu  & 2.21$\pm$ 0.87& 1.19$\pm$ 0.02&0.0488$\pm$0.0051&0.312&0.580$\pm$0.096&0.00/0.00& --&--&${1.5\ 10^{-1}}$/${5.9\ 10^{-4}}$\\
$A2$a-$Y$   & 3.46$\pm$ 0.95& 1.24$\pm$ 0.01&0.0472$\pm$0.0040&0.270&0.682$\pm$0.130&0.00/0.00& --&--&${3.3\ 10^{-1}}$/${1.8\ 10^{-4}}$\\
$A2$b-$Y$   & 4.27$\pm$ 1.32& 1.22$\pm$ 0.03&0.0465$\pm$0.0047&0.270&0.744$\pm$0.148&0.00/0.00&5.5&-1.3/1.00&${3.6\ 10^{-1}}$/${1.3\ 10^{-4}}$\\
$A2$c-$Y$   & 2.63$\pm$ 1.68& 1.25$\pm$ 0.04&0.0447$\pm$0.0046&0.270&0.604$\pm$0.060&0.00/0.00&5.5&-3.3/1.00&${5.3\ 10^{-1}}$/${2.3\ 10^{-1}}$\\
$A5$-allfreq& 2.25$\pm$ 0.30& 1.27$\pm$ 0.04&0.0461$\pm$0.0056&0.265$\pm$0.030&0.698$\pm$0.081&0.00/0.00&--&--&${8.7\ 10^{-7}}$/${3.3\ 10^{-3}}$\\
$A6$-$YM$254& 2.45$\pm$ 0.13& 1.29$\pm$ 0.01&0.0451$\pm$0.0027&0.254$\pm$0.008&0.683$\pm$0.032&0.00/0.00&--&--&${7.3\ 10^{-2}}$/${8.1\ 10^{-1}}$\\
$A6$-$YM$317& 2.18$\pm$ 0.11& 1.18$\pm$ 0.01&0.0489$\pm$0.0025&0.317$\pm$0.009&0.591$\pm$0.026&0.00/0.00&--&--&${6.7\ 10^{-2}}$/${8.9\ 10^{-1}}$\\
$A6$-intrr  & 2.20$\pm$ 0.11& 1.17$\pm$ 0.03&0.0494$\pm$0.0033&0.324$\pm$0.017&0.603$\pm$0.031&0.00/0.00&--&--&${1.9\ 10^{-2}}$/${8.6\ 10^{-1}}$\\
$A6$-nocorrel& 2.28$\pm$ 0.21& 1.21$\pm$ 0.04&0.0497$\pm$0.0051&0.302$\pm$0.024&0.595$\pm$0.052&0.00/0.00&--&--&${1.8\ 10^{-1}}$/${7.2\ 10^{-1}}$\\
$A6$-ov& 2.32$\pm$ 0.14& 1.25$\pm$ 0.01&0.0481$\pm$0.0021&0.277$\pm$0.006&0.687$\pm$0.027&0.04/0.90&--&--&${2.0\ 10^{-1}}$/${7.8\ 10^{-1}}$\\
$A7$-bSE4.9& 2.18$\pm$ 0.03& 1.27$\pm$ 0.00&0.0487$\pm$0.0006&0.274$\pm$0.001&0.601$\pm$0.004&0.00/0.00&4.9&-4.1/1.00&${5.3\ 10^{-1}}$/${1.8\ 10^{0}}$\\
$A7$-Renu& 2.18$\pm$ 0.02& 1.27$\pm$ 0.00&0.0489$\pm$0.0007&0.274$\pm$0.001&0.599$\pm$0.004&0.00/0.00&4.5&-4.7/1.00&${4.9\ 10^{-1}}$/${1.7\ 10^{0}}$\\
$A7$-$YM$-noSE& 2.98$\pm$ 0.03& 1.24$\pm$ 0.00&0.0628$\pm$0.0007&0.306$\pm$0.001&0.725$\pm$0.005&0.00/0.00&--&--&${3.3\ 10^{0}}$/${6.8\ 10^{0}}$\\
$A7$-ov& 1.93$\pm$ 0.01& 1.23$\pm$ 0.00&0.0481$\pm$0.0007&0.291$\pm$0.001&0.564$\pm$0.003&0.00/1.25&3.6&-7.7/1.00&${7.8\ 10^{-1}}$/${2.2\ 10^{0}}$\\
$A7$-$YM$-ov& 2.15$\pm$ 0.03& 1.23$\pm$ 0.00&0.0485$\pm$0.0007&0.289$\pm$0.001&0.581$\pm$0.003&0.00/0.99&4.1&-5.9/1.00&${5.7\ 10^{-1}}$/${1.8\ 10^{0}}$\\
$A7$-noSE& 3.14$\pm$ 0.03& 1.30$\pm$ 0.00&0.0690$\pm$0.0008&0.276$\pm$0.001&0.731$\pm$0.005&0.00/0.00&--&--&${4.7\ 10^{0}}$/${4.0\ 10^{0}}$\\
$A7$-pms& 2.18$\pm$ 0.02& 1.27$\pm$ 0.00&0.0486$\pm$0.0006&0.274$\pm$0.001&0.601$\pm$0.004&0.00/0.00&3.8&-6.2/1.00&${5.2\ 10^{-1}}$/${1.7\ 10^{0}}$\\
$A7$-$YM$244& 2.14$\pm$ 0.02& 1.32$\pm$ 0.00&0.0460$\pm$0.0008&0.244$\pm$0.001&0.588$\pm$0.004&0.00/0.00&4.5&-4.4/1.00&${7.7\ 10^{-1}}$/${1.6\ 10^{0}}$\\
$A7$-$YM$260& 2.18$\pm$ 0.03& 1.28$\pm$ 0.00&0.0467$\pm$0.0006&0.260$\pm$0.001&0.583$\pm$0.004&0.00/0.00&3.7&-6.4/1.00&${5.9\ 10^{-1}}$/${1.6\ 10^{0}}$\\
$A7$-$YM$300& 2.14$\pm$ 0.01& 1.21$\pm$ 0.00&0.0478$\pm$0.0006&0.300$\pm$0.001&0.585$\pm$0.004&0.00/0.00&3.8&-7.3/1.00&${1.7\ 10^{-1}}$/${1.8\ 10^{0}}$\\
$A7$-$YM$310& 2.16$\pm$ 0.02& 1.20$\pm$ 0.00&0.0484$\pm$0.0004&0.310$\pm$0.001&0.584$\pm$0.004&0.00/0.00&3.5&-8.2/1.00&${1.2\ 10^{-1}}$/${1.9\ 10^{0}}$\\
$A7$-$YM$320& 2.17$\pm$ 0.02& 1.18$\pm$ 0.00&0.0501$\pm$0.0004&0.320$\pm$0.001&0.585$\pm$0.004&0.00/0.00&3.5&-8.3/1.00&${1.3\ 10^{-1}}$/${2.0\ 10^{0}}$\\
$A7$-$YM$335& 2.15$\pm$ 0.02& 1.15$\pm$ 0.00&0.0491$\pm$0.0003&0.335$\pm$0.001&0.585$\pm$0.004&0.00/0.00&3.5&-2.0/1.00&${2.6\ 10^{-2}}$/${2.1\ 10^{0}}$\\
$A7$-rot& 2.19$\pm$ 0.03& 1.27$\pm$ 0.00&0.0488$\pm$0.0004&0.275$\pm$0.001&0.595$\pm$0.004&0.00/0.00&5.5&-3.3/1.00&${1.3\ 10^{0}}$/${2.6\ 10^{0}}$\\
\hline
\end{tabular}
\label{resultsAA1}
\end{table*}

%%%%%%%%%%%%%%%%%%%%%%%%%%%%%%%%%%%%%%%%%%%%%%%%%%%%%%%%%%%%%%%
\begin{table*}  %[t]
\caption{Same as Table~\ref{resultsA2}, but for different optimization options (see Sect.~\ref{LM} and Tables~\ref{modelinputsA} and \ref{cases})}
\begin{tabular}{lcccccccccccccc}
\hline\hline
Model &   $T_\mathrm{{eff}}$ &$L$ & [Fe/H] & $\log g$ &  $R$& 
 $\langle\Delta\nu\rangle$ & $\nu_\mathrm{max}$& $\langle r_{02}\rangle$ &$\langle rr_{01/10}\rangle$  &$X_C$ & $\frac{\Delta Y}{\Delta Z}$& $M_\mathrm{{cc}}$ & $R_\mathrm{{zc}}$ & $M_\mathrm{{p}}\sin i$  \\
 & {[K]} & [$L_\odot$] & [dex] & [dex] & [$R_\odot$]  &   [$\mu$Hz] &  [$\mu$Hz] & -- &   --&-- & --&  [$M_\star$] &  [$R_\star$] &[$M_\mathrm{{Jupiter}}$] \\
  \hline
$A1$-$Y$&6116.&2.053& 0.22&4.35&1.28&107.45&2393.& 0.091& 0.033&0.49& 0.0&0.002&0.788&1.27$\pm$0.04\\
$A1$-al0.550&6116.&2.053& 0.22&4.31&1.28&101.88&2210.& 0.097& 0.035&0.51& 2.0&0.016&0.836&1.20$\pm$0.04\\
$A1$-al0.826&6116.&2.053& 0.22&4.29&1.28&101.09&2081.& 0.053& 0.030&0.16& 2.0&0.047&0.713&1.15$\pm$0.03\\
$A1$-ov0.3&6116.&2.053& 0.22&4.30&1.28&101.27&2131.& 0.072& 0.011&0.43& 2.0&0.140&0.769&1.17$\pm$0.04\\
$A1$-Renu&6116.&2.054& 0.22&4.29&1.28&101.29&2125.& 0.074& 0.033&0.28& 2.0&0.029&0.769&1.17$\pm$0.04\\
$A2$a-Renu&6046.&2.062& 0.22&4.28&1.31& 98.13&2064.& 0.083& 0.035&0.35& 2.0&0.023&0.806&1.18$\pm$0.03\\
$A2$a-$Y$&6012.&2.069& 0.21&4.28&1.33& 98.13&2096.& 0.070& 0.033&0.27& 0.7&0.024&0.759&1.21$\pm$0.03\\
$A2$b-$Y$&6008.&2.070& 0.21&4.28&1.33& 98.13&2063.& 0.060& 0.032&0.21& 0.7&0.036&0.731&1.20$\pm$0.04\\
$A2$c-$Y$&6001.&2.060& 0.19&4.29&1.33& 98.26&2106.& 0.080& 0.034&0.33& 0.7&0.010&0.791&1.22$\pm$0.04\\
$A5$-allfreq&6116.&2.053& 0.22&4.33&1.28&105.37&2299.& 0.083& 0.033&0.39& 0.5&0.007&0.778&1.23$\pm$0.04\\
$A6$-$YM$254&6074.&2.058& 0.21&4.32&1.30&103.67&2263.& 0.081& 0.032&0.37& 0.2&0.005&0.777&1.24$\pm$0.03\\
$A6$-$YM$317&6068.&2.048& 0.22&4.28&1.30& 98.69&2087.& 0.084& 0.034&0.35& 2.2&0.022&0.804&1.18$\pm$0.03\\
$A6$-intrr&6094.&2.060& 0.22&4.29&1.29& 99.07&2085.& 0.083& 0.034&0.34& 2.4&0.026&0.802&1.17$\pm$0.04\\
$A6$-nocorrel&6037.&2.054& 0.23&4.28&1.31& 98.29&2092.& 0.083& 0.034&0.35& 1.7&0.020&0.799&1.19$\pm$0.04\\
$A6$-ov&6094.&2.039& 0.26&4.32&1.28&104.20&2252.& 0.082& 0.032&0.39& 0.9&0.017&0.729&1.22$\pm$0.03\\
$A7$-bSE4.9&6019.&2.101& 0.23&4.29&1.34& 98.29&2123.& 0.084& 0.034&0.39& 0.8&0.014&0.800&1.23$\pm$0.03\\
$A7$-Renu&6013.&2.092& 0.23&4.29&1.34& 98.29&2125.& 0.084& 0.034&0.39& 0.8&0.013&0.800&1.23$\pm$0.03\\
$A7$-$YM$-noSE&6069.&2.156& 0.34&4.28&1.33& 98.30&2076.& 0.070& 0.030&0.29& 1.4&0.046&0.754&1.21$\pm$0.03\\
$A7$-ov&5989.&2.012& 0.25&4.29&1.32& 98.28&2112.& 0.087& 0.034&0.41& 1.3&0.012&0.753&1.21$\pm$0.03\\
$A7$-$YM$-ov&5997.&2.030& 0.25&4.29&1.32& 98.28&2107.& 0.085& 0.034&0.38& 1.3&0.013&0.756&1.21$\pm$0.03\\
$A7$-noSE&5983.&2.114& 0.39&4.29&1.36& 98.33&2122.& 0.070& 0.030&0.31& 0.6&0.041&0.745&1.26$\pm$0.03\\
$A7$-pms&6021.&2.101& 0.23&4.29&1.33& 98.29&2125.& 0.084& 0.034&0.39& 0.8&0.013&0.800&1.23$\pm$0.03\\
$A7$-$YM$244&5958.&2.073& 0.21&4.30&1.35& 98.29&2160.& 0.086& 0.033&0.42& 0.0&0.003&0.800&1.27$\pm$0.03\\
$A7$-$YM$260&5971.&2.048& 0.21&4.29&1.34& 98.28&2141.& 0.085& 0.033&0.40& 0.4&0.006&0.801&1.24$\pm$0.03\\
$A7$-$YM$300&6047.&2.067& 0.21&4.29&1.31& 98.28&2096.& 0.084& 0.034&0.37& 1.7&0.017&0.806&1.20$\pm$0.03\\
$A7$-$YM$310&6047.&2.046& 0.21&4.28&1.31& 98.28&2087.& 0.084& 0.034&0.36& 2.0&0.019&0.805&1.19$\pm$0.03\\
$A7$-$YM$320&6055.&2.038& 0.23&4.28&1.30& 98.28&2077.& 0.084& 0.034&0.35& 2.3&0.023&0.804&1.17$\pm$0.03\\
$A7$-$YM$335&6090.&2.045& 0.21&4.28&1.29& 98.29&2058.& 0.083& 0.035&0.34& 2.9&0.027&0.807&1.15$\pm$0.03\\
$A7$-rot&6034.&2.124& 0.28&4.29&1.34& 98.32&2121.& 0.085& 0.034&0.40& 0.8&0.014&0.806&1.23$\pm$0.03\\
\hline
\end{tabular}
\label{resultsAA2}
\end{table*}

\subsection{Optimization with different input physics}
\label{subsets}

In the following, we present optimization models based on the different choices of input physics 
listed in Table~\ref{modelinputs}. These models were optimized following Table~\ref{cases}.
 The results are listed in Tables~\ref{resultsB1} and \ref{resultsB2}. Discussions are found in the main text.
% In the text, at the place where the large table should appear
% add the command:
%\addtocounter{table}{1}
% Tables counters will be well numbered.
% If table 2
\longtab{4}{
\begin{longtable}{lccccccccl}
\caption{Same as Table~\ref{resultsA1}, but for different input physics of the models (see Sect.~\ref{LM} and Tables~\ref{modelinputs} and \ref{cases}).}
\label{resultsB1}
\\
\hline\hline
Model &   Age (Gyr) & $M/M_\odot$ & $ (Z/X)_0$ &  $Y_0$&  $\alpha_\mathrm{mlt}$ &  $\alpha_\mathrm{ov}$ / $\xi_\mathrm{PC}$ & $b_\mathrm{SE}$ & $a_\mathrm{SE}$/$r_\mathrm{SE}$& $\chi^2_\mathrm{R, classic}$ / $\chi^2_\mathrm{R, seism}$\\
\hline
\endfirsthead
\caption{continued.}\\
\hline\hline
Model &   Age (Gyr) & $M/M_\odot$ & $ (Z/X)_0$ &  $Y_0$&  $\alpha_\mathrm{mlt}$ &  $\alpha_\mathrm{ov}$ / $\xi_\mathrm{PC}$ & $b_\mathrm{SE}$ & $a_\mathrm{SE}$/$r_\mathrm{SE}$& $\chi^2_\mathrm{R, classic}$ / $\chi^2_\mathrm{R, seism}$\\
\hline
\endhead
\hline
\endfoot
$B1$& 2.62$\pm$ 1.22& 1.18$\pm$ 0.03&0.0483$\pm$0.0053&0.311&1.762&0.00/0.00& --&--&${4.6\ 10^{-7}}$/\  \  --\\
$B2$a& 1.80$\pm$ 0.85& 1.20$\pm$ 0.02&0.0493$\pm$0.0054&0.312&1.473$\pm$0.222&0.00/0.00& --&--&${1.8\ 10^{-1}}$/${6.2\ 10^{-4}}$\\
$B2$b& 2.18$\pm$ 0.78& 1.19$\pm$ 0.02&0.0491$\pm$0.0051&0.312&1.562$\pm$0.252&0.00/0.00&5.5&-4.8/1.00&${1.2\ 10^{-1}}$/${1.9\ 10^{-4}}$\\
$B2$c& 2.18$\pm$ 0.55& 1.19$\pm$ 0.01&0.0491$\pm$0.0051&0.312&1.562$\pm$0.242&0.00/0.00&5.5&-4.8/1.00&${2.2\ 10^{-1}}$/${7.9\ 10^{-6}}$\\
$B3$& 1.91$\pm$ 1.12& 1.23$\pm$ 0.01&0.0489$\pm$0.0053&0.299$\pm$0.019&1.500$\pm$0.241&0.00/0.00&--&--&${2.6\ 10^{-1}}$/${1.7\ 10^{-2}}$\\
$B4$& 2.06$\pm$ 0.24& 1.12$\pm$ 0.02&0.0511$\pm$0.0047&0.355$\pm$0.017&1.549$\pm$0.096&0.00/0.00&5.5&-5.4/1.00&${3.2\ 10^{-5}}$/${7.0\ 10^{-7}}$\\
$B5$& 2.17$\pm$ 0.32& 1.24$\pm$ 0.05&0.0467$\pm$0.0061&0.281$\pm$0.040&1.769$\pm$0.229&0.00/0.00&--&--&${7.4\ 10^{-7}}$/${4.0\ 10^{-3}}$\\
$B6$& 2.26$\pm$ 0.12& 1.24$\pm$ 0.02&0.0470$\pm$0.0020&0.283$\pm$0.009&1.571$\pm$0.065&0.00/0.00&--&--&${3.0\ 10^{-1}}$/${8.3\ 10^{-1}}$\\
$B7$& 2.49$\pm$ 0.02& 1.23$\pm$ 0.00&0.0494$\pm$0.0005&0.289$\pm$0.001&1.615$\pm$0.009&0.00/0.00&3.5&-8.3/1.00&${3.2\ 10^{-1}}$/${1.8\ 10^{0}}$\\
\hline
$C1$& 2.98$\pm$ 1.33& 1.18$\pm$ 0.03&0.0358$\pm$0.0038&0.296&0.688&0.00/0.00& --&--&${1.1\ 10^{-8}}$/\  \  --\\
$C2$a& 3.22$\pm$ 1.08& 1.18$\pm$ 0.02&0.0363$\pm$0.0037&0.297&0.656$\pm$0.111&0.00/0.00& --&--&${1.0\ 10^{-1}}$/${9.8\ 10^{-5}}$\\
$C2$b& 1.64$\pm$ 0.67& 1.22$\pm$ 0.02&0.0363$\pm$0.0053&0.297&0.544$\pm$0.074&0.00/0.00&5.5&-5.0/1.00&${1.4\ 10^{-1}}$/${6.2\ 10^{-4}}$\\
$C2$c& 1.64$\pm$ 0.66& 1.22$\pm$ 0.02&0.0363$\pm$0.0047&0.297&0.544$\pm$0.083&0.00/0.00&5.5&-5.0/1.00&${1.5\ 10^{-1}}$/${3.6\ 10^{-4}}$\\
$C3$& 2.56$\pm$ 1.18& 1.23$\pm$ 0.03&0.0361$\pm$0.0038&0.280$\pm$0.023&0.601$\pm$0.092&0.00/0.00&--&--&${2.7\ 10^{-1}}$/${8.8\ 10^{-3}}$\\
$C4$& 2.22$\pm$ 0.27& 1.14$\pm$ 0.02&0.0380$\pm$0.0032&0.336$\pm$0.016&0.587$\pm$0.048&0.00/0.00&5.5&-4.9/1.00&${1.9\ 10^{-3}}$/${2.5\ 10^{-6}}$\\
$C5$& 2.28$\pm$ 0.31& 1.24$\pm$ 0.03&0.0346$\pm$0.0038&0.272$\pm$0.019&0.670$\pm$0.075&0.00/0.00&--&--&${8.1\ 10^{-5}}$/${2.5\ 10^{-3}}$\\
$C6$& 2.44$\pm$ 0.13& 1.24$\pm$ 0.02&0.0363$\pm$0.0016&0.275$\pm$0.011&0.606$\pm$0.025&0.00/0.00&--&--&${2.4\ 10^{-1}}$/${8.3\ 10^{-1}}$\\
$C7$& 2.43$\pm$ 0.03& 1.24$\pm$ 0.00&0.0358$\pm$0.0004&0.272$\pm$0.001&0.598$\pm$0.004&0.00/0.00&3.5&-7.6/1.00&${2.9\ 10^{-1}}$/${1.5\ 10^{0}}$\\
\hline
$D1$& 2.72$\pm$ 1.08& 1.18$\pm$ 0.02&0.0478$\pm$0.0055&0.310&0.688&0.00/0.00& --&--&${8.2\ 10^{-7}}$/\  \  --\\
$D2$a& 2.66$\pm$ 0.90& 1.19$\pm$ 0.02&0.0485$\pm$0.0051&0.311&0.630$\pm$0.104&0.00/0.00& --&--&${1.3\ 10^{-1}}$/${2.1\ 10^{-6}}$\\
$D2$b& 1.72$\pm$ 0.87& 1.21$\pm$ 0.02&0.0492$\pm$0.0050&0.312&0.560$\pm$0.114&0.00/0.00&5.5&-4.4/1.00&${1.5\ 10^{-1}}$/${1.4\ 10^{-3}}$\\
$D2$c& 2.09$\pm$ 1.02& 1.20$\pm$ 0.02&0.0488$\pm$0.0052&0.312&0.585$\pm$0.093&0.00/0.00&5.5&-3.9/1.00&${2.4\ 10^{-1}}$/${5.0\ 10^{-4}}$\\
$D3$& 2.01$\pm$ 1.08& 1.23$\pm$ 0.03&0.0487$\pm$0.0068&0.300$\pm$0.026&0.574$\pm$0.083&0.00/0.00&--&--&${2.6\ 10^{-1}}$/${1.6\ 10^{-2}}$\\
$D4$& 2.03$\pm$ 0.24& 1.14$\pm$ 0.02&0.0501$\pm$0.0049&0.348$\pm$0.017&0.583$\pm$0.040&0.00/0.00&5.5&-4.4/1.00&${5.6\ 10^{-3}}$/${6.2\ 10^{-6}}$\\
$D5$& 2.08$\pm$ 0.25& 1.18$\pm$ 0.06&0.0484$\pm$0.0050&0.323$\pm$0.045&0.620$\pm$0.159&0.00/0.00&--&--&${4.2\ 10^{-5}}$/${7.6\ 10^{-3}}$\\
$D6$& 2.28$\pm$ 0.12& 1.23$\pm$ 0.02&0.0452$\pm$0.0017&0.287$\pm$0.008&0.588$\pm$0.022&0.00/0.00&--&--&${3.9\ 10^{-1}}$/${8.1\ 10^{-1}}$\\
$D7$& 2.05$\pm$ 0.02& 1.23$\pm$ 0.00&0.0485$\pm$0.0007&0.297$\pm$0.001&0.594$\pm$0.004&0.00/0.00&4.3&-5.3/1.00&${3.0\ 10^{-1}}$/${2.0\ 10^{0}}$\\
\hline
$E1$& 4.50$\pm$ 1.70& 1.13$\pm$ 0.06&0.0405$\pm$0.0046&0.302&0.688&0.00/0.00& --&--&${5.6\ 10^{-3}}$/\  \  --\\
$E2$a& 1.40$\pm$ 0.99& 1.21$\pm$ 0.02&0.0404$\pm$0.0046&0.302&0.466$\pm$0.082&0.00/0.00& --&--&${1.9\ 10^{-1}}$/${3.9\ 10^{-4}}$\\
$E2$b& 1.26$\pm$ 0.89& 1.21$\pm$ 0.02&0.0405$\pm$0.0046&0.302&0.479$\pm$0.085&0.00/0.00&5.5&-5.7/1.00&${8.8\ 10^{-2}}$/${6.1\ 10^{-4}}$\\
$E2$c& 1.90$\pm$ 1.06& 1.19$\pm$ 0.02&0.0406$\pm$0.0046&0.302&0.516$\pm$0.072&0.00/0.00&5.5&-5.1/1.00&${1.8\ 10^{-1}}$/${2.2\ 10^{-3}}$\\
$E3$& 1.61$\pm$ 1.44& 1.22$\pm$ 0.03&0.0406$\pm$0.0046&0.290$\pm$0.025&0.488$\pm$0.093&0.00/0.00&--&--&${2.6\ 10^{-1}}$/${2.5\ 10^{-2}}$\\
$E4$& 1.79$\pm$ 0.28& 1.22$\pm$ 0.02&0.0416$\pm$0.0047&0.291$\pm$0.018&0.515$\pm$0.034&0.00/0.00&5.5&-4.9/1.00&${1.7\ 10^{-1}}$/${2.2\ 10^{0}}$\\
$E5$& 2.33$\pm$ 0.40& 1.24$\pm$ 0.04&0.0406$\pm$0.0047&0.269$\pm$0.023&0.633$\pm$0.093&0.00/0.00&--&--&${2.0\ 10^{-5}}$/${3.3\ 10^{-3}}$\\
$E6$& 2.46$\pm$ 0.08& 1.19$\pm$ 0.01&0.0462$\pm$0.0018&0.309$\pm$0.004&0.591$\pm$0.027&0.00/0.00&--&--&${5.8\ 10^{-1}}$/${9.4\ 10^{-1}}$\\
$E7$& 1.70$\pm$ 0.01& 1.23$\pm$ 0.00&0.0417$\pm$0.0007&0.287$\pm$0.001&0.506$\pm$0.003&0.00/0.00&3.8&-9.0/1.00&${4.3\ 10^{-1}}$/${7.1\ 10^{0}}$\\
\hline
$F1$& 2.85$\pm$ 1.18& 1.18$\pm$ 0.02&0.0482$\pm$0.0052&0.311&2.000&0.00/0.00& --&--&${1.5\ 10^{-6}}$/\  \  --\\
$F2$a& 3.52$\pm$ 0.96& 1.17$\pm$ 0.01&0.0482$\pm$0.0052&0.311&2.057$\pm$0.472&0.00/0.00& --&--&${6.7\ 10^{-2}}$/${1.2\ 10^{-2}}$\\
$F2$b& 2.38$\pm$ 0.93& 1.19$\pm$ 0.02&0.0491$\pm$0.0050&0.312&1.732$\pm$0.342&0.00/0.00&5.5&-4.1/1.00&${1.2\ 10^{-1}}$/${6.7\ 10^{-2}}$\\
$F2$c& 3.33$\pm$ 0.91& 1.17$\pm$ 0.02&0.0488$\pm$0.0050&0.312&2.008$\pm$0.331&0.00/0.00&5.5&-2.7/1.00&${2.0\ 10^{-1}}$/${1.1\ 10^{-5}}$\\
$F3$& 1.06$\pm$ 1.13& 1.23$\pm$ 0.02&0.0487$\pm$0.0079&0.311$\pm$0.027&1.397$\pm$0.270&0.00/0.00&--&--&${2.6\ 10^{-1}}$/${1.5\ 10^{-2}}$\\
$F4$& 2.08$\pm$ 0.25& 1.13$\pm$ 0.02&0.0508$\pm$0.0043&0.350$\pm$0.016&1.674$\pm$0.142&0.00/0.00&5.5&-4.9/1.00&${1.9\ 10^{-3}}$/${8.5\ 10^{-6}}$\\
$F5$& 2.17$\pm$ 0.31& 1.25$\pm$ 0.05&0.0466$\pm$0.0061&0.281$\pm$0.035&1.955$\pm$0.244&0.00/0.00&--&--&${1.1\ 10^{-6}}$/${4.0\ 10^{-3}}$\\
$F6$& 2.29$\pm$ 0.13& 1.25$\pm$ 0.02&0.0456$\pm$0.0017&0.277$\pm$0.007&1.690$\pm$0.090&0.00/0.00&--&--&${4.1\ 10^{-1}}$/${8.2\ 10^{-1}}$\\
$F7$& 2.67$\pm$ 0.02& 1.18$\pm$ 0.00&0.0477$\pm$0.0004&0.310$\pm$0.001&1.848$\pm$0.013&0.00/0.00&3.5&-7.2/1.00&${2.7\ 10^{-1}}$/${2.2\ 10^{0}}$\\
\hline
$G1$& 2.94$\pm$ 1.33& 1.18$\pm$ 0.03&0.0488$\pm$0.0057&0.312&0.688&0.00/0.00& --&--&${4.0\ 10^{-7}}$/\  \  --\\
$G2$a& 1.39$\pm$ 0.84& 1.22$\pm$ 0.02&0.0503$\pm$0.0088&0.313&0.514$\pm$0.073&0.00/0.00& --&--&${2.3\ 10^{-1}}$/${1.5\ 10^{-3}}$\\
$G2$b& 1.21$\pm$ 0.59& 1.22$\pm$ 0.02&0.0494$\pm$0.0186&0.312&0.515$\pm$0.050&0.00/0.00&5.5&-5.2/1.00&${1.4\ 10^{-1}}$/${4.3\ 10^{-4}}$\\
$G2$c& 2.51$\pm$ 0.99& 1.19$\pm$ 0.02&0.0499$\pm$0.0051&0.313&0.608$\pm$0.093&0.00/0.00&5.5&-3.7/1.00&${2.0\ 10^{-1}}$/${1.7\ 10^{-3}}$\\
$G3$& 1.12$\pm$ 1.15& 1.22$\pm$ 0.00&0.0502$\pm$0.0239&0.313$\pm$0.035&0.493$\pm$0.115&0.00/0.00&--&--&${2.6\ 10^{-1}}$/${2.0\ 10^{-2}}$\\
$G4$& 2.09$\pm$ 0.25& 1.13$\pm$ 0.02&0.0508$\pm$0.0056&0.348$\pm$0.021&0.581$\pm$0.028&0.00/0.00&3.8&-8.5/1.00&${7.5\ 10^{-3}}$/${6.4\ 10^{-1}}$\\
$G5$& 2.17$\pm$ 0.31& 1.25$\pm$ 0.04&0.0471$\pm$0.0059&0.281$\pm$0.033&0.667$\pm$0.082&0.00/0.00&--&--&${3.9\ 10^{-7}}$/${4.2\ 10^{-3}}$\\
$G6$& 2.33$\pm$ 0.12& 1.26$\pm$ 0.02&0.0477$\pm$0.0023&0.274$\pm$0.010&0.597$\pm$0.019&0.00/0.00&--&--&${3.6\ 10^{-1}}$/${8.0\ 10^{-1}}$\\
$G7$& 2.16$\pm$ 0.02& 1.27$\pm$ 0.00&0.0474$\pm$0.0007&0.271$\pm$0.001&0.586$\pm$0.004&0.00/0.00&4.0&-5.7/1.00&${4.2\ 10^{-1}}$/${1.7\ 10^{0}}$\\
\hline
$H1$& 2.90$\pm$ 1.77& 1.18$\pm$ 0.04&0.0483$\pm$0.0063&0.311&0.688&0.00/0.00& --&--&${7.4\ 10^{-8}}$/\  \  --\\
$H2$a& 1.63$\pm$ 0.86& 1.21$\pm$ 0.02&0.0494$\pm$0.0055&0.312&0.534$\pm$0.087&0.00/0.00& --&--&${2.0\ 10^{-1}}$/${4.7\ 10^{-4}}$\\
$H2$b& 1.87$\pm$ 0.78& 1.20$\pm$ 0.02&0.0491$\pm$0.0052&0.312&0.563$\pm$0.096&0.00/0.00&5.5&-4.5/1.00&${1.2\ 10^{-1}}$/${5.0\ 10^{-4}}$\\
$H2$c& 2.12$\pm$ 0.98& 1.20$\pm$ 0.02&0.0493$\pm$0.0052&0.312&0.582$\pm$0.087&0.00/0.00&5.5&-4.1/1.00&${2.2\ 10^{-1}}$/${9.4\ 10^{-4}}$\\
$H3$& 1.80$\pm$ 1.14& 1.23$\pm$ 0.01&0.0490$\pm$0.0053&0.301$\pm$0.020&0.550$\pm$0.097&0.00/0.00&--&--&${2.6\ 10^{-1}}$/${1.7\ 10^{-2}}$\\
$H4$& 2.09$\pm$ 0.25& 1.13$\pm$ 0.02&0.0509$\pm$0.0045&0.349$\pm$0.016&0.581$\pm$0.040&0.00/0.00&5.5&-4.7/1.00&${1.5\ 10^{-3}}$/${1.6\ 10^{-6}}$\\
$H5$& 2.17$\pm$ 0.34& 1.25$\pm$ 0.06&0.0467$\pm$0.0066&0.280$\pm$0.048&0.671$\pm$0.110&0.00/0.00&--&--&${1.4\ 10^{-6}}$/${4.0\ 10^{-3}}$\\
$H6$& 2.29$\pm$ 0.12& 1.27$\pm$ 0.02&0.0481$\pm$0.0023&0.273$\pm$0.010&0.605$\pm$0.021&0.00/0.00&--&--&${4.1\ 10^{-1}}$/${8.1\ 10^{-1}}$\\
$H7$& 2.19$\pm$ 0.02& 1.27$\pm$ 0.00&0.0486$\pm$0.0007&0.273$\pm$0.001&0.598$\pm$0.004&0.00/0.00&4.3&-5.0/1.00&${4.6\ 10^{-1}}$/${1.7\ 10^{0}}$\\
\hline
$I1$& 2.95$\pm$ 1.23& 1.18$\pm$ 0.03&0.0485$\pm$0.0053&0.311&0.688&0.15/0.00& --&--&${3.2\ 10^{-12}}$/\  \  --\\
$I2$a& 2.49$\pm$ 0.85& 1.19$\pm$ 0.02&0.0494$\pm$0.0051&0.312&0.602$\pm$0.096&0.15/0.00& --&--&${1.3\ 10^{-1}}$/${5.7\ 10^{-4}}$\\
$I2$b& 1.85$\pm$ 0.77& 1.20$\pm$ 0.02&0.0492$\pm$0.0055&0.312&0.560$\pm$0.094&0.15/0.00&5.5&-4.5/1.00&${1.3\ 10^{-1}}$/${4.8\ 10^{-4}}$\\
$I2$c& 2.07$\pm$ 0.97& 1.20$\pm$ 0.02&0.0494$\pm$0.0058&0.312&0.576$\pm$0.084&0.15/0.00&5.5&-4.2/1.00&${2.4\ 10^{-1}}$/${9.9\ 10^{-4}}$\\
$I3$& 1.99$\pm$ 1.33& 1.23$\pm$ 0.03&0.0490$\pm$0.0053&0.299$\pm$0.023&0.564$\pm$0.092&0.15/0.00&--&--&${2.6\ 10^{-1}}$/${1.5\ 10^{-2}}$\\
$I4$& 2.08$\pm$ 0.24& 1.18$\pm$ 0.02&0.0497$\pm$0.0049&0.320$\pm$0.016&0.577$\pm$0.038&0.15/0.00&5.5&-4.3/1.00&${8.0\ 10^{-2}}$/${9.3\ 10^{-4}}$\\
$I5$& 2.03$\pm$ 0.37& 1.39$\pm$ 0.02&0.0425$\pm$0.0047&0.204$\pm$0.004&0.703$\pm$0.121&0.15/0.00&--&--&${2.2\ 10^{-1}}$/${8.5\ 10^{-1}}$\\
$I6$& 2.22$\pm$ 0.11& 1.20$\pm$ 0.01&0.0482$\pm$0.0024&0.303$\pm$0.006&0.591$\pm$0.029&0.15/0.00&--&--&${2.9\ 10^{-1}}$/${4.5\ 10^{0}}$\\
$I7$& 2.17$\pm$ 0.02& 1.27$\pm$ 0.00&0.0488$\pm$0.0007&0.274$\pm$0.001&0.596$\pm$0.004&0.15/0.00&4.4&-4.8/1.00&${4.7\ 10^{-1}}$/${3.0\ 10^{0}}$\\
\hline
$J1$& 3.02$\pm$ 1.39& 1.18$\pm$ 0.03&0.0487$\pm$0.0056&0.311&0.688&1.80/0.00& --&--&${7.0\ 10^{-6}}$/\  \  --\\
$J2$a& 1.66$\pm$ 0.85& 1.21$\pm$ 0.02&0.0497$\pm$0.0066&0.313&0.532$\pm$0.082&1.80/0.00& --&--&${2.1\ 10^{-1}}$/${5.8\ 10^{-4}}$\\
$J2$b& 1.73$\pm$ 0.74& 1.21$\pm$ 0.02&0.0492$\pm$0.0061&0.312&0.548$\pm$0.089&1.80/0.00&5.5&-4.8/1.00&${1.4\ 10^{-1}}$/${4.1\ 10^{-4}}$\\
$J2$c& 2.06$\pm$ 0.91& 1.20$\pm$ 0.02&0.0495$\pm$0.0070&0.312&0.571$\pm$0.078&1.80/0.00&5.5&-4.4/1.00&${2.5\ 10^{-1}}$/${4.7\ 10^{-4}}$\\
$J3$& 1.99$\pm$ 0.95& 1.23$\pm$ 0.01&0.0493$\pm$0.0057&0.301$\pm$0.019&0.560$\pm$0.099&1.80/0.00&--&--&${2.6\ 10^{-1}}$/${1.4\ 10^{-2}}$\\
$J4$& 2.02$\pm$ 0.22& 1.13$\pm$ 0.02&0.0517$\pm$0.0048&0.356$\pm$0.015&0.569$\pm$0.041&1.80/0.00&5.5&-5.1/1.00&${5.5\ 10^{-7}}$/${1.2\ 10^{-5}}$\\
$J5$& 0.80$\pm$ 0.24& 1.43$\pm$ 0.07&0.0442$\pm$0.0052&0.200$\pm$0.048&0.890$\pm$0.282&1.80/0.00&--&--&${5.3\ 10^{-1}}$/${6.8\ 10^{0}}$\\
$J6$& 1.73$\pm$ 0.09& 1.21$\pm$ 0.02&0.0528$\pm$0.0022&0.315$\pm$0.008&0.585$\pm$0.020&1.80/0.00&--&--&${2.4\ 10^{-1}}$/${8.3\ 10^{0}}$\\
$J7$& 2.02$\pm$ 0.01& 1.23$\pm$ 0.00&0.0489$\pm$0.0008&0.296$\pm$0.001&0.566$\pm$0.003&1.80/0.00&4.2&-6.5/1.00&${3.1\ 10^{-1}}$/${5.5\ 10^{0}}$\\
\hline
$K1$& 2.87$\pm$ 1.22& 1.17$\pm$ 0.03&0.0448$\pm$0.0049&0.307&0.688&0.00/1.30& --&--&${1.2\ 10^{-5}}$/\  \  --\\
$K2$a& 2.43$\pm$ 0.86& 1.18$\pm$ 0.02&0.0452$\pm$0.0049&0.307&0.600$\pm$0.095&0.00/1.30& --&--&${1.5\ 10^{-1}}$/${1.1\ 10^{-3}}$\\
$K2$b& 1.70$\pm$ 0.74& 1.20$\pm$ 0.02&0.0450$\pm$0.0052&0.307&0.547$\pm$0.093&0.00/1.30&5.5&-4.7/1.00&${1.3\ 10^{-1}}$/${4.8\ 10^{-4}}$\\
$K2$c& 2.32$\pm$ 0.96& 1.19$\pm$ 0.02&0.0450$\pm$0.0049&0.307&0.592$\pm$0.090&0.00/1.30&5.0&-4.7/1.00&${2.4\ 10^{-1}}$/${5.4\ 10^{-4}}$\\
$K3$& 2.11$\pm$ 1.06& 1.22$\pm$ 0.02&0.0450$\pm$0.0049&0.293$\pm$0.019&0.575$\pm$0.109&0.00/1.30&--&--&${2.7\ 10^{-1}}$/${1.2\ 10^{-2}}$\\
$K4$& 2.12$\pm$ 0.24& 1.12$\pm$ 0.02&0.0464$\pm$0.0044&0.347$\pm$0.016&0.577$\pm$0.044&0.00/1.30&5.5&-4.7/1.00&${1.3\ 10^{-3}}$/${1.6\ 10^{-6}}$\\
$K5$& 2.18$\pm$ 0.31& 1.25$\pm$ 0.03&0.0438$\pm$0.0050&0.275$\pm$0.023&0.682$\pm$0.075&0.00/1.30&--&--&${3.2\ 10^{-6}}$/${1.6\ 10^{-3}}$\\
$K6$& 1.92$\pm$ 0.07& 1.24$\pm$ 0.01&0.0489$\pm$0.0014&0.292$\pm$0.003&0.603$\pm$0.012&0.00/1.30&--&--&${1.4\ 10^{0}}$/${8.9\ 10^{-1}}$\\
$K7$& 2.14$\pm$ 0.01& 1.22$\pm$ 0.00&0.0494$\pm$0.0006&0.301$\pm$0.001&0.592$\pm$0.003&0.00/1.30&3.8&-6.5/1.00&${6.1\ 10^{-1}}$/${2.7\ 10^{0}}$\\
\hline
\end{longtable}
}

%---------------------------------------------------------------------------
% In the text, at the place where the large table should appear
% add the command:
%\addtocounter{table}{1}
% Tables counters will be well numbered.
% If table 2
\longtab{5}{
\begin{longtable}{lcccccccccccccc}
\caption{Same as Table~\ref{resultsA2}, but for different input physics of the models (see Sect.~\ref{LM} and Tables~\ref{modelinputs} and \ref{cases}).}
\label{resultsB2}
\\
\hline\hline
Model &   $T_\mathrm{{eff}}$ &$L$ & [Fe/H] & $\log g$ &  $R$& 
 $\langle\Delta\nu\rangle$ & $\nu_\mathrm{max}$& $\langle r_{02}\rangle$ &$\langle rr_{01/10}\rangle$  &$X_C$ & $\frac{\Delta Y}{\Delta Z}$& $R_\mathrm{{cc}}$ & $R_\mathrm{{zc}}$ & $M_\mathrm{{p}}\sin i$  \\
 & {[K]} & [$L_\odot$] & [dex] & [dex] & [$R_\odot$]  &   [$\mu$Hz] &  [$\mu$Hz] & -- &   --&-- & --&  [$R_\star$] &  [$R_\star$] &[$M_\mathrm{{Jupiter}}$] \\
\hline
\endfirsthead
\caption{continued.}\\
\hline\hline
Model &   $T_\mathrm{{eff}}$ &$L$ & [Fe/H] & $\log g$ &  $R$& 
 $\langle\Delta\nu\rangle$ & $\nu_\mathrm{max}$& $\langle r_{02}\rangle$ &$\langle rr_{01/10}\rangle$  &$X_C$ & $\frac{\Delta Y}{\Delta Z}$& $M_\mathrm{{cc}}$ & $R_\mathrm{{zc}}$ & $M_\mathrm{{p}}\sin i$  \\
 & {[K]} & [$L_\odot$] & [dex] & [dex] & [$R_\odot$]  &   [$\mu$Hz] &  [$\mu$Hz] & -- &   --&-- & --&  [$M_\star$] &  [$R_\star$] &[$M_\mathrm{{Jupiter}}$] \\
\hline
\endhead
\hline
\endfoot
$B1$&6116.&2.053& 0.22&4.30&1.28&101.45&2138.& 0.078& 0.034&0.31& 2.0&0.027&0.779&1.18$\pm$0.03\\
$B2$a&6038.&2.064& 0.22&4.28&1.32& 98.13&2072.& 0.088& 0.035&0.40& 2.0&0.020&0.821&1.19$\pm$0.03\\
$B2$b&6053.&2.063& 0.22&4.28&1.31& 98.13&2073.& 0.084& 0.034&0.36& 2.0&0.023&0.806&1.18$\pm$0.03\\
$B2$c&6053.&2.063& 0.22&4.28&1.31& 98.21&2073.& 0.084& 0.034&0.36& 2.0&0.023&0.806&1.18$\pm$0.03\\
$B3$&6022.&2.066& 0.22&4.28&1.32& 98.14&2087.& 0.087& 0.034&0.40& 1.6&0.017&0.815&1.20$\pm$0.03\\
$B4$&6116.&2.054& 0.22&4.27&1.28& 98.14&2028.& 0.084& 0.035&0.33& 3.4&0.034&0.812&1.14$\pm$0.03\\
$B5$&6116.&2.053& 0.22&4.32&1.28&104.12&2250.& 0.084& 0.033&0.38& 1.1&0.011&0.786&1.22$\pm$0.05\\
$B6$&6016.&2.062& 0.21&4.29&1.32& 98.28&2109.& 0.084& 0.034&0.37& 1.1&0.013&0.802&1.21$\pm$0.03\\
$B7$&6012.&2.046& 0.23&4.29&1.32& 98.28&2099.& 0.081& 0.034&0.34& 1.3&0.019&0.792&1.21$\pm$0.03\\
\hline
$C1$&6116.&2.053& 0.22&4.30&1.28&101.39&2139.& 0.104& 0.034&0.26& 2.0&0.011&0.784&1.18$\pm$0.04\\
$C2$a&6057.&2.062& 0.22&4.28&1.31& 98.13&2054.& 0.074& 0.035&0.24& 2.0&0.016&0.785&1.17$\pm$0.03\\
$C2$b&6048.&2.063& 0.22&4.29&1.31& 98.14&2111.& 0.091& 0.035&0.43& 2.0&0.007&0.837&1.20$\pm$0.03\\
$C2$c&6048.&2.063& 0.22&4.29&1.31& 98.20&2112.& 0.091& 0.035&0.43& 2.0&0.007&0.837&1.20$\pm$0.03\\
$C3$&6021.&2.066& 0.22&4.28&1.32& 98.14&2088.& 0.081& 0.034&0.31& 1.3&0.004&0.807&1.21$\pm$0.04\\
$C4$&6108.&2.055& 0.22&4.28&1.28& 98.13&2046.& 0.084& 0.036&0.32& 3.6&0.018&0.819&1.15$\pm$0.03\\
$C5$&6117.&2.053& 0.22&4.32&1.28&103.73&2244.& 0.083& 0.033&0.34& 1.0&0.000&0.798&1.21$\pm$0.04\\
$C6$&6024.&2.062& 0.23&4.29&1.32& 98.45&2119.& 0.083& 0.034&0.33& 1.1&0.003&0.807&1.21$\pm$0.03\\
$C7$&6013.&2.053& 0.22&4.29&1.32& 98.27&2118.& 0.083& 0.034&0.33& 1.0&0.001&0.808&1.22$\pm$0.03\\
\hline
$D1$&6116.&2.053& 0.22&4.30&1.28&101.73&2143.& 0.075& 0.032&0.32& 2.0&0.043&0.769&1.18$\pm$0.03\\
$D2$a&6051.&2.063& 0.22&4.28&1.31& 98.13&2060.& 0.077& 0.032&0.33& 2.0&0.044&0.782&1.18$\pm$0.03\\
$D2$b&6050.&2.068& 0.22&4.29&1.31& 98.14&2093.& 0.089& 0.034&0.43& 2.0&0.036&0.818&1.19$\pm$0.03\\
$D2$c&6049.&2.065& 0.22&4.28&1.31& 98.20&2080.& 0.084& 0.034&0.39& 2.0&0.040&0.804&1.19$\pm$0.03\\
$D3$&6022.&2.066& 0.22&4.28&1.32& 98.14&2087.& 0.086& 0.034&0.41& 1.6&0.034&0.808&1.20$\pm$0.04\\
$D4$&6103.&2.056& 0.22&4.28&1.29& 98.13&2039.& 0.084& 0.033&0.36& 3.2&0.049&0.808&1.15$\pm$0.03\\
$D5$&6115.&2.053& 0.22&4.29&1.28&100.74&2128.& 0.084& 0.033&0.38& 2.4&0.040&0.798&1.17$\pm$0.05\\
$D6$&6020.&2.058& 0.19&4.29&1.32& 98.31&2100.& 0.083& 0.034&0.39& 1.3&0.030&0.800&1.21$\pm$0.03\\
$D7$&6047.&2.091& 0.22&4.29&1.32& 98.30&2102.& 0.085& 0.033&0.41& 1.5&0.036&0.803&1.21$\pm$0.03\\
\hline
$E1$&6103.&2.051& 0.22&4.27&1.28& 99.04&2030.& 0.063& 0.040&0.00& 2.0&0.032&0.756&1.14$\pm$0.05\\
$E2$a&6036.&2.064& 0.22&4.28&1.32& 98.13&2073.& 0.095& 0.035&0.46& 2.0&0.007&0.861&1.19$\pm$0.03\\
$E2$b&6062.&2.061& 0.22&4.29&1.30& 98.14&2114.& 0.096& 0.035&0.48& 2.0&0.008&0.859&1.19$\pm$0.03\\
$E2$c&6069.&2.062& 0.22&4.29&1.30& 98.20&2091.& 0.089& 0.035&0.38& 2.0&0.005&0.839&1.18$\pm$0.03\\
$E3$&6023.&2.065& 0.22&4.28&1.32& 98.14&2086.& 0.092& 0.035&0.44& 1.5&0.004&0.850&1.20$\pm$0.04\\
$E4$&6044.&2.063& 0.23&4.29&1.31& 98.17&2108.& 0.090& 0.035&0.41& 1.5&0.004&0.838&1.20$\pm$0.03\\
$E5$&6116.&2.053& 0.22&4.32&1.28&103.98&2251.& 0.083& 0.033&0.34& 0.7&0.000&0.799&1.22$\pm$0.04\\
$E6$&6107.&2.098& 0.28&4.29&1.30& 99.19&2096.& 0.082& 0.035&0.32& 2.0&0.016&0.810&1.18$\pm$0.03\\
$E7$&6015.&2.019& 0.23&4.29&1.31& 98.37&2127.& 0.091& 0.034&0.43& 1.4&0.003&0.839&1.20$\pm$0.03\\
\hline
$F1$&6116.&2.053& 0.22&4.29&1.28&101.37&2128.& 0.075& 0.033&0.29& 2.0&0.028&0.771&1.17$\pm$0.03\\
$F2$a&6069.&2.061& 0.22&4.28&1.30& 98.13&2043.& 0.066& 0.033&0.23& 2.0&0.039&0.751&1.16$\pm$0.03\\
$F2$b&6054.&2.063& 0.22&4.28&1.31& 98.14&2067.& 0.081& 0.034&0.33& 2.0&0.024&0.798&1.18$\pm$0.03\\
$F2$c&6071.&2.084& 0.22&4.27&1.31& 98.17&2035.& 0.069& 0.034&0.24& 2.0&0.038&0.759&1.17$\pm$0.03\\
$F3$&6022.&2.065& 0.22&4.28&1.32& 98.14&2087.& 0.097& 0.035&0.51& 2.0&0.017&0.852&1.20$\pm$0.03\\
$F4$&6108.&2.055& 0.22&4.27&1.28& 98.13&2034.& 0.084& 0.035&0.33& 3.2&0.033&0.811&1.14$\pm$0.03\\
$F5$&6116.&2.053& 0.22&4.32&1.28&104.16&2251.& 0.083& 0.033&0.38& 1.0&0.011&0.786&1.22$\pm$0.04\\
$F6$&6009.&2.061& 0.19&4.29&1.33& 98.28&2110.& 0.083& 0.034&0.37& 0.9&0.010&0.802&1.22$\pm$0.03\\
$F7$&6085.&2.097& 0.21&4.28&1.31& 98.29&2061.& 0.077& 0.034&0.30& 2.0&0.027&0.786&1.18$\pm$0.03\\
\hline
$G1$&6116.&2.053& 0.22&4.29&1.28&101.33&2126.& 0.073& 0.033&0.28& 2.0&0.030&0.767&1.17$\pm$0.04\\
$G2$a&6029.&2.065& 0.22&4.28&1.32& 98.14&2080.& 0.093& 0.035&0.46& 2.0&0.019&0.838&1.20$\pm$0.03\\
$G2$b&6049.&2.064& 0.22&4.29&1.31& 98.14&2113.& 0.095& 0.035&0.48& 2.0&0.018&0.842&1.20$\pm$0.03\\
$G2$c&6056.&2.063& 0.22&4.28&1.31& 98.19&2064.& 0.080& 0.034&0.32& 2.0&0.026&0.792&1.18$\pm$0.03\\
$G3$&6022.&2.066& 0.22&4.28&1.32& 98.14&2086.& 0.096& 0.035&0.49& 2.0&0.019&0.849&1.20$\pm$0.03\\
$G4$&6114.&2.055& 0.21&4.28&1.28& 98.30&2046.& 0.084& 0.035&0.34& 3.2&0.033&0.811&1.14$\pm$0.03\\
$G5$&6116.&2.054& 0.22&4.32&1.28&104.10&2251.& 0.083& 0.033&0.38& 1.0&0.012&0.787&1.22$\pm$0.04\\
$G6$&6010.&2.073& 0.21&4.29&1.33& 98.29&2122.& 0.083& 0.034&0.37& 0.8&0.012&0.798&1.23$\pm$0.03\\
$G7$&6007.&2.080& 0.21&4.29&1.33& 98.28&2128.& 0.085& 0.034&0.39& 0.7&0.011&0.804&1.23$\pm$0.03\\
\hline
$H1$&6116.&2.054& 0.22&4.29&1.28&101.32&2126.& 0.074& 0.033&0.28& 2.0&0.030&0.769&1.17$\pm$0.04\\
$H2$a&6035.&2.065& 0.22&4.28&1.32& 98.13&2075.& 0.090& 0.035&0.43& 2.0&0.019&0.828&1.19$\pm$0.03\\
$H2$b&6053.&2.062& 0.22&4.28&1.31& 98.14&2087.& 0.087& 0.034&0.39& 2.0&0.020&0.817&1.19$\pm$0.03\\
$H2$c&6054.&2.064& 0.22&4.28&1.31& 98.20&2077.& 0.084& 0.034&0.36& 2.0&0.022&0.807&1.18$\pm$0.03\\
$H3$&6022.&2.066& 0.22&4.28&1.32& 98.14&2086.& 0.089& 0.034&0.41& 1.6&0.017&0.820&1.20$\pm$0.03\\
$H4$&6109.&2.054& 0.22&4.28&1.28& 98.13&2036.& 0.084& 0.035&0.33& 3.2&0.032&0.810&1.14$\pm$0.03\\
$H5$&6116.&2.053& 0.22&4.32&1.28&104.17&2253.& 0.083& 0.033&0.38& 1.0&0.011&0.786&1.22$\pm$0.05\\
$H6$&6014.&2.086& 0.22&4.29&1.33& 98.38&2125.& 0.083& 0.034&0.38& 0.8&0.013&0.797&1.23$\pm$0.03\\
$H7$&6011.&2.089& 0.23&4.29&1.34& 98.29&2125.& 0.084& 0.034&0.39& 0.8&0.013&0.800&1.23$\pm$0.03\\
\hline
$I1$&6116.&2.053& 0.22&4.29&1.28&101.29&2127.& 0.072& 0.021&0.38& 2.0&0.088&0.769&1.17$\pm$0.03\\
$I2$a&6050.&2.063& 0.22&4.28&1.31& 98.13&2061.& 0.078& 0.024&0.42& 2.0&0.086&0.796&1.18$\pm$0.03\\
$I2$b&6051.&2.063& 0.22&4.28&1.31& 98.14&2089.& 0.087& 0.027&0.48& 2.0&0.082&0.818&1.19$\pm$0.03\\
$I2$c&6053.&2.064& 0.22&4.28&1.31& 98.19&2080.& 0.084& 0.026&0.46& 2.0&0.083&0.810&1.19$\pm$0.03\\
$I3$&6022.&2.066& 0.22&4.28&1.32& 98.14&2087.& 0.085& 0.027&0.48& 1.6&0.078&0.813&1.20$\pm$0.04\\
$I4$&6065.&2.062& 0.22&4.28&1.30& 98.14&2072.& 0.083& 0.026&0.45& 2.3&0.086&0.810&1.18$\pm$0.03\\
$I5$&6044.&2.066& 0.20&4.34&1.31&106.11&2392.& 0.086& 0.031&0.49& 0.0&0.000&0.775&1.31$\pm$0.04\\
$I6$&6041.&2.020& 0.22&4.29&1.30& 99.37&2119.& 0.082& 0.026&0.46& 1.7&0.078&0.801&1.19$\pm$0.03\\
$I7$&6011.&2.090& 0.23&4.29&1.34& 98.30&2126.& 0.083& 0.027&0.49& 0.8&0.071&0.801&1.23$\pm$0.03\\
\hline
$J1$&6116.&2.053& 0.22&4.30&1.28&101.35&2130.& 0.071& 0.012&0.42& 2.0&0.125&0.769&1.17$\pm$0.04\\
$J2$a&6031.&2.065& 0.22&4.28&1.32& 98.13&2078.& 0.090& 0.024&0.52& 2.0&0.121&0.829&1.20$\pm$0.03\\
$J2$b&6048.&2.061& 0.22&4.29&1.31& 98.14&2097.& 0.089& 0.023&0.51& 2.0&0.121&0.824&1.19$\pm$0.03\\
$J2$c&6051.&2.063& 0.22&4.28&1.31& 98.19&2085.& 0.085& 0.021&0.49& 2.0&0.122&0.812&1.19$\pm$0.03\\
$J3$&6022.&2.066& 0.22&4.28&1.32& 98.14&2087.& 0.086& 0.022&0.51& 1.6&0.119&0.815&1.20$\pm$0.03\\
$J4$&6116.&2.053& 0.22&4.28&1.28& 98.13&2037.& 0.084& 0.020&0.44& 3.4&0.134&0.817&1.14$\pm$0.03\\
$J5$&6228.&2.022& 0.24&4.42&1.22&120.50&2807.& 0.097& 0.031&0.70& 0.0&0.085&0.760&1.34$\pm$0.05\\
$J6$&6081.&2.057& 0.26&4.30&1.29&100.14&2149.& 0.089& 0.023&0.51& 2.0&0.122&0.813&1.20$\pm$0.03\\
$J7$&6014.&2.039& 0.22&4.29&1.32& 98.28&2113.& 0.086& 0.022&0.51& 1.5&0.116&0.811&1.21$\pm$0.03\\
\hline
$K1$&6116.&2.054& 0.22&4.29&1.28&101.54&2124.& 0.076& 0.034&0.29& 2.0&0.021&0.687&1.17$\pm$0.04\\
$K2$a&6046.&2.063& 0.22&4.28&1.31& 98.13&2064.& 0.086& 0.034&0.37& 2.0&0.015&0.746&1.18$\pm$0.03\\
$K2$b&6051.&2.064& 0.22&4.28&1.31& 98.14&2085.& 0.090& 0.035&0.42& 2.0&0.014&0.762&1.19$\pm$0.03\\
$K2$c&6055.&2.062& 0.22&4.28&1.31& 98.20&2063.& 0.082& 0.034&0.34& 2.0&0.018&0.740&1.18$\pm$0.03\\
$K3$&6022.&2.066& 0.22&4.28&1.32& 98.14&2087.& 0.086& 0.034&0.39& 1.4&0.011&0.744&1.20$\pm$0.03\\
$K4$&6109.&2.054& 0.22&4.27&1.28& 98.13&2025.& 0.084& 0.035&0.33& 3.4&0.027&0.745&1.14$\pm$0.03\\
$K5$&6116.&2.053& 0.22&4.32&1.28&104.40&2247.& 0.083& 0.033&0.38& 1.0&0.007&0.700&1.22$\pm$0.04\\
$K6$&6061.&2.121& 0.26&4.29&1.32& 98.46&2111.& 0.086& 0.034&0.40& 1.4&0.018&0.745&1.22$\pm$0.03\\
$K7$&6011.&2.033& 0.26&4.28&1.32& 98.31&2089.& 0.084& 0.034&0.37& 1.6&0.017&0.727&1.20$\pm$0.03\\
\hline
\end{longtable}
}
\end{appendix}
\end{document}